\documentclass[twoside,12pt]{article}
\usepackage{epsfig}
\input boxedeps.tex
\SetRokickiEPSFSpecial  
\HideDisplacementBoxes

\def\Journal#1#2#3#4{{#1} {#2} (#4) #3 }
\def\NCA{{\em Nuovo Cimento} A}

\def\PRO{{\em Prog. Theor. Phys.}}
\def\NPB{{\em Nucl. Phys.} B}

\def\PLB{{\em Phys. Lett.} B}

\def\PRL{\em Phys. Rev. Lett.}
\def\PREV{\em Phys. Rev.}
\def\PREP{\em Phys. Rep.}

\def\PRD{{\em Phys. Rev.} D}
\def\PRC{{\em Phys. Rev.} C}
\def\PRB{{\em Phys. Rev.} B}
\def\ZPC{{\em Z. Phys.} C}

\def\ANNP{\em Ann. Phys. (N.Y.)}
\def\RMP{{\em Rev. Mod. Phys.}}

\def\Prod{\Pi }

\newcommand{\D}{\overline{D}}
\newcommand{\da}{\dagger}  
\newcommand{\be}{\begin{equation}}
\newcommand{\eq}{\end{equation}}
\newcommand{\ee}{\end{equation}}
\newcommand{\bea}{\begin{eqnarray}}
\newcommand{\eea}{\end{eqnarray}}
   
\newcommand{\Tr}{{\rm \, Tr \!}}    
\newcommand{\newl}{l}               
\newcommand{\newtau}{t}             
  
\begin{document}
\title{
The Relativistic Bound State Problem in QCD:\\ 
Transverse Lattice Methods\\
}

\author{Matthias Burkardt
\\
Department of Physics,\\ New Mexico State University
, Las Cruces, NM 88003, U.S.A.
\\
\\
Simon Dalley\\
Centre for Mathematical Sciences,\\ 
Cambridge University, Wilberforce Road
Cambridge CB3 0WA, England}

\maketitle

\begin{abstract} 
The formalism for describing hadrons using 
a light-cone Hamiltonian of $SU(N)$ gauge theory
on a coarse transverse lattice is reviewed. Physical gauge
degrees of freedom are represented by disordered flux fields on the
links of the lattice.
A renormalised light-cone
Hamiltonian is obtained by making a colour-dielectric
expansion for the link-field interactions.
Parameters in the Hamiltonian are renormalised
non-perturbatively  by
seeking regions in parameter space with enhanced
Lorentz symmetry. In the case of pure gauge theories
to lowest non-trivial order of the colour-dielectric expansion,
this is sufficient to determine accurately
all parameters  in the large-$N$ limit.
We summarize results from applications to  glueballs.
After quarks are added, the Hamiltonian and Hilbert space
are expanded in both dynamical fermion and link fields. 
Lorentz and chiral symmetry are not sufficient to 
accurately determine all parameters
to lowest non-trivial order of these expansions.   
However, Lorentz symmetry and   one phenomenological 
input, a chiral symmetry breaking scale, are enough
to fix all parameters unambiguously.
Applications to light-light and
heavy-light mesons are described.

\end{abstract}
\tableofcontents
\section{Introduction}
Relativistic strongly-bound states in generic
four-dimensional gauge theories represent a formidable theoretical challenge.
Future progress in particle physics is likely
to hinge upon a detailed theoretical understanding of these questions,
since they are both of intrinsic interest in hadronic and nuclear
physics, and important for the proper identification of new physics
beyond the Standard Model.
This has led  to the development of 
Hamiltonian quantisation on a light-front.
In the presence of
suitable high-energy cut-offs, the light-cone vacuum state is trivial and  
constituent-like wavefunctions can be built upon it in a 
Lorentz boost-invariant fashion.
In fact, these light-cone wavefunctions 
carry all the non-perturbative, coherent
information about bound and scattering states in a general 
Lorentz frame. 
This review is about a particular non-perturbative 
method of formulating and solving
the light-cone Hamiltonian problem of non-abelian
gauge theories --- the transverse lattice method \cite{bard1} 
--- with emphasis on
QCD and hadronic structure. It is particularly well suited to
the scale of the strong interactions between short distances described by 
asymptotically free QCD and the relatively long range phenomena of
nuclear forces. A successful description
of this intermediate region should not only include `global' properties
of hadrons, such as masses and decay constants,
but also the wealth of 
experimental data on hadronic sub-structure, such as structure
functions, form factors, and distribution amplitudes.
All of these observables are simply related to light-cone wavefunctions.

The transverse lattice method brings together two powerful ideas
in quantum field theory: light-cone Hamiltonian quantisation
and lattice gauge theory. For readers unfamiliar with the light-cone 
framework, we provide a brief introduction to its salient points in
the rest of this chapter. There exist a number of recent and more extensive
reviews on the theory and applications of light-cone quantisation
\cite{review,mb:adv}.  
For this reason we omit any detailed description of light-cone 
approaches other
than the transverse lattice in this article. 
No special knowledge of lattice gauge theory is needed for
this review, although an understanding at the level of introductory
chapters of textbooks \cite{books} would no doubt be useful.

An important ingredient in the formulation of transverse lattice 
gauge theory we expound here, is the colour-dielectric expansion.
We characterise a dielectric formulation as one in which physical gluon fields,
or rather the $SU(N)$ group elements they generate, are replaced by 
collective variables which represent an average over the 
fluctuations on short distance scales.
These dielectric variables carry colour and form an effective gauge
field theory
with classical action minimised at zero field, 
meaning that colour flux  is expelled from the vacuum at the classical level. 
The price one pays for starting with a simple vacuum structure,
which  arises only for a rather low momentum cut-off on the effective theory,
is that the effective cut-off Hamiltonian is initially poorly
constrained. We will demonstrate, however, 
that the colour-dielectric expansion,
together with requirements of symmetry restoration, is sufficient to
organise the interactions in the Hamiltonian in a way suitable for
practical solution. To date, all such solutions have been obtained in
the limit of large number of colours $N \to \infty$ of $SU(N)$ gauge
theories. For this reason, the kinds of bound state that have been
investigated have been limited to glueballs and mesons. 
Although the $N \to \infty$ limit  simplifies calculations, it is not
necessary, and in principle baryons can also be treated in the same
framework.

A number of more well-established techniques exist for studying
boundstate problems of quarks and gluons in non-abelian gauge theory,
not to mention the many possible approaches that treat some or all
of the boundstates as effective degrees of freedom in themselves.
Typically, there is a (computational) 
trade-off between the output detail that one can 
predict and the amount of information one must input from experiment;
one is always hoping for a high output/input ratio.
Lattice gauge theory, of which there are many variants,
 treats the problems from first
principles. This means that symmetry alone determines the output of
the calculation and no data are taken from experiment, aside from
parameters that are undetermined in QCD (such as the confinement scale or quark
masses). However, the number of things one can reasonably compute in 
practice is much less than one would like.
One advantage of a lattice method that computes light-cone
wavefunctions, such as the transverse lattice gauge theory, is that
a single calculation leads to estimates of a wide range of observables.

Boundstate calculations are made easier if one is prepared to take some
discrete information from experiment, such as values of condensates as
used in QCD sum rules \cite{cz}.  Still greater 
predictive power is obtained if aspects of the
gauge theory are modelled, for example by using models of the vacuum
\cite{vac} to 
determine condensates used in sum rules, or models of vertex
functions and propagators to be used in truncations of 
Dyson-Schwinger equations \cite{ds}.
We will suggest that the colour-dielectric formulation
of transverse lattice gauge theory can, with reasonable effort,
 yield fairly accurate output
from first principles, or at worst from a small amount  of discrete
experimental input. On the other hand, it contains within its
solutions the entire amplitude structure of hadrons at resolutions of
order 1/2 fermi and above; it has a high output/input ratio.

\subsection{\it Boundstates in light-cone co-ordinates}
\label{lccoord}
We denote the co-ordinate four vector
$(t,{\vec x}) = (x^0,(x^1,x^2,x^3))$.
In light cone co-ordinates $x^\pm=x_{\mp}=(x^0\pm x^3)/\sqrt{2}$,
${\bf x}=(x^1, x^2)$,  and $x^+$ is treated as the canonical `time' variable,
i.e. the (light-cone) wavefunction is defined on a null-plane
$x^+ = {\rm constant}$. The Lorentz indices $\mu, \nu \in \{ 0,1,2,3 \}$ are
split into light-cone indices 
$\alpha,\beta \in \{+,-\}$ and transverse indices 
$r,s\in \{1,2\}$ (we automatically sum over repeated 
raised and lowered indices). 
Since $x^3$ is an arbitrarily chosen space
direction, this formulation lacks manifest rotational
invariance. However,
the null-plane possesses manifest boost invariance, this being
ultimately more useful for relativistic problems.
One might imagine that the light-cone framework therefore bears no
relation to nonrelativistic quantum 
mechanics (NRQM). However, there is an isomorphism between the
structure of many-particle states in both cases \cite{susskind}, which is the
first hint of the utility of light-cone co-ordinates in the
description of relativistic boundstates.

Boost (Galilei) transformations in NRQM
\begin{eqnarray}
{\vec x}^\prime &=&{\vec x} + {\vec v}t
\\
t^\prime&=&t
\end{eqnarray}
are purely kinematic because they leave the
quantization surface $t=0$ invariant. This property
has important consequences  for a many body system. For example,
wavefunctions in the rest 
frame and in a boosted frame are related by a simple
shift of (momentum) variables ${\vec p}$, e.g. for a two-body system boosted
by velocity ${\vec v}$ the wavefunction behaves as
\be
\psi_{\vec v}({\vec p}_1,{\vec p}_2)
=\psi_{\vec 0}({\vec p}_1-m_1{\vec v},
{\vec p}_2-m_2{\vec v}).
\label{eq:nrboost}
\eq
Furthermore, if the Hamiltonian is translationally
invariant, the dynamics of the {\sl center of mass}
\be
{\vec R}=\sum_i x_i {\vec r}_i,
\eq
with $x_i = m_i/M$ the mass fraction of particle $i$ and $M=\sum_i m_i$,
separates from the intrinsic variables, making
it possible to work in the center of mass frame.

One of the features that normally complicates the 
description of relativistic bound states is that
equal $t$ hyperplanes are not invariant under
relativistic boosts
\begin{eqnarray}
{\bf x}^\prime &=&{\bf x}
\\
x^{3\prime} &=&
\gamma\left(x^3  + v x^0\right)
\\
x^{0\prime}&=&\gamma\left(x^0+\frac{v}{c^2}x^3\right),
\eea
with $\gamma^{-2}=1-\frac{v^2}{c^2}$.
As a result,
boosts are in general a dynamical operation; the
generator of boost transformations contains
interactions and there exists no simple
generalization of Eq. (\ref{eq:nrboost}) to
a relativistic system quantized at equal $x^0$.
Furthermore, the notion of the center of mass has
no useful generalization in such an equal-$x^0$
quantized relativistic framework.

One of the advantages of the light-cone framework, where quantisation
is performed at equal-$x^+$ null-planes,
is that
there is a subgroup of kinematical boosts among
the generators of the Poincar\'e group \cite{soper}. 
To see this
let us start from the usual Poincar\'e algebra
\begin{eqnarray}
\left[P^\mu,P^\nu\right]&=&0 \label{eq:poincare}\\
\left[M^{\mu\nu},P^\rho\right] &=& i\left(
g^{\nu \rho}P^\mu - g^{\mu \rho}P^\nu\right)
\\
\left[M^{\mu \nu}, M^{\rho \lambda}\right]
&=&i\left(g^{\mu \lambda}M^{\nu \rho}
+g^{\nu \rho}M^{\mu \lambda} 
-g^{\mu \rho}M^{\nu \lambda}
-g^{\nu \lambda}M^{\mu \rho}\right) 
\eea
where the generators of rotations and boosts are
respectively
$J_i={1 \over 2} \varepsilon_{ijk}M_{jk}$ and $K_i=M_{i0}$.
We now introduce transverse boost-rotation 
operators
\begin{eqnarray}
B_1 &=& \frac{1}{\sqrt{2}}\left(K_1+J_2\right) 
\\ 
B_2 &=& \frac{1}{\sqrt{2}}\left(K_2-J_1\right) 
\label{eq:perpboost}
\eea
so $B_r = -M_{-r}$. It follows from the Poincar\'e algebra 
 that these satisfy commutation relations
\begin{eqnarray}
\left[J_3,B_r\right]&=& i\varepsilon_{rs}B_s
\\
\left[P_r,B_s\right] &=& -i\delta_{rs}P^+
\\
\left[P^-,B_r\right] &=& -iP_r 
\\
\left[P^+,B_r\right]&=& 0
\eea
where $\varepsilon_{12}=-\varepsilon_{21}=1$, and $\varepsilon_{11}=
\varepsilon_{22}=0$.
Together with the well known commutation relations
\begin{eqnarray}
\left[J_3,P_r\right] &=& i\varepsilon_{rs}P_s
\\
\left[P^-,P_r\right] &=& \left[P^-,P^+\right]
=\left[P^-,J_3\right] = 0 \\
\left[P^+,P_r\right] &=& \left[P^+,B_r\right]
=\left[P^+,J_3\right]=0 
\eea
these are the same relations as
the commutators among the generators of 
the Galilei transformations for NRQM in the plane, 
provided we make the identifications
\begin{eqnarray}
P^-&\longrightarrow& \mbox{Hamiltonian}
 \\
{\bf P}=(P^1,P^2) &\longrightarrow& \mbox{momentum in the
plane}
\\
 P^+ &\longrightarrow& \mbox{mass}
\\
J_3 &\longrightarrow& \mbox{rotations around $x^3$-axis}
\\
{\bf B}=(B^1,B^2) &\longrightarrow&
\mbox{generator of boosts in the plane}
 ,
\eea	
i.e. $e^{i{\bf v} \cdot {\bf B}}
{\bf P} e^{-i{\bf v} \cdot {\bf B}}
= {\bf P}+P^+{\bf v}$.
Because of this isomorphism between transverse boost-rotations
in light-cone coordinates and boosts in the
context of NRQM in the plane, many familiar results
from NRQM can be directly carried over to 
relativistic systems.
In particular, since the longitudinal
momentum fractions $x_i \equiv k^+_i/P_{total}^+$
play a role very similar to the mass fractions
$m_i/M_{total}$ in NRQM, it is very natural that
we find as a reference point for distributions
in the transverse plane the {\sl transverse center
of momentum}
\be
{\bf R} = \frac{\sum_i k^+_i {\bf r}_{i}} 
{P^+_{total}}  = \sum_i x_i {\bf r}_{i},
\ee
where $k^{+}_{i}$ and ${\bf r}_{i}$ are the longitudinal momentum
and transverse position of the $i^{\rm th}$ particle respectively.
Moreover, the fractions $x_i$, like masses in NRQM, are invariant
under all the boosts $K_i$ ($K_3$ rescales longitudinal momentum).

The   Hamiltonian, which  
evolves the  light-cone wavefunction in light-cone time $x^+$, is
$P^- = (P^0 - P^3)/\sqrt{2}$, where $P^0$ is the usual energy.
Both $P^-$ and the light-cone momentum $P^+ = (P^0 + P^3)/\sqrt{2}$
are positive definite for massive particles. This simple observation
has the important consequence that the vacuum state, which has $P^+ =
0$, cannot mix with massive particle (parton) degrees of freedom. If a theory
can be formulated in terms of massive degrees of freedom, the
light-cone vacuum is therefore trivially empty. Usually it is possible to
achieve this situation in gauge theories by applying  high-energy
cut-offs.\footnote{A caveat involving the light-cone gauge fixing is
described later.} 

In the Schrodinger picture, a stable relativistic boundstate (such as 
a hadron) can be
represented as a state  $|\psi(P)\rangle$ at a particular canonical
time,
which is an eigenvector of the Hamiltonian generator of time
evolution.
In light-cone quantisation, $|\psi(P)\rangle$ is an eigenvector of
$P^-$, defined at $x^+ = {\rm constant}$, and is labelled by momenta
$(P^+, {\bf P})$.
It can be
expanded in the light-cone Fock space of its parton constituents.
An important consideration is the extent to which this expansion
converges, since in a highly relativistic system particle production can be
copious.
If the  Fock space 
amplitude is denoted $\psi_n(x_i,{\bf k}_{ i},s_i)$
for $n$ partons carrying longitudinal momentum fractions
$x_i = k^{+}_{i}/P^+$, transverse momenta  ${\bf k}_{ i}$,
 and helicities $s_{i}$, then
\be
|\psi(P)\rangle = \sum_n  
\int \left[ \Prod_{i=1}^n d^2{\bf k}_{i} dx_i \right]
\delta\left(1-\sum_{i} x_i\right) 
\delta\left({\bf P} -\sum_{i}{\bf k}_{i}\right)
\sum_{s_i} \psi_n(x_i,{\bf k}_{i},s_i) 
|n;x_1,{\bf k}_{1},s_1,..., x_n, {\bf k}_{n}, s_n
\rangle \ .
\label{fockexp}
\ee
The multi-parton structure of this wavefunction  is much simpler than in other
quantization schemes. As indicated above, one can choose high-energy
cut-offs such that it contains no disconnected vacuum
contributions. The convergence with $n$ of the Fock expansion
Eq.~(\ref{fockexp}) is usually very fast in this case
for low-mass states $|\psi(P) \rangle$.
The parton constituents appearing in
Eq.~(\ref{fockexp})
are on their mass shell
\be
k^{-}_{i} = {{\bf k}^{2}_{i} + m^{2}_{i} \over 2k^{+}_{i}}
\eq
but off the light-cone energy shell $P^- < \sum_{i} k^{-}_{i}$. Since
$P^+ = \sum_{i} k^{+}_{i}$ and ${\bf P} = \sum_{i} {\bf k}_{i}$, the light-cone
energy of a Fock state contribution
typically rises like the square of the number of
constituents rather than the number of constituents. The lowest energy
boundstates are then dominated by just a few constituents.

In order to take advantage of these special 
properties of light-cone coordinates in the description of hadrons,
it is necessary to 
formulate QCD in a light-cone Hamiltonian framework.
In this review, we are going to outline
the concept of the transverse lattice regulator, which is tailored to
suit the manifest symmetries of the light-cone framework.
As in other lattice gauge theories, this high-energy cut-off 
naturally exhibits linear confinement in the
bare Hamiltonian. It  will also facilitate the introduction of massive
elementary degrees of freedom.
The rapid convergence of the Fock space expansion in constituents
is explicitly realised (for the pure glue theory at least).
A number of illustrative applications to the boundstate physics of glueballs,
light and heavy mesons will be given.

\subsection{\it Physical motivation for light-cone framework}
The physical origin of interest in light-cone formulations of
quantum field theory can be traced to the fact that 
many high-energy scattering experiments probe hadron
structure exceedingly close to the light-cone. 
For example, in deep inelastic
electron-nucleon scattering experiments 
(DIS) one measures the inclusive cross section 
(Fig.~\ref{fig:dis})
\be
\frac{d^2\sigma}{dE^\prime d\Omega}
= \frac{\alpha^2}{4E^2 \sin^4\frac{\theta}{2}}\left[ 
    W_2(Q^2,\nu) \cos^2 \frac{\theta}{2} 
+ 2 W_1(Q^2,\nu) \sin^2\frac{\theta}{2}\right], 
\label{eq:dis}
\eq
where $P$ is the momentum of the nucleon before
the scattering, $q$ is the momentum transfer, $\nu$ is the energy 
transfer in the lab frame, $\theta$ the lepton scattering angle in
this frame and $E$, $E'$ the lepton initial and final energies.

Since the momentum transfer is always space-like
in these experiments, it is convenient to introduce
$Q^2=-q^2>0$. The functional dependence on the
kinematical variables in  Eq.~(\ref{eq:dis}) is the 
most general one permitted by Lorentz invariance. 
In general $W_{1,2}(Q^2,\nu)$ are complicated
functions of two variables ($Q^2$ and $\nu$) which 
parameterize the non-trivial structure of the nucleon
target. However, in the Bjorken limit, where
both the momentum transfer $Q^2\equiv-q^2$ and the 
energy transfer $P\cdot q \equiv M\nu$ become very 
large, such that $x_{Bj}=\frac{Q^2}{2M\nu}$ stays 
finite, where $M$ is nucleon mass, one finds that the structure functions 
satisfy approximate `scaling'
\be
M W_1(Q^2,\nu) \approx F_1(x_{Bj})
\quad
\quad
\quad
\quad
\nu W_2(Q^2,\nu) \approx F_2(x_{Bj}).
\label{parton}
\eq
These scaling functions $F_i$ probe light-like 
correlations in the target. To see this, one first
uses the optical theorem to relate the inclusive
lepton-hadron cross section to the imaginary
part of the forward Compton amplitude (Fig.~\ref{fig:dis}).
In the Bjorken limit, only those contributions to
the forward Compton amplitude survive where the
incoming and the outgoing photons couple to the
same quark line. All other contributions are
suppressed because they require the exchange of 
additional gluons to route the large momentum from
the incoming photon to the outgoing photon. This
is the physics reason for the dominance of so called
`handbag diagrams' in DIS (Fig.~\ref{fig:fcomptb}).  

Asymptotic freedom implies that one can
neglect final state interactions of the struck quark
in DIS. Applied to the forward Compton amplitude, 
this means that one can neglect interactions of
the `hard' quark which transfers the momentum in
the handbag diagram.
Finally, being a high-energy quark, this `active'
quark in DIS moves with nearly the speed of light
in a direction prescribed by the kinematics of the
scattering event (Fig.~\ref{fig:compst}).
Since the struck quark does not interact\footnote{
This is strictly true in light-cone gauge $A^+=0$. 
In other
gauges, the interaction of the struck quark is 
described by a `Glauber phase factor', which can
be identified with a straight line gauge string.},
it is as if the quark had been removed from the 
target at one space time point and is then replaced
at another space time point displaced by a 
light-like distance. It should thus not come as a 
surprise that the physics
which the parton distributions Eq.~(\ref{parton}) probe is related to
light-like correlation functions of quark fields of the form
\be
\langle \psi(P) | \bar{\psi}_\alpha(0, {\bf x}) \psi_\beta (x^-, {\bf x})
| \psi(P) \rangle ,
\label{eq:llcf}
\ee
where $|\psi(P) \rangle$ is the target state. 
The `$3$' direction in this case is the
direction of the space component of the momentum
transfer in the rest frame of the target.
\begin{figure}
\centering
\begin{picture}(300,200)(100,560)
\epsfig{file=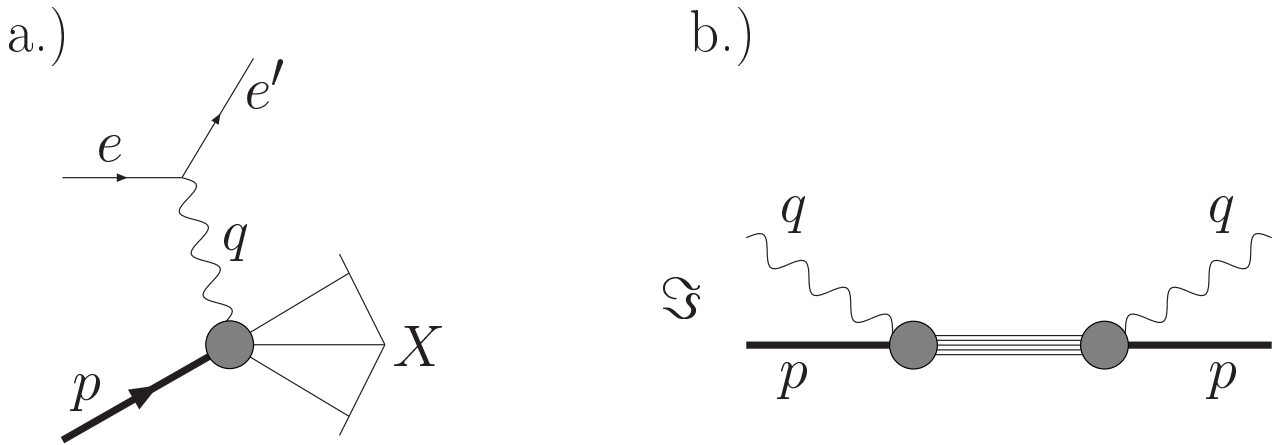, scale=0.9}
\end{picture}
\caption{a.)
Deep inelastic lepton-hadron scattering.
$X$ represents an arbitrary (unmeasured)
hadronic final state.
The inclusive cross section can be expressed in
terms of the imaginary part of the forward
Compton amplitude b.).
\label{fig:dis}}
\end{figure}
\begin{figure}
\centering
\begin{picture}(300,200)(100,550)
\epsfig{file=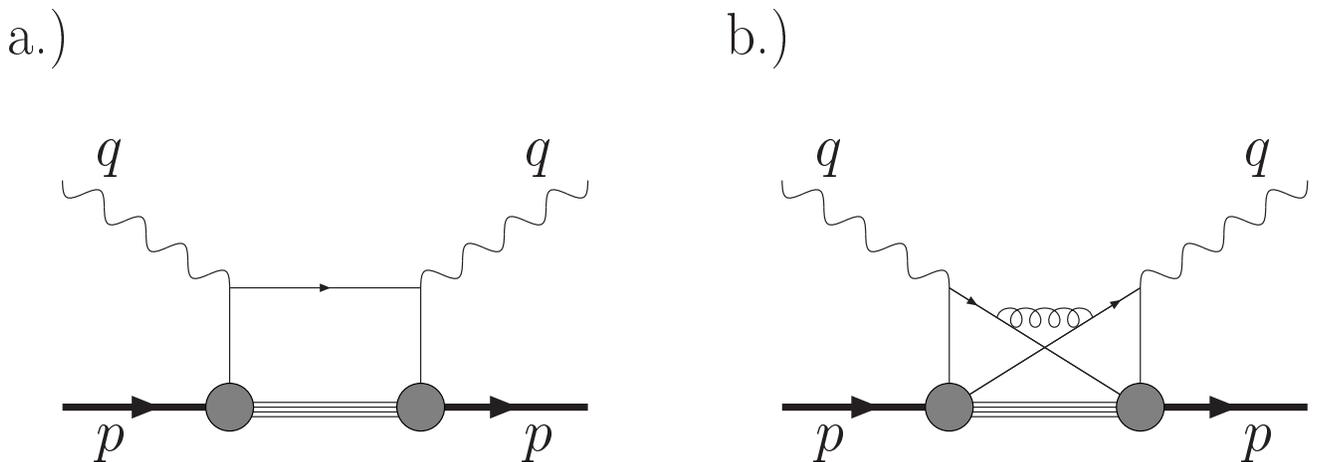, 
scale=0.9}
\end{picture}
\caption{Contributions to the forward Compton
amplitude where the two virtual photons couple to
a.) the same quark (`handbag' diagram), b.) 
different quarks (`cat's ears' diagram).
Diagrams where the two photons couple to different
quarks require at least one additional hard gluon
exchange in order to route the large momentum $q$
from the incoming to the outgoing photon-quark 
vertex.
\label{fig:fcomptb}}
\end{figure}
\begin{figure}
\centering
\begin{picture}(300,350)(100,265)
\epsfig{file=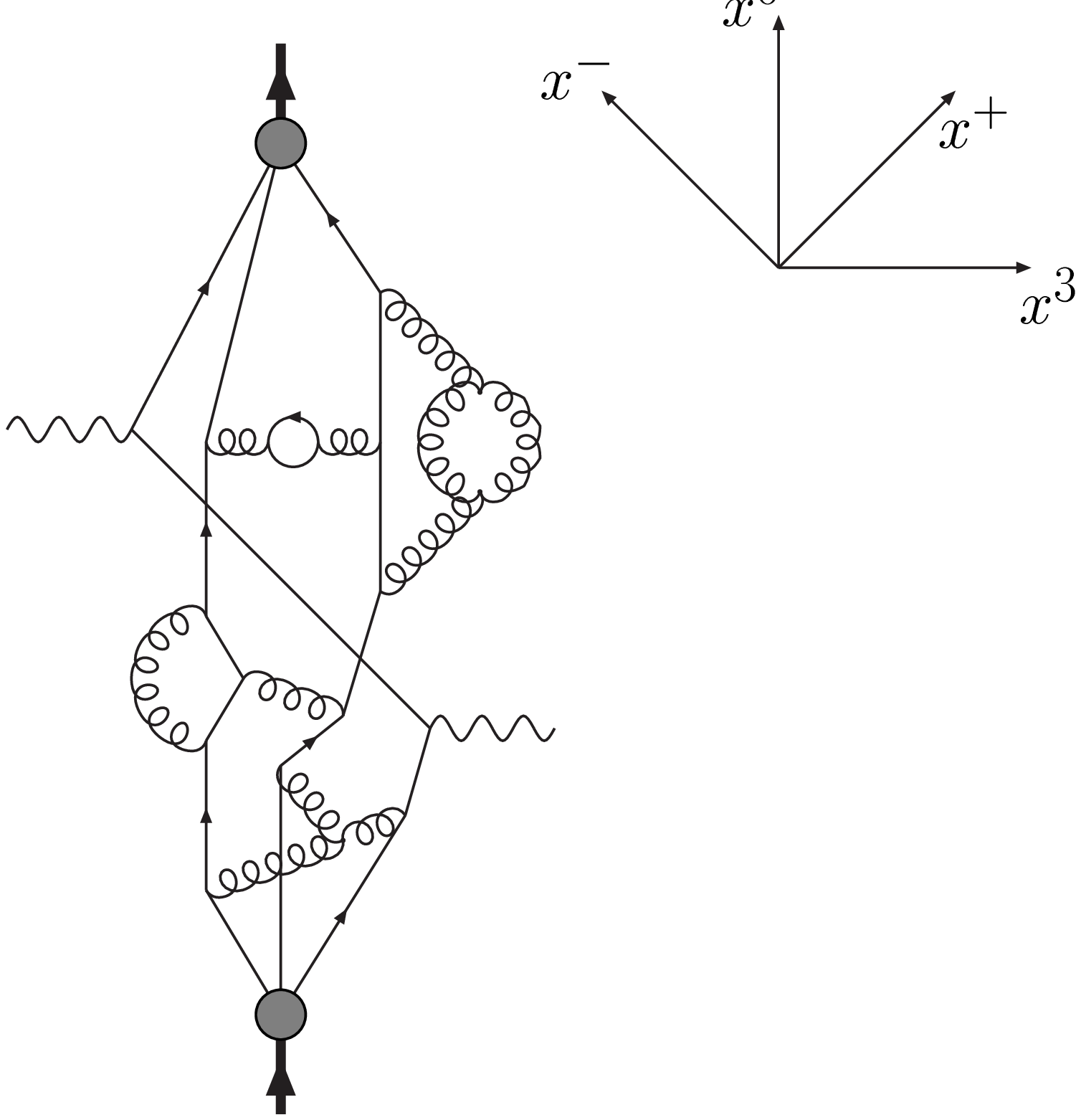, scale=0.8}
\end{picture}
\caption{Space-time cartoon illustrating Compton
scattering in the Bjorken limit, where the struck
quark propagates without interactions along a 
light-like direction.
\label{fig:compst}}
\end{figure}
The first important ramification of these
simple observations is that only
in the light-cone framework
can one express parton distributions probed in
DIS as a ground state property of the nucleon.
In all other frames the light-like correlation
function Eq.~(\ref{eq:llcf}) involves correlations
in the time direction and therefore knowledge
of the ground state wavefunction of the target
is not sufficient to describe parton distributions
--- one also needs to know the time evolution of the
target with one quark replaced by a quark that 
moves with nearly the speed of light along a
straight line.

In terms of the Fock expansion (\ref{fockexp}), the parton
distribution 
function is defined as
\be
q(x) = \sum_n 
\int \left[\Prod_{i=1}^n  dx_i d^2{\bf k}_{i} \right]
\delta\left(1-\sum_{i} x_i\right) 
\delta\left({\bf P} -\sum_{i}{\bf k}_{i}\right)
\sum_{s_i} \left| \psi_n(x_i,{\bf k}_{i},s_i) \right|^2
\sum_j \delta(x-x_j),
\label{eq:momden}
\ee
where the $\sum_j$ extends over quarks with flavor
$q$. Then
\be
2F_{1}(x) = {F_{2}(x) \over x} = \sum_{q} e_{q}^{2} q(x)
\eq
for electric charges $e_q$ of flavour $q$.
In all other frames there is no such simple 
interpretation of structure functions and one has to use less direct
methods to compute parton distributions (e.g. in
Euclidean lattice gauge theory one calculates moments and 
inverts the moment expansion).

Although DIS is perhaps the most prominent example
of the applications of the light-cone framework, there are
many other examples for high-energy scattering
experiments where light-cone coordinates play a distinguished
role. 
The underlying physics reason why they play
such a role is the simple fact that
constituents travel along a nearly light-like
direction after receiving a high energy-momentum
transfer. 

\section{Pure transverse lattice gauge theory}
\subsection{\it Colour-dielectric formulation of the light-cone Hamiltonian}
\label{sec:glue}
In this section, we discuss the ideas that lead one 
to tackle
light-cone Hamiltonian quantization of gauge theories with a
lattice cut-off. We outline some of the possible approaches
to their construction. Particular attention will be paid
to a method --- the  colour-dielectric formulation ---
that has led to a number of explicit results in hadronic
physics, and which are described in more detail 
later. We begin with pure gauge theories, leaving 
the treatment of fermions on the transverse lattice to 
section \ref{sec:ferm}.

To proceed to the solutions of a quantum field 
theory, 
with its continuously infinite degrees of freedom, 
one must put kinematical cut-offs or other restrictions on the Hilbert space.
To remove the errors this introduces, one may then extrapolate 
these cut-offs, provided a continuum limit exists, or  renormalise 
observables to account for degrees of freedom above the cut-off.
Often one does both of these things,
with an approximate renormalisation at a given cut-off
designed to improve convergence of the
extrapolation. Alternatively, instead of extrapolating the cut-off,
one may try to do a systematically better approximate renormalisation
at a given cut-off, for example by allowing gradually more complicated
forms for renormalised operators appearing in observables. 

For the study of QCD, we take the view that some residual
gauge invariance, left over after any gauge fixing and imposition of
cut-offs, is probably essential to produce generic linear confinement. By
`generic',
we mean before any renormalisation of couplings and/or extrapolation
to the continuum limit has been made. In this way, even a zeroth
approximation to  the cut-off gauge
theory will automatically produce strong interaction
physics  close to that of the
real world. A lattice cut-off is one means of retaining the required
residual gauge invariance \cite{wilson,kogut}.

For the purposes of Hamiltonian quantization, one must
have a continuous time direction. In the case of light-cone
Hamiltonian quantization, in addition to continuous light-cone time
$x^+$, light-cone space $x^-$ should not have a short distance
cut-off either. 
This is because $x^-$ is conjugate to $p^+$ and a large $p^+$ cut-off
excludes small not large light-cone energies $p^-$,
and therefore artefacts introduced by a lattice in
$x^-$ do not necessarily disappear in the formal 
continuum limit \cite{mb:adv}. 
Furthermore, it is necessary to keep the $x^-$ 
direction continuous if one wants to preserve the 
manifest boost invariance in one direction, which 
is one of the advantages of the light-cone 
formulation.
Therefore, at most we can impose a lattice cut-off
on transverse directions. The transverse lattice was  suggested by
Bardeen and Pearson \cite{bard1}, shortly after 
the discovery of lattice gauge theories. We will leave the question
about how to cut off large $x^-$ until later.

In $3+1$ spacetime dimensions we introduce a square 
lattice of spacing $a$
in the `transverse' directions ${\bf x}=\{x^1,x^2\}$ 
and a continuum in the $\{x^0,x^3\}$ directions. 
$SU(N)$ gauge field 
degrees of freedom are represented by continuum 
Hermitian gauge potentials 
$A_{\alpha}({\bf x},x^+,x^-)$ and 
$N$ x $N$ link matrices $M_r({\bf x},x^+,x^-)$. 
$A_{\alpha}({\bf x})$ resides on the plane
${\bf x} = {\rm constant}$, while $M_r({\bf x})$ is associated
with a link from ${\bf x}$ to ${\bf x} + a \hat{\bf r}$, where
$\hat{\bf r}$ is a unit vector in direction $r$. $M_r({\bf x})^{\dagger}$
goes from ${\bf x} + a \hat{\bf r}$ to ${\bf x}$.
These variables map under transverse lattice gauge
transformations $U({\bf x},x^+,x^-) \in SU(N)$ as
\begin{eqnarray}
        A_{\alpha}({\bf x}) & \to & U({\bf x}) A_{\alpha}({\bf x}) 
        U^{\da}({\bf x}) + {\rm i} \left(\partial_{\alpha} U({\bf x})\right) 
        U^{\da}({\bf x})  \nonumber \\
        M_r({\bf x}) &  \to & U({\bf x}) M_r({\bf x})  
        U^{\da}({\bf x} + a\hat{\bf r})  \ .
\end{eqnarray}
The simplest gauge-covariant combinations are 
$M_r$, $F_{\alpha \beta}= \partial_{\alpha} A_{\beta} -
\partial_{\beta} A_{\alpha} - {\rm i} [A_{\alpha}, A_{\beta}]$, 
${\rm det} M_r$, $\D^{\alpha} M_r$, etc.,
where the  covariant derivative is
\begin{eqnarray}
        \D_{\alpha} M_r({\bf x})  
        & =  & \left(\partial_{\alpha} +i A_{\alpha} ({\bf x})\right)
        M_r({\bf x})  
-  i M_r({\bf x})   A_{\alpha}({{\bf x}+a \hat{\bf r}}) \ .
\label{covdiv}
\end{eqnarray}
Note that $M$ itself need not be restricted to $SU(N)$, although
it is necessary to do so when approaching the transverse
continuum limit $a \to 0$.

If we do limit ourselves to  $M \in SU(N)$,
the simplest transverse lattice action that can produce the correct
continuum limit is
\begin{eqnarray}
A & = & \sum_{\bf x}  \int dx^+ dx^-  -{a^2\over 2g^2} 
\Tr \left\{ F^{\alpha\beta}
F_{\alpha\beta} \right\}  +
{1\over g'^2} \sum_{r}\Tr \left\{
\D_{\alpha} M_r({\bf x}) (\D^{\alpha} M_r({\bf x}))^\da \right\} 
\nonumber \\ && +
{1\over 2 g''^2 a^2} \sum_{r \neq s} 
\left(   \Tr\left\{ M_{r} ({\bf x}) 
M_{s} ({\bf x} + a \hat{\bf r})
M_{r}^{\da}
({\bf x} + a \hat{\bf s} )  
M_{s}^{\da}({\bf x})
\right\} -1\right)  \ . 
\label{action}
\end{eqnarray}
As the lattice spacing $a$ is taken to zero,
the interaction terms will select smooth configurations as the
dominant contributions to the quantum path integral;
both the interactions mediated by the local two-dimensional gauge
fields $A_{\alpha}$ and
the plaquette interactions $g''$ will
generate large potentials, unless the
link configurations are smooth.  
Inserting the Bloch-wave expansion
\be
M_{r}({\bf x})= {\rm exp} \left[ {\rm i}
a A_{r} ({\bf x} + a\hat{\bf r}/2) \right]\ ,
\eq
and displaying  only the lowest order contributions in powers of $a$, 
one obtains 
\begin{eqnarray}
A  & = &  - \int dx^0 dx^3 \  \sum_{\bf x}
{a^2\over 2g^2} \Tr \left\{ F^{\alpha\beta}
F_{\alpha\beta} \right\}
+ {a^2\over 2g'^2}  \Tr \left\{ F^{\alpha r}
F_{\alpha r} + F^{r \alpha} F_{r \alpha}\right\}
\nonumber \\
&& +{a^2\over 2g''^2}  \Tr \left\{ F^{r s}
F_{r s} \right\}
+  O(a^4) \ .
\label{cont}
\end{eqnarray}
Tuning $g=g'=g''$ yields the Lorentz-invariant 
classical continuum limit.
In deriving Eq.~(\ref{cont}), the fields 
$A_{\pm}({\bf x})$ were also assumed to be slowly 
varying on the lattice.

{}From Eq.~(\ref{action}), we see that the basic 
action for each link on the transverse lattice is 
the two-dimensional unitary $SU(N)$ principal 
chiral non-linear sigma model. The models are gauged 
and connected to one another through the plaquette 
interaction and the covariant derivative 
Eq.~(\ref{covdiv}). In an ideal world, there would be an 
exact solution to the primary chiral sigma model, 
which could be used as a kernel to solve the entire 
theory perturbatively in the interactions. The idea 
would be to use the states that are diagonal with 
respect to the two-dimensional non-linear sigma 
model Hamiltonian as a basis for construction of the 
gauge-singlet bound states of the full 
higher-dimensional theory. Griffin \cite{grif} has 
suggested that by introducing Wess-Zumino \cite{wz}
terms into the sigma model action, the non-linear 
dynamics can be studied in the basis of (linear) 
currents given by the well-studied and exactly 
solvable Wess-Zumino-Witten (WZW) model \cite{wzw}. 
The Wess-Zumino terms in the
action will become irrelevant operators, suppressed by powers of $a$,
 in the transverse continuum limit.
Although promising, the technical details of carrying
through this approach \cite{grif2} have proved sufficiently formidable that no
realistic calculations have yet been performed.

If we relax the $SU(N)$ constraint, allowing the 
link variables $M$ to be general complex matrices, 
we must add to the action Eq.~(\ref{action}) a 
gauge-invariant potential $V(M)$, with the minimal 
requirement that it constrains $M$ to the $SU(N)$
group manifold as $a \to 0$. The study of such 
linearized, or `colour-dielectric' lattice gauge 
theories at finite $a$ has some history in the case 
of four-dimensional Euclidean lattices; we refer to 
the review of Pirner \cite{hans}. Physically, 
linearized  variables $M$ may be thought of as a 
being obtained by a smearing of all flux lines over 
paths ${\cal C}$ between two points 
$({\bf x}, {\bf x} + a \hat{\bf r})$, with some
weight $\rho ({\cal C})$;
\be
M_r({\bf x}) = \sum_{\cal C} \rho ({\cal C}) {\rm P}\  {\rm exp}\left\{ {\rm i}
\int_{\cal C} A_{\mu} dx^{\mu} \right\} \ .
\eq
Different weights give rise to different potentials 
$V(M)$, related by reparameterization invariance.
A simple potential that produces the correct 
$a \to 0$ behaviour is
\be
V(M) = {N  \over \lambda} \left( \Tr \left\{
(1 - M^{\da}_r ( {\bf x})  M_r ({\bf x}))^2 \right\} 
+ (\det{M} - 1)^2 \right)
\ ,
\eq 
where $\lambda \to 0$ as $a \to 0$. However, it is 
not equivalent to a smeared continuum theory when 
$\lambda$ is finite --- such a 
potential would be infinitely more complicated.
Now we deal with  two-dimensional {\em linear} sigma models at each ${\bf
x}$ and, provided $M=0$ is the groundstate,
a simple basis for solving the full
four-dimensional theory presents itself. However, it is easy to see
that any $V(M)$ with the correct $a \to 0$ properties will have a
tachyonic mass term for $M$ near this limit. 
The non-trivial vacuum structure that
must be present to stabilize the groundstate  severely complicates
matters, particularly in light-cone quantization. One may perform
quantization of the lattice theory with couplings chosen so that
$M=0$ {\em is} the groundstate, but one cannot approach the
usual continuum limit in this case. 

At this point, it is worth recalling what is known about Euclidean
colour-dielectric lattice gauge theory.
If the link matrix $M$ on a 4-dimensional Euclidean lattice is not
tachyonic, 
a `strong coupling' expansion of
the path integral  about 
$M=0$ may be performed \cite{wein}. 
Mack has shown that a colour-dielectric picture of confinement
results \cite{mack,kogsus}. Accordingly, we shall refer to this as the {\em 
colour-dielectric regime}. Moreover, there is   
evidence for $SU(2)$ that the partition
function, subject to `block-spin' transformations,
has renormalisation group trajectories  which pass into
the colour-dielectric regime at short enough correlation length \cite{hans}. 
Therefore, the picture one should keep in mind is the following. 
Suppose  one decomposes $M=HU$ into a Hermitian matrix $H$ and a unitary 
matrix $U$. As the continuum limit $a \to 0$ is approached,
$H = H_{0} + \tilde{H}$ gets a VEV $H_{0}$, while the fluctuation
$\tilde{H}$ becomes very heavy and decouples. Near the continuum limit, $H_0$
appears in the equations of motion like a generalised dielectric
constant \cite{mack}. In this regime the field $M$ is tachyonic.
As the lattice spacing $a$ is
increased,  scaling trajectories may push  one into a region of
positive mass squared 
for $M$, where $H_0$ vanishes and $\tilde{H}$ is fully 
dynamical. The mass of $M$ then increases with $a$. 
This suggests it may be possible to obtain results relevant
to the continuum limit by studying the colour-dielectric regime.
The more detailed investigation of this idea on the transverse lattice
will be presented later. In the remainder of this section we will set up the
details of the light-cone quantization on the transverse lattice,
with potentials $V(M)$ chosen so that $M=0$ is the minimum.

For definiteness, consider the Lagrangian
\be
L  =   \sum_{{\bf x}} \int dx^- 
-{1 \over 2 G^2} \Tr \left\{ F^{\alpha \beta} F_{\alpha \beta} \right\}
- U_{\bf x}(M) +
  \sum_{r}\Tr \left\{
\D_{\alpha} M_r({\bf x}) (\D^{\alpha} M_r({\bf x}))^\da \right\} \ ,
 \label{lag}
\eq
where 
\be
U_{\bf x}(M)   =   \mu_{b}^2  \sum_{r} 
\Tr\left\{M_r({\bf x}) M_r^{\da}({\bf x})\right\}  -
{\beta\over N a^2} \sum_{r \neq s} 
  \Tr\left\{ M_{r} ({\bf x}) 
M_{s} ({\bf x} + a \hat{\bf r})
M_{r}^{\da}
({\bf x} + a \hat{\bf s} )  
M_{s}^{\da}({\bf x})
\right\}  \ . \label{pot}
\eq  
Since $M$ is now a linear variable, we are free to rescale it so that
it has a canonically normalised kinetic term. We will 
choose $\mu_b^{2}$ sufficiently large that we may quantize
about $M_r=0$. 

The gauge invariance is partially fixed by setting $A_{-} = 0$.
This axial gauge allows us to construct a Hilbert space of
positive norm states that is diagonal in light-cone momentum space.
$A_{+}$ then satisfies a  constraint equation of motion, 
which can be used to eliminate it at the classical level;
\begin{eqnarray}
(\partial_{-})^2 A_{+} &  = & {G^2 \over 2} \left( J^+ - {1 \over N} \Tr
\ J^+ \right) \ , \label{cons} \\
J^{+}({\bf x}) &=& {\rm  i} \sum_{r}
\left(
M_r ({\bf x}) \stackrel{\leftrightarrow}{\partial}_{-} 
M_r^{\da}({\bf x})  + M_r^{\da}({\bf x} - a\hat{\bf r}) 
\stackrel{\leftrightarrow}{\partial}_{-} M_r({\bf x} - a\hat{\bf r})
\right) \ . \label{current} 
\end{eqnarray}
Introducing gauge indices $\{ i,j \in \{1,2, \cdots N \}\}$, the
canonical momenta are found to be 
$\pi_{ij}(M_r) = \partial_{-}M_{r,ij}^{*}$. It is then straightforward
to canonically derive
the generators of translations in $x^+$ and $x^-$.
At time $x^+ = 0$ these are respectively
\begin{eqnarray}
 P^-  &  = &  \int dx^- \sum_{{\bf x}} 
   {G^2 \over 4} \left(\Tr\left\{ 
              J^{+} \frac{1}{({\rm i} \partial_{-})^{2}} J^{+} \right\}
            -{1 \over N} 
        \Tr\left\{ J^+  \right\} {1 \over ({\rm i}\partial_{-})^{2} }
     \Tr\left\{ J^+ \right\} \right) + U_{\bf x}(M)  
\label{lcham}\\
P^+ & = &  \int dx^- \sum_{{\bf x}, s} 2 \Tr  
                        \left\{ \partial_- M_s({\bf x})  
                \partial_- M_s({\bf x})^{\da} \right\}
\label{lcmom}
\end{eqnarray}
$P^-$ is the light-cone Hamiltonian. 
Of course, the transverse translation generator 
${\bf P}$ does not exist because of the short-distance
lattice cut-off on ${\bf x}$. It is nevertheless possible to 
construct states boosted in the transverse  direction, since
at time $x^+ = 0$  one may canonically derive the
further light-cone generators
\begin{eqnarray}
M^{-+} & = &   \int dx^- \sum_{{\bf x}, s}  x^- 
             2 \Tr\left\{ \partial_{-} M_s({\bf x}) 
             \partial_{-} M_s({\bf x})^{\da} \right\}  \ ,
\label{boost1}   \\
M^{+r} & = & -  \int dx^- \sum_{{\bf x}, s} 
      2 \left( x^r + \frac{a}{2} \delta^{rs}\right) 
                      \Tr  \left\{ \partial_- M_s({\bf x})  
                \partial_- M_s({\bf x})^{\da} \right\} \ . \label{boost2}
\end{eqnarray}
For the last expression, the transverse co-ordinate of a link has by
convention been taken at the centre of the link.
$M^{-+}$ generates boosts in the $x^-$  direction, while $M^{+r}$ is
a combination of boost in the $x^r$ direction and rotation \cite{soper}.

\subsection{\it Determination of Fock space eigenfunctions}
In the quantum theory, commutation relations at fixed $x^+$ are
\begin{eqnarray}
        \left[M_{r,ij}(x^-,{\bf x}), 
        \partial_- M_{s,kl}^{*}(y^-,{\bf y})\right]
       &  = & {{\rm i} \over 2} \delta_{ik}\,\delta_{jl}\, \delta (x^- -y^-)
        \, \delta_{\bf x y} \,\delta_{rs} \ .
\end{eqnarray}
A convenient Fock space representation  employs longitudinal
momentum space but transverse position space;
\begin{eqnarray}
 M_r(x^+=0,x^-,{\bf x})   &=&  
        \frac{1}{\sqrt{4 \pi }} \int_{0}^{\infty} {dk^+ \over \sqrt{ k^+}}
        \left( a_{-r}(k^+,{\bf x})\, e^{ -{\rm i} k^+ x^-}  +  
        a^{\da}_r(k^+,{\bf x})\, e^{ {\rm i} k^+ x^-} \right)   \; ,
        \label{expand}
\\
   \left[a_{\lambda,ij}(k^+,{\bf x}), 
        a_{\rho,kl}^{*}(\tilde{k}^+, {\bf y})\right] 
        & = & \delta_{ik}\, \delta_{jl}\, \delta_{\lambda \rho}\, 
        {\bf \delta_{x y}}\,\delta(k^+-\tilde{k}^+) \;, \\
   \left[a_{\lambda,ij}(k^+,{\bf x}),
        a_{\rho,kl}(\tilde{k}^+, {\bf y})\right] & = & 0 \;.
\end{eqnarray}
Here, $\lambda$ and $\rho \in \{ \pm 1, \pm 2\}$, $a_{\lambda,ij}^{*} = 
a_{\lambda,ji}^{\dagger}$. We define a Fock vacuum state
via  $a_{\lambda,ij} |0 \rangle = 0$ $\forall \lambda,i,j$.
The  operator $a_{ r,ij}^{\dagger}(k^+,{\bf x})$ then
creates a link-parton with longitudinal momentum
$k^+$, carrying colour $i$ at ${\bf x}$ to $j$ at 
 ${\bf x} + a\widehat{\bf r}$, while $a_{-r,ij}^{\dagger}(k^+,{\bf
x})$ creates a link with opposite orientation.
This Fock space is  diagonal in $P^+$ Eq.~(\ref{lcmom}) and serves as a basis
for finding the eigenvalues of the matrix $P^-$.

The only infinite renormalisations required in $P^-$ come from 
normal-orderings due to infinite self-energies. Here we may follow the
same procedure as in two-dimensional gauge theories 
\cite{hoof1,shei,bard2,tomaras}. With normal-ordered currents $J^+$, the
current-current interaction in Eq.~(\ref{lcham}) produces linear and
logarithmically divergent self-energies 
\be
{\bar{G}^2 \over 4 \pi } \int_{0}^{\infty} {dp^+ \over p^+}
+ {\bar{G}^2 \over  \pi }\int_{0}^{k^+} dp^+ {k^+ \over (k^+ -p^+)^2}
\ , \label{renorm}
\eq
where $\bar{G}=G\sqrt{(N^2-1)/N}$ with the dimensions of mass. The
logarithmic divergence may be absorbed by a renormalisation
$\mu_{b}^2 \to \mu_{R}^{2}$. The linear divergence must be retained
to cancel a similar divergence appearing at small momentum 
transfer \cite{hoof1}.
With this prescription, the Fock vacuum is also the physical vacuum 
$P^+ |0\rangle = P^- |0\rangle =
0$, provided we may neglect $k^+ = 0$ modes. Since the quadratic term
in $U_{\bf x}(M)$ (see Eq.~(\ref{pot})) contributes
energy $\mu_{R}^{2}/k^+$ for each mode of momentum $k^+$, it 
follows  that zero modes of link fields with $\mu_{R}^{2}>0$
have infinite energy
and decouple from physical
states. This is the transverse lattice version of the
colour-dielectric regime --- a Fock space expansion about $M_r=0$
is energetically justified.

The charges $\tilde{J}^+(0,{\bf x})= \int dx^- J^{+}(x^-,{\bf x})$ 
are the generators of residual $x^-$-independent 
gauge symmetries. From the singular behaviour of 
$P^-$ Eq.~(\ref{lcham}), one sees that the energy of a 
Fock state cannot be finite unless it is annihilated 
by $:\tilde{J}^+(0,{\bf x}):$ for any ${\bf x}$. 
This is the light-cone version of the Gauss law 
constraint (see Eq.~(\ref{cons})), which forces the
net flux into any point to be zero \cite{bard1}. 
The basis of finite-energy Fock
states are thus global gauge singlets, such as 
\be
         \Tr\left\{ a_{-1}^{\da}(k_1^{+},{\bf x})\, 
        a_2^{\da}(k_2^{+},{\bf x})\, a_{-1}^{\da}(k_3^{+},
{\bf x}+a \widehat{2}-a\widehat{1}) 
        a_{-2}^{\da}(k_4^{+},{\bf x}-a\widehat{1}) \,
        a_{1}^{\da}(k_5^{+},{\bf x}-a \widehat{1})\,
        a_{1}^{\da}(k_6^{+},{\bf x})\right\}  |0\rangle \ ,
         \label{typ}
\eq
as illustrated in Fig.~\ref{egg}.
The longitudinal momenta $k_m^{+}$ are
constrained  to  $\sum_{m=1}^{6} k_m^{+} = P^+$ in this example. 
Viewed in the position space $(x^-,{\bf x})$, typical gauge singlets
are illustrated in Fig.~\ref{lattice1}.

\begin{figure}
\centering
\BoxedEPSF{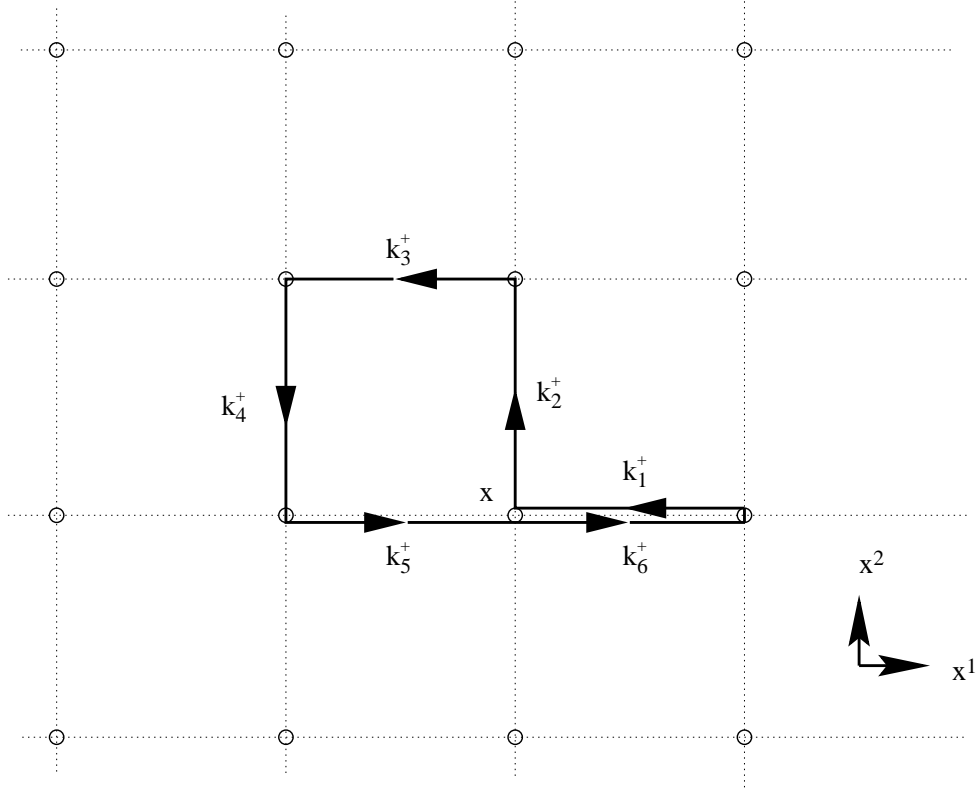 scaled 600}
\caption{ An example of a length-6 loop on the transverse lattice, showing
also the longitudinal momentum carried by  each link,
corresponding to the Fock state Eq.~(\ref{typ}).
\label{egg}}
\end{figure}

\begin{figure}
\centering
\BoxedEPSF{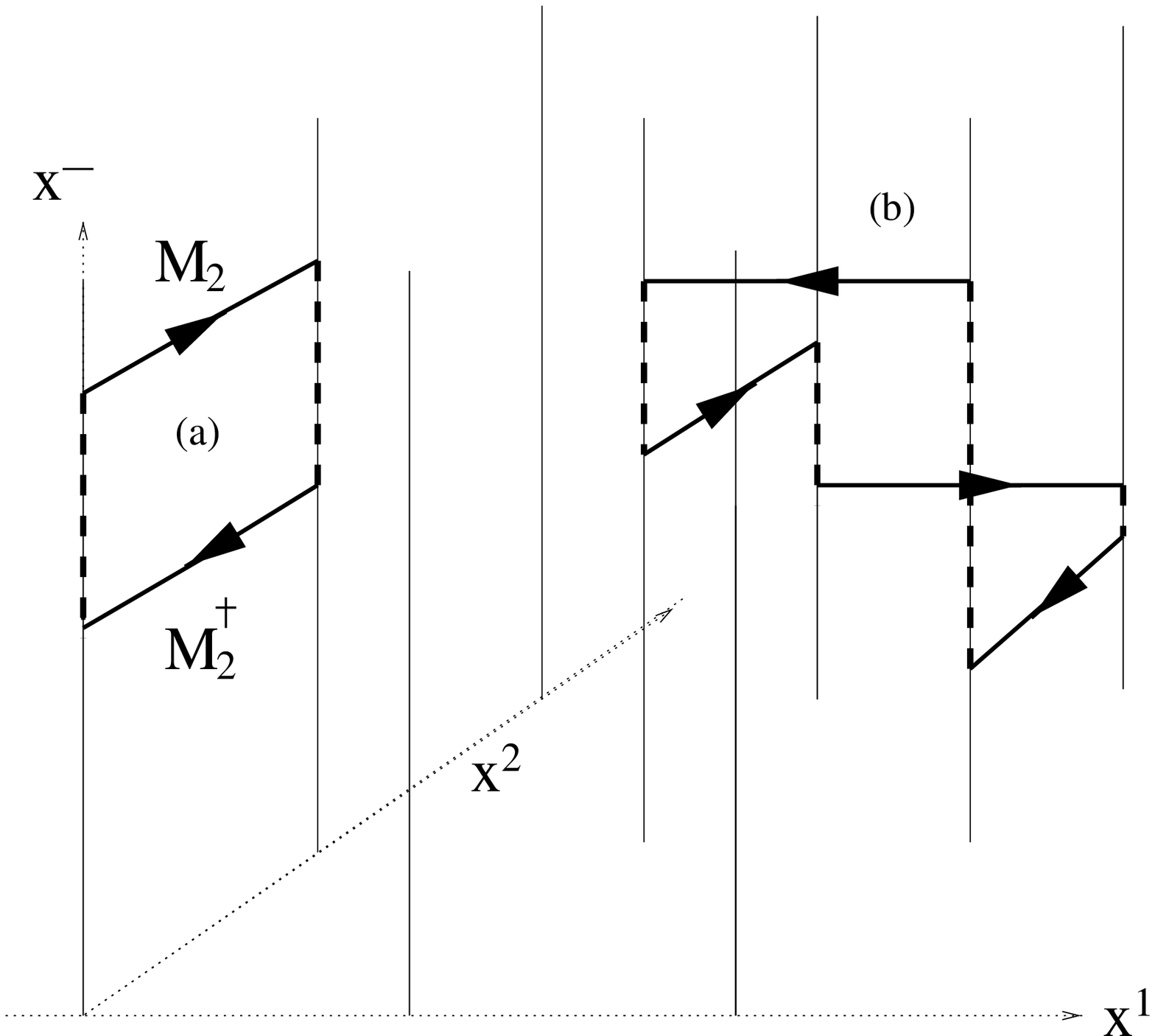 scaled 600}
\caption{Gauge singlet configurations on the transverse lattice 
at fixed $x^+$. Solid arrowed line represents a link matrix, chain
dark
lines represent $P \ {\rm exp} \int dx^- A_{-}$ insertions
required for gauge invariance (these become trivial in $A_{-}=0$ gauge).
\label{lattice1}}
\end{figure}

The transverse lattice theory 
possesses some discrete symmetries. There is charge conjugation
\be
{\cal C} \left( a_{+r,ij}^{\da}({\bf x}) |0\rangle \right) =   a_{-r,ji}^{\da}
({\bf x})|0\rangle \; .
\eq
There are two orthogonal reflection symmetries ${\cal
P}_r$ such that  $P_r(x^s) = -\delta_{rs} x^s$ and 
\be
{\cal P}_r   \, \left( a_{+r,ij}^{\da} ({\bf x}) |0\rangle \right) = 
 a_{-r,ij}^{\da}(P_r({\bf x}) - a\hat{\bf r}) |0\rangle  \; .
\eq
The operation ${\cal P}_3(x^3) = -x^3$, and therefore the parity operator
${\cal P}={\cal P}_3 {\cal P}_2 {\cal P}_1$, is more subtle in light-cone
quantization since it is dynamical. On a set of $p$ 
free particles of equal mass, the free particle 
limit of ${\cal P}_3$ acts as
\cite{horn}
\be
{\cal P}_{3}^{\rm free} \left( {k_m^{+} \over P^+} \right) 
= \left( k_m^{+} \sum_{m'=1}^{p} {1 \over k_{m'}^{+}}
\right)^{-1} \label{pfree} \; .
\eq
This expression is sometimes useful for estimating the parity of a state, for
example by tracking the state from the heavy particle limit, where ${\cal
P}_{3}^{\rm free}$ should coincide with ${\cal P}_3$.
90-degree rotations $x^1 \to x^2$ are exact and
can be used to distinguish the angular momentum
projections $ J_3 = 0, \pm 1, \pm 2$ from each other. 
Together these discrete symmetries form the group
$D_4$ \cite{bard2}, with 
one-dimensional representations and a single two-dimensional
irreducible representation.
The one-dimensional representations corresponds to $ J_3
= 0$ or symmetric and antisymmetric combinations of $ J_3= \pm 2$.
The two-dimensional
contains $J_3 = \pm 1$.

The simplest possible gauge-singlet Fock states consist of a
link---anti-link
pair (see Fig.~\ref{lattice1}(a)). A boundstate made from just these
configurations can be written
\be
|\psi(P^+)\rangle = \sum_{\bf x}
\int_{0}^{P^+} {dk_1 dk_2\over N} \;
 \delta \! \left(P^+ - k_1^{+} - k_{2}^{+} \right) \,
f_{r}(k_1^{+}/P^+,k_2^{+}/P^+) 
\Tr \left\{ a_{r}^{\da}(k_1^{+},{\bf x}) 
a_{-r}^{\da}(k_2^{+},{\bf x})
\right\} \,  |0\rangle  \ . \label{wf}
\eq
$|\psi (P^+)\rangle$ is a ${\bf P}=0$ state oriented in direction $r$. Let
us use this subspace to illustrate the boundstate problem. We drop the
index $r$, since it simply labels a doublet degeneracy, and introduce
momentum fraction $x = k_{1}^{+}/P^{+}$.
Projecting the eigenvalue equation $2P^+ P^- |\psi (P^+)\rangle =
{\cal M}^2 |\psi (P^+)\rangle$
onto Fock basis states, one derives the following integral
equation for individual Fock components \cite{bard1}
\begin{eqnarray}
{{\cal M}^2 \over  \overline{G}^2} f(x,1-x) & = & 
\left( {m_{b}^{2} \over x}+ {m_{b}^{2} \over 1-x } +{1 \over
4\sqrt{x (1-x)}}\right)
f(x,1-x)  \nonumber \\
&& + { 1 \over 4 \pi} \int_{0}^{1} dy {(x + y)(2-y-x) \over
\sqrt{x (1-x) y (1-y)}}
\left\{ {f(x,1-x) - f(y,1-y) \over (y-x)^2 }
	\right\}  \ . \label{bound}
\end{eqnarray}
${\cal M}$ is the invariant mass squared of the boundstate 
and $m_b = {\mu_{R} \over \bar{G}}$. 
The plaquette term $\beta$ in Eq.~(\ref{pot}) does not 
contribute in this subspace. We notice that the 
boundstate problem is equivalent in this case to that
of two-dimensional QCD with complex scalar particles 
of matter \cite{shei,tomaras}. The spectrum of 
eigenvalues ${\cal M}$ is infinite and discrete, 
corresponding to boundstate excitations
of the double flux line connecting the adjoint scalars in the $x^-$ direction.
Eigenfunctions $f(x,1-x)$ are of the form
\be
f(x,1-x)  =  x^{\alpha} (1-x)^{\alpha} P(x) \; \; \; ; \; \;
\alpha \tan (\pi \alpha)  =  m_{b}^2 \ ,\label{wform}
\eq
such that $P(0)  >    0$ and $P(1)  >  0$. The 
endpoint index $\alpha$ is determined by consistency 
of the limit $x \to 0$ in Eq.~(\ref{bound}), and is a 
simple example of a high-energy boundary condition 
for finite ${\cal M}$.  
The spectrum  can be labelled by the number of zeros in
$P(x)$.
The groundstate is a symmetric function of $x$ with 
no zeros; it has quantum numbers $|J_{3}|^{{\cal P} {\cal C}} =
0^{++},2^{++}$,
where we use ${\cal P}^{\rm free}$ to find ${\cal P}$. The first
excited state has one zero and $J_{3}^{{\cal P} {\cal C}} =
\pm 1^{+-}$.
These states have some of the quantum numbers expected of the lightest
glueballs, although the full wavefunction
can be very different from that in the link---anti-link
truncation of Fock space.

If we allow more links in Fock space 
(see Fig.~\ref{lattice1}(b) for example), the 
plaquette term $\beta$ and the gauge kinetic term 
can now mix sectors differing by two links. This 
provides a mechanism for propagation of states
on the transverse lattice. The wavefunction becomes 
more complicated than the form Eq.~(\ref{wf}). As well 
as four- and higher-link components, the endpoint 
indices such as $\alpha$ get  renormalised \cite{sd1}
(contrary to assumptions often made in the 
literature). It quickly becomes formidable to deal 
with an analytic basis of functions. As a result, 
most calculations with many particles have been 
performed with discrete numerical bases.

The transverse lattice does not completely regulate 
a light-cone quantum field theory because typically 
infra-red divergences  appear in the $x^-$ direction,
after non-dynamical fields have been eliminated. 
This has been discussed earlier, concerning  the
current-current interaction 
$J^+ (\partial_{-})^{-2} J^+$ in Eq.~(\ref{lcham}) 
which  contains a small $k^+$ singularity. Cutting 
out this region, the eigenvalues of the Hamiltonian 
can be made finite in the principle-value sense by
including linearly-divergent self-energies \cite{hoof1}.

Unfortunately, by putting a cut-off on small $k^+$, 
by making $x^-$ periodic for example \cite{old}, it appears 
impossible to eliminate or gauge away the $k^+ = 0$ 
modes of $A_{\pm}$. In fact, it is possible to gauge 
away the zero mode of $A_{+}$ at a particular $x^+$,
but the zero mode of $A_{-}$ will always remain as a
dynamical quantum-mechanical degree of freedom in 
Fock space \cite{fix}. 
There is some confusion in the literature as to
how much this single mode can affect physical 
observables. A numerical estimate in 
ref.\cite{pauli} claimed an effect for the spectrum 
of two-dimensional gauge theory with adjoint matter. 
Even here, it is not known to what extent the
dynamics of the zero mode can be accounted for in 
renormalisation of existing couplings in the 
Hamiltonian. All transverse lattice calculations to 
date have made the approximation of explicitly
omitting this zero mode from the Fock space.

Two basic techniques have been applied in the 
literature for performing Fock space calculations, 
which we now briefly describe. Both have their 
advantages and shortcomings. In general it is safest
to use both methods, comparing results for
consistency.

The first method uses a finite basis of wavefunctions
$\psi_S(x_1,x_2,\cdots,x_n)$, where $S$ labels the 
shape in the transverse direction while $x_i$ labels 
the $P^+$ momentum fraction carried by each link. 
The singular behaviour described above requires 
non-analytic behaviour of $\psi$ when one or more of
its arguments vanish. The simplest case has already 
been described when only the $n=2$ sector is 
retained Eq.~(\ref{wf}). A complete set of polynomials 
or trig functions would form a suitable basis for 
the analytic function $P(x)$ and, with the correct 
$\alpha$, low-lying eigenfunctions converge rapidly 
in truncations of the basis. These forms are usually 
generalised to higher $n$ for use in diagonalising
the full Fock space Hamiltonian,
\be
\psi_S(x_1,x_2,\cdots,x_n) \sim x_{1}^{\alpha} x_{2}^{\alpha} \cdots 
x_{n}^{\alpha} P(x_1,x_2,\cdots,x_n) \ , \label{fact}
\eq
They have the advantage that all integrals can be 
performed analytically, using variations on the 
identity \cite{bard2}
\be
\int_{0}^{1} dx dy { [x(1-x)]^{\alpha} [y(1-y)]^{\beta} \over
(x-y)^2 } = -{\alpha \beta \over 2(\alpha + \beta) } B(\alpha,\alpha)
B(\beta,\beta)
\label{eq:integral}
\eq
However, in general  $\alpha$ becomes renormalised 
in all but the highest Fock state \cite{sd1} and 
there is no simple analytic formula for it analogous 
to Eq.~(\ref{wf}). Moreover, the factorized form 
Eq.~(\ref{fact}) is incorrect at corners of phase space.
When two or more momenta vanish, the wavefunction in
general does not. With the incorrect
endpoint behaviour, one may get slow 
convergence in the truncation of basis functions 
$P$, and an independent check is desirable.

An alternative basis uses the Discrete Light-Cone 
Quantization (DLCQ)
\cite{dlcq}. This uses the fact that the Fock space splits up into
disjoint sectors of fixed $P^+$. By making $x^- = x^- + {\cal L}$ 
periodic with a momentum-dependent period  ${\cal L} = 2 \pi K/ P^+$,
for some integer $K$, parton momentum fractions $k^+/P^+$ take the form 
$i/K$ for positive integers $i < K$, independent of ${\cal L}$. 
In other words, they are simply partitions of $K$, divided
by $K$. For given $K$ this produces a finite-dimensional approximation
to the Fock space, and answers can be extrapolated to $K = \infty$. 
It is often expedient to take anti-periodic boundary conditions, so
that $k^+/P^+ = j/2K$ where $j$ is odd. This allows a better sampling
of the small $k^+$ region, leading to faster convergence when 
observables are extrapolated in $K$. 
For example, at $K=4$, the allowed  colour-singlet
states would be
\begin{eqnarray}
&&\left\{ \Tr \ \{ a^{\dagger}_{1} (P^+/4) a^{\dagger}_{-1} (3P^+/4) \}
|0\rangle,
\Tr \ \{ a^{\dagger}_{-1} (P^+/4) a^{\dagger}_{1} (3P^+/4)\}
|0\rangle , \right.  
\nonumber \\
&& \left. \Tr \ \{ a^{\dagger}_{2} (P^+/4) a^{\dagger}_{-2} (3P^+/4) \}
|0\rangle,
\Tr \ \{ a^{\dagger}_{-2} (P^+/4) a^{\dagger}_{2} (3P^+/4)\}
|0\rangle , \right.   
\nonumber \\
&& \left. \Tr \ \{ a^{\dagger}_{1} (P^+/4) a^{\dagger}_{1} (P^+/4) 
a^{\dagger}_{-1} (P^+/4) a^{\dagger}_{-1} (P^+/4) \} |0\rangle,
\Tr \ \{ a^{\dagger}_{2} (P^+/4) a^{\dagger}_{2} (P^+/4) 
a^{\dagger}_{-2} (P^+/4) a^{\dagger}_{-2} (P^+/4) \} |0\rangle
, \right. 
\nonumber \\
&& \left. \Tr \ \{ a^{\dagger}_{1} (P^+/4) a^{\dagger}_{2} (P^+/4) 
a^{\dagger}_{-2} (P^+/4) a^{\dagger}_{-1} (P^+/4) \} |0\rangle,
\Tr \ \{ a^{\dagger}_{2} (P^+/4) a^{\dagger}_{1} (P^+/4) 
a^{\dagger}_{-1} (P^+/4) a^{\dagger}_{-2} (P^+/4) \} |0\rangle
, \right. 
\nonumber \\
&& \left. \Tr \ \{ a^{\dagger}_{1} (P^+/4) a^{\dagger}_{-2} (P^+/4) 
a^{\dagger}_{2} (P^+/4) a^{\dagger}_{-1} (P^+/4) \} |0\rangle,
\Tr \ \{ a^{\dagger}_{-2} (P^+/4) a^{\dagger}_{1} (P^+/4) 
a^{\dagger}_{-1} (P^+/4) a^{\dagger}_{2} (P^+/4) \} |0\rangle , \right. 
\nonumber \\
&& \left. \Tr \ \{ a^{\dagger}_{1} (P^+/4) a^{\dagger}_{2} (P^+/4) 
a^{\dagger}_{-1} (P^+/4) a^{\dagger}_{-2} (P^+/4) \} |0\rangle,
\Tr \ \{ a^{\dagger}_{2} (P^+/4) a^{\dagger}_{1} (P^+/4) 
a^{\dagger}_{-2} (P^+/4) a^{\dagger}_{-1} (P^+/4) \} |0\rangle
\right\}
\nonumber
\end{eqnarray}
The number of Fock states increases exponentially with $K$. This
method has the advantage that it is easy to write down a basis
and evaluate the interactions for many particles, without having
to make any ansatz. However, because
the discretization is not very sensitive to the small $k^+$ 
singularities in the interactions, convergence in
$K$ is quite slow. 
This can be partially overcome by using a continuous wavefunction
basis for particles that interact and a DLCQ basis for spectators,
when evaluating matrix elements of $P^-$. This improvement
technique is described
in refs.\cite{dv1,dv2}.

So far we have considered states that are translationally
invariant on the transverse lattice. States
with non-zero ${\bf P}$ may be obtained using the boost
operator $M_{-r}$  Eq.~(\ref{boost2}), whose action is
\begin{eqnarray}
{\rm e}^{-{\rm i}b^r M_{-r}} \, a_\lambda^\da(k^+, {\bf x}) \, 
{\rm e}^{{\rm i}b^r M_{-r}} & = & a_\lambda^\da(k^+, {\bf x})\,
{\rm e}^{-{\rm i}k^+ {\bf b}\cdot ( {\bf x} + a {\rm Sgn}(\lambda)
\hat{\bf \lambda}/2)}  \ ; \\
{\rm e}^{-{\rm i}b^r M_{-r}}\, |\psi(P^+, {\bf P})\rangle & = & 
|\psi(P^+, {\bf P}- {\bf b} P^+)\rangle \; \label{boost}.
\end{eqnarray}
In this way an arbritary connected  $p$-link state when boosted
becomes
\be
    \sum_{\bf y}
        {\rm e}^{i \bf{P}\cdot \left({\bf y}+{\bf \bar x}\right)} 
      \Tr\left\{ a_{\lambda_1}^\da(k_{1}^{+},{\bf x}_1+{\bf y}) \, 
        a_{\lambda_2}^\da(k_{2}^{+},{\bf x}_2+{\bf y})
        \cdots  a_{\lambda_p}^\da(k_{p}^{+},{\bf x}_p+{\bf y})\right\} 
        \left|0\right\rangle
        \; ,  \label{dfunct}
\eq
where 
\be
\sum_{i=1}^p \widehat{\lambda}_i  =  0  \ \ ;  \ \
\sum_{i=1}^p k_{i}^{+}  =  P^{+} \ \  ; \ \
        {\bf \bar x} = \frac{1}{P^+}\sum_{i=1}^p k_{i}^{+} \left({\bf x}_i+
                   \frac{a {\rm Sgn}(\lambda) 
\, \widehat{\bf \lambda}_i}{2}\right) \ .
               \label{com}
\eq

Longitudinal boosts, generated by $M_{-+}$ Eq.~(\ref{boost1}), simply rescale
$P^{+}$. For this reason it is often convenient to use boost-invariant
$P^+$ momentum fractions.

\subsection{\it Confinement of heavy sources}

We now show how generic linear confinement arises in the
transverse lattice theory in the colour-dielectric regime.
An unambiguous criterion for this is provided by the
potential between static colour sources \cite{matthias}. 
We introduce a heavy scalar field $\phi(x^+,x^-,{\bf x})$ of
mass $\rho$, in the fundamental representation of the gauge group,
which is associated to lattice plane ${\bf x}={\rm constant}$. 
The simplest gauge invariant terms one can add to the action
are
\be
  \sum_{\bf x} \int dx^- \, dx^+\, 
        \left(D_\alpha \phi\right)^\da D^\alpha \phi 
        - \rho^2 \phi^\da \phi  \ , \label{heavy}
\eq
where
\be
                D_\alpha \phi = \partial_\alpha \phi
+i A_\alpha \phi \; .
\eq
The heavy field contributes to the gauge current $J^\alpha$ 
in Eqn.~(\ref{current}) a new term
\be
        J^\alpha_{\rm heavy} = -i \left(D^\alpha \phi\right)\phi^\da
         +i \phi \left(D^\alpha \phi\right)^\da \; .
\eq
Note that $\phi \phi^\da$ is an $N\times N$ colour matrix.

Let $P^\alpha$ represent the full 2-momentum
of a system containing $h$ heavy particles.  
It is convenient to split the full momentum into a ``heavy'' part
plus a ``residual'' part $P^\alpha_{\rm res}$,
\be
        P^\alpha = \rho h v^\alpha + P^\alpha_{\rm res} \; , 
        \label{resid}
\eq
where $v^\alpha$ is the covariant velocity of the
heavy quarks, $v^\alpha v_\alpha =1$.  
The full invariant mass squared (at ${\bf P} = 0$) is
\be
 {\cal M}^2 = 2 P^+ P^- =
      \left(h\rho\right)^2 +2 h \rho v^+ P^{-}_{\rm res} +
        2 P^{+}_{\rm res}\left( P^{-}_{\rm res} + h \rho v^- \right)
\eq
The  choice of $v^+$ is arbitrary and it is convenient to choose it
such that $P^{+}_{\rm res}=0$.  
Consequently, $v^+ P^{-}_{\rm res}$ is just the shift of the 
full invariant mass $\cal M$ due to the interactions:
\be
  {\cal M} = h \rho +v^+ P^{-}_{\rm res} +O\left(1/\rho\right)\; .
\eq
The minimum eigenvalue of the operator $v^+ P^{-}_{\rm res}$ is the 
usual energy associated with
the heavy quark potential.
It may be computed by extending Fock space
to include the heavy sources.
Define
\be
  \phi = \frac{1}{\sqrt{4 \pi}}\int_{-\infty}^\infty 
        \frac{dk}{\sqrt{\rho v^+ + k}} \left(
         b(k)\,{\rm  e}^{-i v_\alpha x^\alpha \rho-i k x^-}
         + d^\da(k)\,{\rm  e}^{i v_\alpha x^\alpha \rho+i k x^-}\right)
           \label{phi} \; .
\eq
The ${\rm  e}^{i \rho v_\alpha x^\alpha}$ term removes an 
overall momentum $\rho v^\alpha$ from the 2-momentum.
Again, residual gauge invariance in $A_{-} = 0$ gauge leads to
only singlet states having finite energy.
As an example, let us construct a state with two co-moving 
infinitely heavy sources on the same site ${\bf x}$ maintaining a fixed 
$x^3$-separation $L$ (this corresponds to an $x^-$ separation 
of $2 v^- L$):
\begin{eqnarray}
  |2\rangle &=& \frac{2 \rho v^+}{\sqrt{N}}\int dx^- \,
      \phi^\da \!\left(x^+,x^- + L v^-\right) \,
      \phi\!\left(x^+,x^- - L v^- \right) |0 \rangle 
      {\rm  e}^{-2 i \rho v_\alpha x^\alpha} \\
    &=& \frac{1}{\sqrt{N}}\int_{-\infty}^\infty dk\, b^\da(k) d^\da(-k)
       {\rm  e} ^{2 i L v^- k} \; .
\end{eqnarray}
See Fig.~\ref{lattice2}(a).
The associated energy shift is found to be 
\be
  \frac{\langle 2|v^+ P^{-}_{\rm res} |2\rangle}{\langle 2|2\rangle} =
      \frac{\overline{G}^2  \left|L\right|}{4}  \; .  \label{heavyheavy}
\eq
Thus, we have linear confinement in this direction with squared string tension
$\overline{G}^2/4$. Physically, it arises because the flux that
passes between the heavy sources in this particular state
cannot spread out in the transverse direction.

\begin{figure}
\centering
\BoxedEPSF{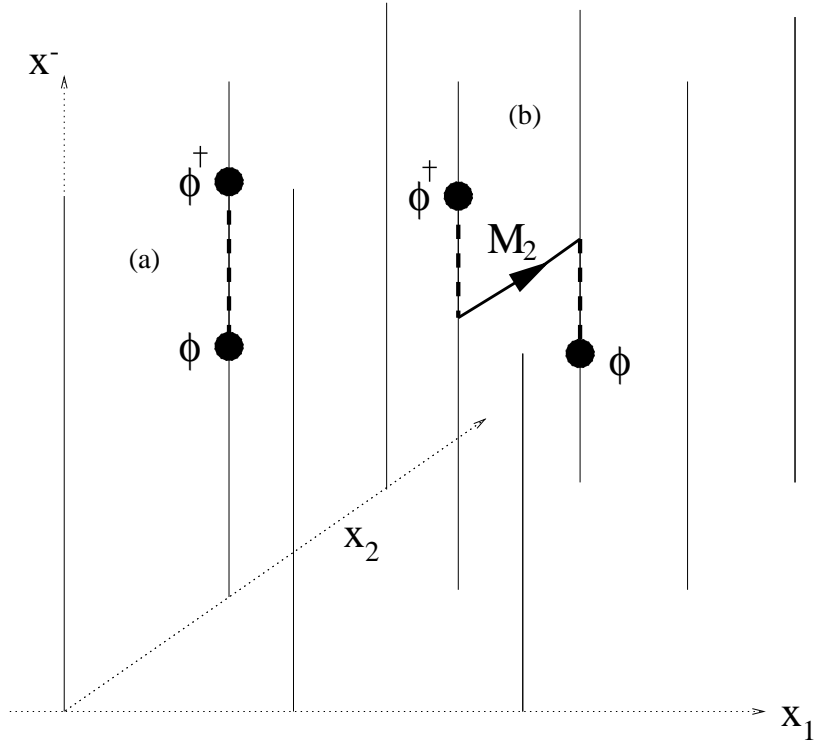 scaled 600}
\caption{Gauge singlet configurations of fundamental representation
sources $\phi$ and link fields $M$ on the transverse lattice 
at fixed $x^+$. Solid arrowed line represents a link matrix, chain
dark
lines represent $P \ {\rm exp} \int dx^- A_{-}$ insertions
required for gauge invariance (these become trivial in $A_{-}=0$ gauge).
\label{lattice2}}
\end{figure}

Sources separated in the transverse direction must be joined by a
string of links, for gauge invariance (see Fig.~\ref{lattice2}(b) for
a one-link example). It is easy to see that,
for sufficiently wide transverse separation and large link mass $m_b$, the 
potential is dominated by the mass of the links forming this
string, whose number will be minimized to form the potential. 
Each additional link increases the energy by $\bar{G} m_b$ and the separation
of sources by $a$, leading to linear confinement with squared string tension
$m_b \bar{G}/ a$. 

If we demand equivalence of the string tension
$\sqrt{\sigma}$ in
continuum and lattice directions, this is one method of fixing the 
lattice spacing
$a$ in units of $\overline{G}$.
The full heavy-source calculation  will involve fluctuations in the
number of links. This will renormalise the string tensions in each
direction. Unless some fine-tuning is
performed,  to completely screen the behaviour above by transverse
fluctuations in the number of links, linear confinement
will still be  present.

Related to the heavy source calculations is the boundstate
problem for winding modes that exist if we  compactify transverse
directions.
By making the transverse lattice compact in direction
$r$ say
\be 
{\bf x} \equiv {\bf x} + a \hat{\bf r} D_{r} \ ,
\eq
where $D_{r}$ is the number of transverse links in direction 
$r$,
we can construct a basis of
Fock states that wind around these directions. For example 
\be
         \Tr\left\{ a_{r_1}^\da(k_{1}^{+},{\bf x}) \, 
        a_{r_2}^\da(k_{2}^{+},{\bf x} + a\hat{\bf r}_1)
        \cdots  a_{r_p}^\da(k_{p}^{+},{\bf x}+
                                     a\hat{\bf r}D_{r}-
                                     a\hat{\bf r}_p)\right\} 
        \left|0\right\rangle
        \;   \label{wind}
\eq
has winding number one in direction $\hat{\bf r}$.
The mass spectrum of such winding modes rises linearly with
$D_{r}$ as $D_{r} \to \infty$. 
Unlike the potential between heavy
sources,
there are no `endpoint' effects, and so the asymptotic linear rise
should set in more quickly, especially for the groundstate in each
$D_r$ sector.
Note that we cannot
construct winding modes around a compactified $x^3$ as this clashes
with the choice of light-cone coordinates.

\subsection{\it Simplification in the Large $N$ limit}
\label{largen}
It is well-known that gauge theories undergo considerable simplification
in the limit of large $N$ \cite{hoof2}. This is especially true of
light-cone gauge theories, since the boundstate equation becomes
a linear Schr\"odinger equation for connected loops of flux of
the form $\Tr \ \{ a^{\da} a^{\da} \cdots a^{\da} \}$ \cite{thorn}. 
Transitions to disconnected forms $\Tr \ \{ a^{\da} a^{\da} \cdots a^{\da} \}
\cdot \Tr \ \{ a^{\da} a^{\da} \cdots a^{\da} \}$ are suppressed
by $1/N$, as are interactions between non-sequential $a^{\da}$ modes
in the colour trace. Gauge singlet states involving the antisymmetric
tensor $\epsilon_{i_1 \cdots i_N}$ involve $O(N)$ links and therefore
have infinite energy for sufficiently large link mass $m_b$.
All this would also be true of other quantization schemes.
However, in light-cone quantization with a trivial vacuum,
there can be no creation or annihilation of loops of
flux from the vacuum. Such processes, when allowed,
are enhanced by powers of $N$,
meaning that only in light-cone field theory does the leading order
in $1/N$ evaluate connected amplitudes. This is illustrated in
Fig.~\ref{sausage} for a loop-loop correlation, which is 
relevant for finding the spectrum of invariant masses ${\cal M}$.
The corresponding boundstates, which are glueballs, 
are absolutely stable in the large-$N$
limit.

\begin{figure}
\centering
\BoxedEPSF{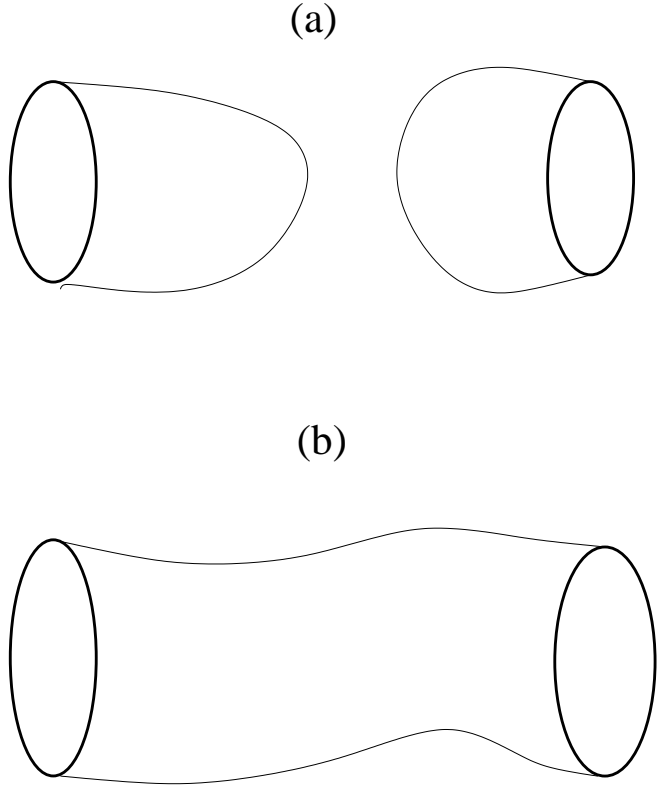 scaled 600}
\caption{Worldsheet swept out by propagation of a flux loop (dark
line). 
(a) The leading
contribution $O(N^2)$, the disconnected amplitude,
that is absent in light-cone quantization with a
trivial vacuum. (b) Leading contribution $O(N^0)$ in light-cone
quantization, the connected amplitude.
\label{sausage}}
\end{figure}

Another simplification afforded by the large-$N$ limit is
Eguchi-Kawai reduction \cite{ek}. 
For the transverse lattice, this means that  
the light-cone Hamiltonian in 
the basis of ${\bf P = 0}$ Fock states 
is the same as the Hamiltonian
for the corresponding problem 
where the transverse lattice is compactified on one 
link in each direction \cite{sd2}. 
In other words, one makes the identification 
\be
M_r({\bf x}) = M_r  \;\;\;\; \mbox{for all ${\bf x}$.} \label{id}
\eq
in both basis states and Hamiltonian. 
For the basis states themselves, the identification Eq.~(\ref{id})
is obviously a one-one
mapping. Any connected flux loop for a state of ${\bf P}=0$ 
is completely
specified by the sequence
of orientations and longitudinal momenta of the link modes 
in the colour trace,  and does not require
knowledge of the absolute positions of the links on the lattice.
In effect, large-$N$ gauge theory has
reduced to 1+1-dimensional large-$N$ gauge theory with two complex
matter fields in the adjoint representation. Powerful 
techniques developed in the study of two-dimensional field theory of
this kind \cite{kleb} can now be brought to bear on the problem. 

\begin{figure}
\centering
\BoxedEPSF{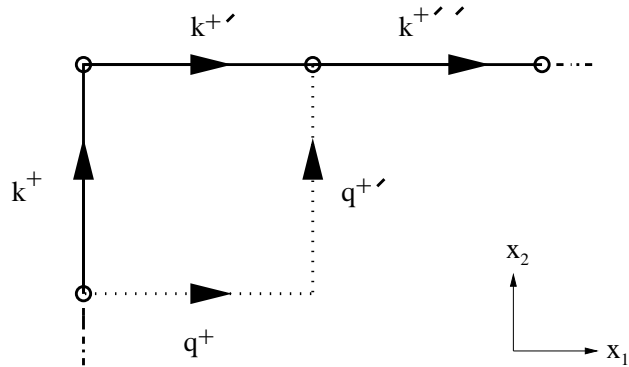 scaled 600}
\caption{Action of the plaquette operator (\ref{plaq}) on successive
links in a state. The dotted line represents the final configuration.
\label{flip}}
\end{figure}

Let us illustrate how the dimensional reduction of the  Hamiltonian 
is justified with an
example involving the plaquette interaction in Eq.(\ref{pot})
\be
 \sum_{r \neq s} 
  \Tr\left\{ M_{r} ({\bf x}) 
M_{s} ({\bf x} + a \hat{\bf r})
M_{r}^{\da}
({\bf x} + a \hat{\bf s} )  
M_{s}^{\da}({\bf x})
\right\} \label{plaq} \ .
\eq
We consider a $2 \to 2$ amplitude which occurs on neighbouring
links in the colour trace of a state (such as Eq.(\ref{typ})). 
The operator (\ref{plaq})
contains in its mode expansion the combination
\be
\Tr \left\{a^{\da}_{+1}(q^+,{\bf x}) 
a^{\da}_{+2}(q^{+ \prime},{\bf x}+a\hat{1}) 
a_{+1}(k^{+\prime},{\bf x}+a\hat{2}) 
a_{+2}(k^+,{\bf x}) \right\} \; , \; \;\;\; q^++q^{+\prime}
= k^+ + k^{+\prime} \; . \label{combin}
\eq
Acting upon the successive links illustrated in 
Fig.~\ref{flip}
\be
\Tr \left\{ \cdots a^{\da}_{+2,kj}(k^+,{\bf x}) 
a^{\da}_{+1,ji}(k^{+\prime},{\bf x}+a\hat{2}) 
a^{\da}_{+1,ih}(k^{+\prime \prime},{\bf x}+a\hat{2}+a\hat{1}) 
\cdots \right\} |0\rangle 
\label{success}
\eq
the result is to move the first two links diagonally and  redistribute
the light-cone momentum, yielding
\be
N \Tr \left\{  \cdots a^{\da}_{+1,kj}(q^+,{\bf x}) 
a^{\da}_{+2,ji}(q^{+ \prime},{\bf x}+a\hat{1})
a^{\da}_{+1,ih}(k^{+\prime \prime},{\bf x}+a\hat{2}+a\hat{1})
\cdots \right\} |0\rangle \label{true}
\eq
In the dimensionally reduced theory the ${\bf x}$ label is dropped.
The result this time has an extra term if $k^{+\prime}=k^{+\prime
\prime}$
since the second and third links are identified. The extra term is
\begin{eqnarray}
&&  \Tr \left\{ a^{\da}_{+1}(k^{+\prime} )a^{\da}_{+1}(q^+) 
a^{\da}_{+2}(q^{+\prime}) 
\right\}
 \Tr \left\{ \cdots \delta_{kh} \cdots \right\} |0\rangle \; , \label{false}
\end{eqnarray}
which is suppressed by $1/N$ compared to $(\ref{true})$.
In fact, the first colour trace in (\ref{false}) represents a winding
mode around the compactified lattice (see (\ref{wind})).
Note that, aside from momentum conservation,
the longitudinal coordinate plays no role in the calculation; it is
only
the colour index structure that is important.
The generalisation of these arguments to other gauge invariant
operators in the Hamiltonian that contain  a finite number number of
link fields is straightforward, showing that additional interactions
that appear due to compactification are suppressed in the $N \to
\infty$ limit.
The Hamiltonian in the basis of reduced states may also be evaluated
at non-zero ${\bf P}$, by boosting with the $M^{+r}$ generator 
without loss of generality \cite{dv2}.

Note that the Eguchi-Kawai reduction is exact, provided that $|0\rangle$ is the
true vacuum. This is only the true vacuum in the colour-dielectric
regime, which does not contain the transverse continuum limit.
This is consistent with the  finding  in other lattice quantization
schemes, where the naive reduction is only valid on coarse lattices
\cite{twist}.

\section{A solution of pure gauge theory at large $N$}
\label{sec:sol}
In this section, we outline an explicit solution of a transverse lattice
gauge theory \cite{dv3,dv4}. It aims to produce results for the low-lying
glueball boundstates of 
QCD to leading order in the $1/N$ expansion. This will be achieved 
by renormalising couplings (non-perturbatively) so as
to approximately restore continuum symmetries violated by the lattice cut-off.
QCD may be {\em defined} as that local
quantum field theory with gauge,
Poincar\'e, chiral, parity, and charge conjugation
symmetries that has the asymptotically free continuum limit. We will
make the additional, but reasonable, assumption of universality. That
is, the continuum limit of QCD is unique in the sense that it
 cannot be continuously 
deformed into another theory, with the same definition,
that gives different answers. In other words, there are 
no adjustable dimensionless
parameters in  QCD, although a dimensionful
confinement scale, such as $\sqrt{\sigma}$, must be taken from
experiment. 
(We regard the quark masses
as part of the electroweak action).
If $P^-$ is the light-cone Hamiltonian of QCD without approximation, its 
eigenvalues should take the relativistic form
\be
P^- = {{\cal M}^2 + |{\bf P}|^2 \over 2 P^+} , \label{exact}
\eq
for eigenstates of momentum $P^+$ and ${\bf P}$.

Once a cut-off is introduced, some of the symmetries of QCD will
be broken. For generality, we must now include  in the action
couplings to terms that preserve only the residual symmetries 
unviolated by the cut-off. Thus,
on a transverse lattice we should include couplings that 
preserve only gauge symmetry, $90^{\rm o}$-rotations about $x^3$, 
boosts along $x^3$, etc. In general, there are an infinite number of
possible terms one could add. 
The idea behind renormalisation of
Hamiltonians is that, as the cut-off varies,
these extra couplings can be tuned so as to maintain the same
eigenvalues of $P^-$, at least for the accessible
momenta in the presence of the cut-off.\footnote{Some of the first attempts to
formulate renormalisation group ideas focussed on Hamiltonians
\cite{wilson1}. An interesting weak-coupling renormalisation scheme for 
light-cone Hamiltonians has been developed in
ref.\cite{flow}. Since it relies 
heavily upon running couplings in the
neighborhood of the continuum limit, so that asymptotic freedom
allows the perturbative renormalisation of  couplings,
it is not suited to the coarse transverse lattice.}
The extra couplings
which are allowed, once one breaks continuum symmetries,
are thus not independent. In particular there should be a particular
region of coupling space, over which the cut-off varies,
which gives the same eigenvalues for
$P^-$ as the continuum limit.

In the previous section we developed a transverse 
lattice gauge theory with linear link
variables $M$ quantized about the $M=0$ vacuum --- the colour-dielectric
regime. There is always a cut-off in this theory, since
the transverse continuum limit cannot be directly accessed
without dealing with a new vacuum problem. 
Treating this as an effective theory, one would normally expect 
to be able to tune some of the couplings to restore symmetry and have
to fix the remaining ones phenomenologically (they would ultimately
be determined by the fundamental physics above the momentum cut-off).
However, at least in the case of pure gauge theories,
evidence has been found that symmetry constraints alone
can accurately determine couplings on the transverse lattice. 
A possible explanation of this is that, 
on the transverse lattice, one always works in the continuum
limit of QCD in the $x^0$ and $x^3$ directions,
meaning that restoration of full space-time symmetry should pick out 
QCD and not some other continuum limit of the lattice gauge theory. 
The requirements of Lorentz covariance and explicit gauge
invariance seem to be very restrictive in this formulation.
In practice one searches the space of couplings
to see if one can reproduce
the continuum momentum dependence of eigenvalues Eq.~(\ref{exact}). The 
reasoning is that if we would follow such a symmetry-restoring
region all the way
to the continuum limit $a=0$, this would be the continuum limit
of QCD if the uniqueness assumptions above are valid.
Note that we do not demand anything of the masses
${\cal M}$ in Eq.~(\ref{exact}), only the relativistic form of dispersion.
These masses should remain invariant in the symmetry-restoring
region of couplings, if QCD is unique, whatever the cut-off,
and represent a prediction of the theory. 

In  order explore coupling constant space, one must first introduce an
approximation scheme for the cut-off Hamiltonian, to render the
number of couplings finite.
So it is
necessary to introduce criteria for deciding which are the
most important for the physics to  be studied and how symmetries will
be restored approximately. 
One searches for a
finite-dimensional approximation ${\cal T}_s$ to a renormalised
trajectory ${\cal T}$
(see Fig.~\ref{topology1}).

\begin{figure}
\centering
\BoxedEPSF{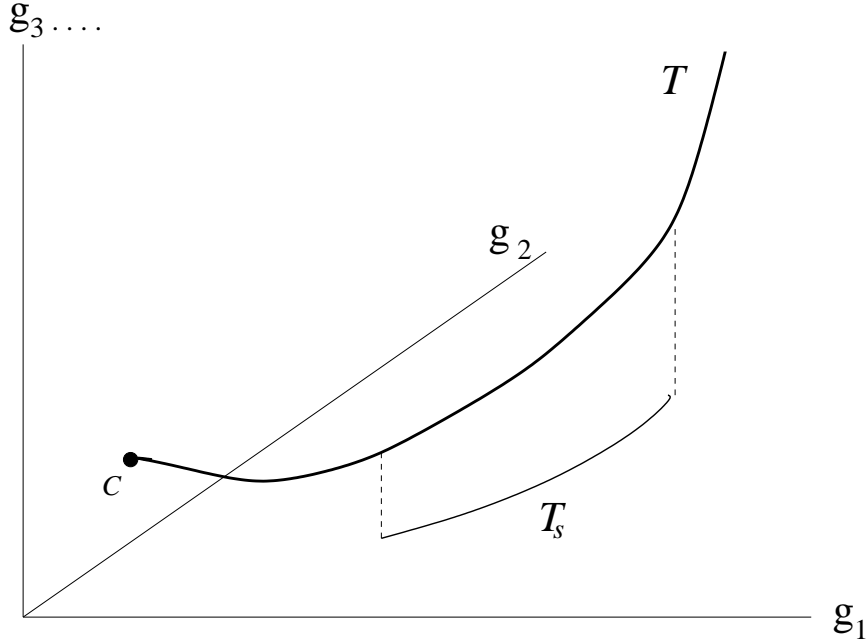 scaled 600}
\caption{$\{g_1,g_2,g_3,\ldots \}$ represents an infinite dimensional
space of allowed couplings in the cut-off Hamiltonian, $\{g_1,g_2\}$ a
finite-dimensional subspace accessible in practice. ${\cal T }$ is
the renormalised trajectory that produces the same eigenvalues as
continuum QCD Eq.~(\ref{exact}), ${\cal T}_s$ an approximation to it,
${\cal C}$ the intersection of ${\cal T}$ with
the continuum $a=0$. 
\label{topology1}}
\end{figure}

\subsection{\it Topography of coupling space}

A systematic framework for studying the pure glue
transverse lattice Hamiltonians was introduced in ref.\cite{bard2}
and developed in refs.\cite{dv1,dv2,matthias,dv3,dv4}. 
Firstly, if we demand that $P^+$
remains a kinematic operator (independent of interactions), 
then the kinetic term for link fields
and sources 
in the continuous directions is restricted to the quadratic forms
given in Eq.~(\ref{lag}) 
and Eq.~(\ref{heavy}). We have much more freedom to choose the
potential $U_{\bf x}(M)$, generalising the form Eq.~(\ref{pot}).
One idea is to expand the Hamiltonian that results from it  in 
powers of the dynamical link-field $M$
(after elimination of non-dynamical fields $A_{\pm}$).
Such a ``colour-dielectric'' power expansion can only be
justified in a  region of coupling
space where these fields are sufficiently heavy that light-cone
wavefunctions
of interest converge quickly in link-field number.
For sufficiently heavy fields, 
the lowest-mass lattice hadrons will  consist of a few partons 
$M$ with little mixing into configurations with many partons. 
Because of the special properties of light-cone co-ordinates, noted
in Sec.~\ref{lccoord}, fields do not have to be very heavy at all for this
to be viable.
The physical motivation for expecting 
a renormalised trajectory to exist in the region of link-fields $M$
with positive mass squared is 
the colour-dielectric picture of confinement
\cite{bard1,mack}. 
Gauge symmetry, residual Poincar\'e symmetry, dimensional counting in the 
continuous directions $(x^+,x^-)$,\footnote{Operators should be of
dimension two or less with respect to scaling of the continuous
co-ordinates $(x^0, x^3 )$, as in a $1+1$ dimensional field theory.
In $1+1$ dimensions, fermions $\Psi$ have dimension 1/2, while link
fields behave like scalars, with dimension 0.}
together with this power expansion in dynamical fields, limits the
number of allowed operators in $U_{\bf x}(M)$. If we also work to
leading order of the $1/N$ expansion, then we can also take 
advantage of the simplifications
noted in section~\ref{largen}. 
In refs.\cite{dv3,dv4} the most general tranvsersly-local
pure glue Hamiltonian to order $M^4$ 
and most general heavy source Hamiltonian to order $M^2$ was studied
in the large-$N$ limit. The potential that produces it is
\begin{eqnarray}
 U_{\bf x} & = & 
 \mu_{b}^2   \sum_r  
 \Tr\left\{M_r M_r^{\da}\right\} 
- {\beta\over N a^2} \sum_{r \neq s} 
  \Tr\left\{ M_{r} ({\bf x}) 
M_{s} ({\bf x} + a \hat{\bf r})
M_{r}^{\da}
({\bf x} + a \hat{\bf s} )  
M_{s}^{\da}({\bf x})
\right\} 
\nonumber \\
&& + {\lambda_1 \over a^{2} N} \sum_r  
\Tr\left\{ M_r M_r^{\da}
M_r M_r^{\da} \right\} 
+  {\lambda_2 \over a^{2} N}\sum_r  
\Tr\left\{ M_r ({\bf x}) M_r({\bf x} + a \hat{\bf r} )
M_r^{\da}({\bf x} + a \hat{\bf r} ) M_r^{\da} ({\bf x})\right\} \nonumber \\
&& + {\lambda_3 \over a^{2} N^2} \sum_r    
\left( \Tr\left\{ M_r M_r^{\da} \right\} \right)^2
+  {\lambda_4 \over a^{2} N}  
\sum_{\sigma=\pm 2, \sigma^\prime = \pm 1}
        \Tr\left\{ 
M_\sigma^{\da} M_\sigma M_{\sigma^\prime}^{\da} M_{\sigma^\prime} \right\} 
        \nonumber\\
&& +  {4 \lambda_5 \over a^{2} N^2} 
\Tr\left\{ M_1 M_1^{\da} \right\}\Tr\left\{ M_2 M_2^{\da} \right\} 
\nonumber \\
&& - \frac{\tau_1}{N G^2} \sum_r   
\Tr \left\{ F^{\alpha \beta}\, F_{\alpha \beta}\, 
\left(M_{r}^{\da} M_{r}+M_{r} M_{r}^{\da}        \right)
 \right\} 
\nonumber\\ &&
 - \frac{\tau_2}{N G^2} \sum_r   
\Tr \left\{ M_{r}^{\da}({\bf x})\, F^{\alpha \beta}({\bf x})\, M_{r} ({\bf x})
 \, F_{\alpha \beta}({\bf x} +
a \hat{\bf r}) \right\}
\; . \label{pot1}
\end{eqnarray}
We have defined $M_{r} = M_{-r}^{\dagger}$ and 
hold $\overline{G} \to {G} \sqrt{N}$ finite in 
the $N \to \infty$ limit, using it to set the single 
dimensionful scale of pure gauge theory.
Dimensionless versions of the other 
couplings can then be constructed
\begin{eqnarray}
        m^{2}_{b} & = & {\mu_{R}^{2} \over \overline{G}^2 }  \; , \;\;
        \newl_i = {\lambda_i \over a^{2} \overline{G}^2 }  \  , \;\;
        \newtau_i  =  \frac{\tau_i}{\overline{G}} \; , \; \;
         b = {\beta \over a^{2} \overline{G}^2 } \; . \label{space}
\end{eqnarray}
Note that the dimensionless mass $m_b$ is made from the renormalised
mass $\mu_R$ (see Eq.(\ref{renorm})). 
One parameter should play the role of a dimensionless lattice spacing,
$a\overline{G}$. This space of couplings Eq.~(\ref{space}) was 
searched and the eigenvalues
of $P^-$ found for various momenta. A $\chi^2$-test of symmetry
restoration, including variables to quantify deviation from
relativistic dispersion Eq.~(\ref{exact}) in low-mass glueballs and
rotational
invariance of the heavy source potential, produced a `topography'
of coupling space partly illustrated in Fig.~\ref{chart}. For each pair of
couplings, all other couplings were varied.

\begin{figure}
\centering
$m_b$\hspace{0.1in}\BoxedEPSF{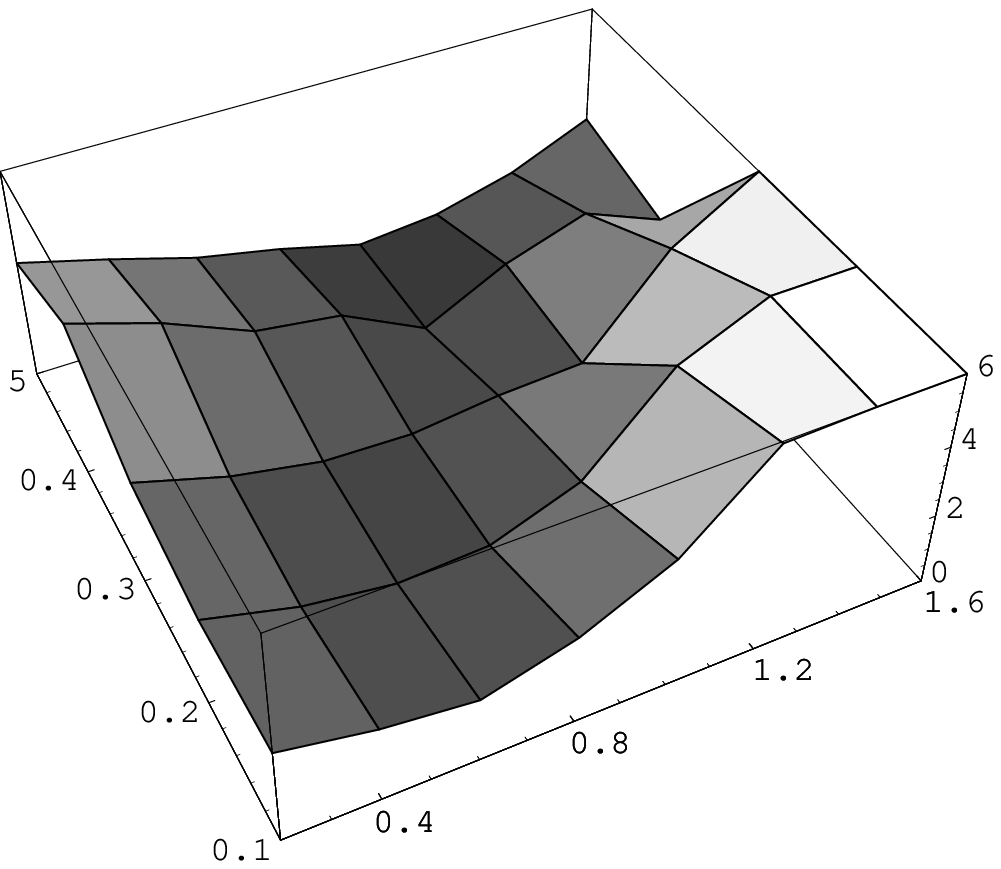 scaled 500}
\BoxedEPSF{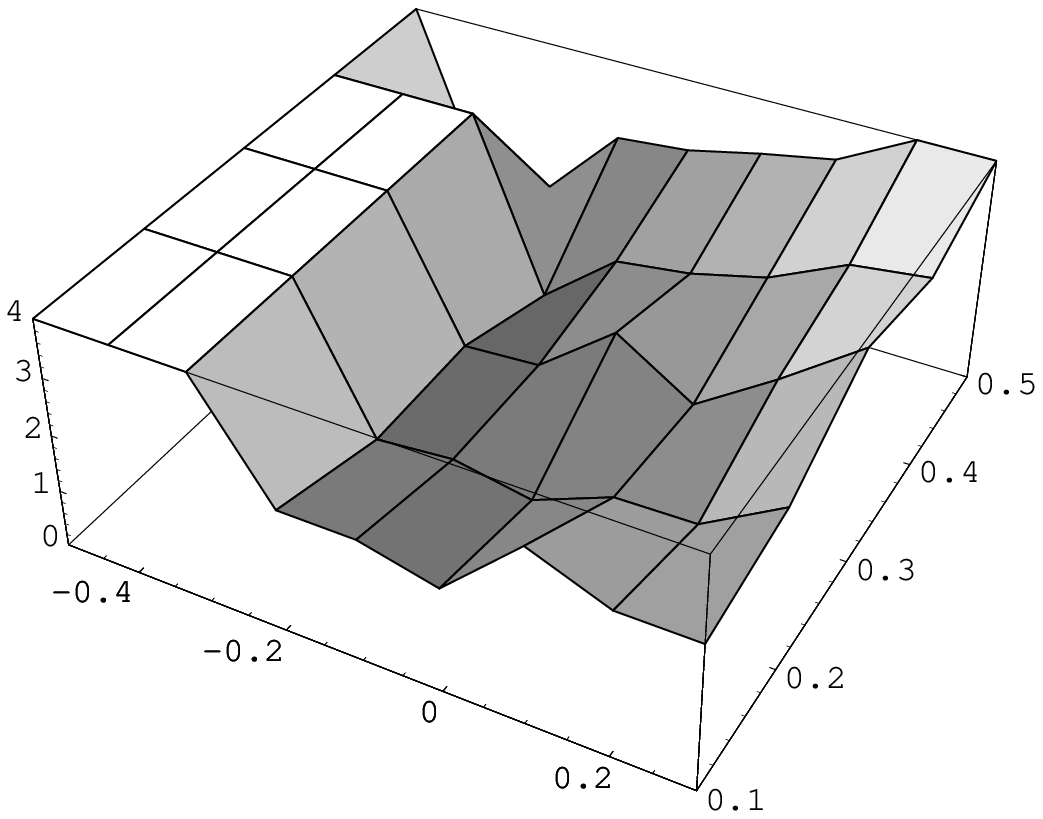 scaled 500}\hspace{0.1in} $m_b$\\[-20pt]
\hspace{-0.2in} $b$ \hspace{1.9in}  $l_1$\\[12pt]
\BoxedEPSF{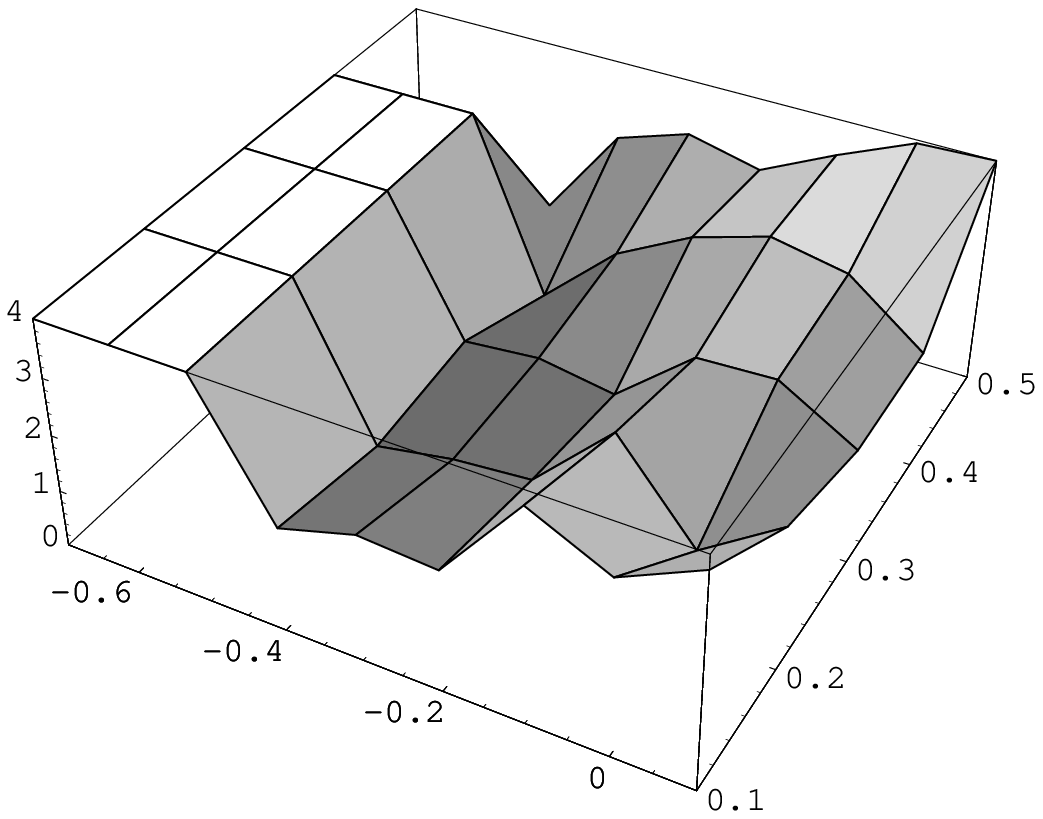 scaled 500}\hspace{5pt}$m_b$
\hspace{0.1in} \BoxedEPSF{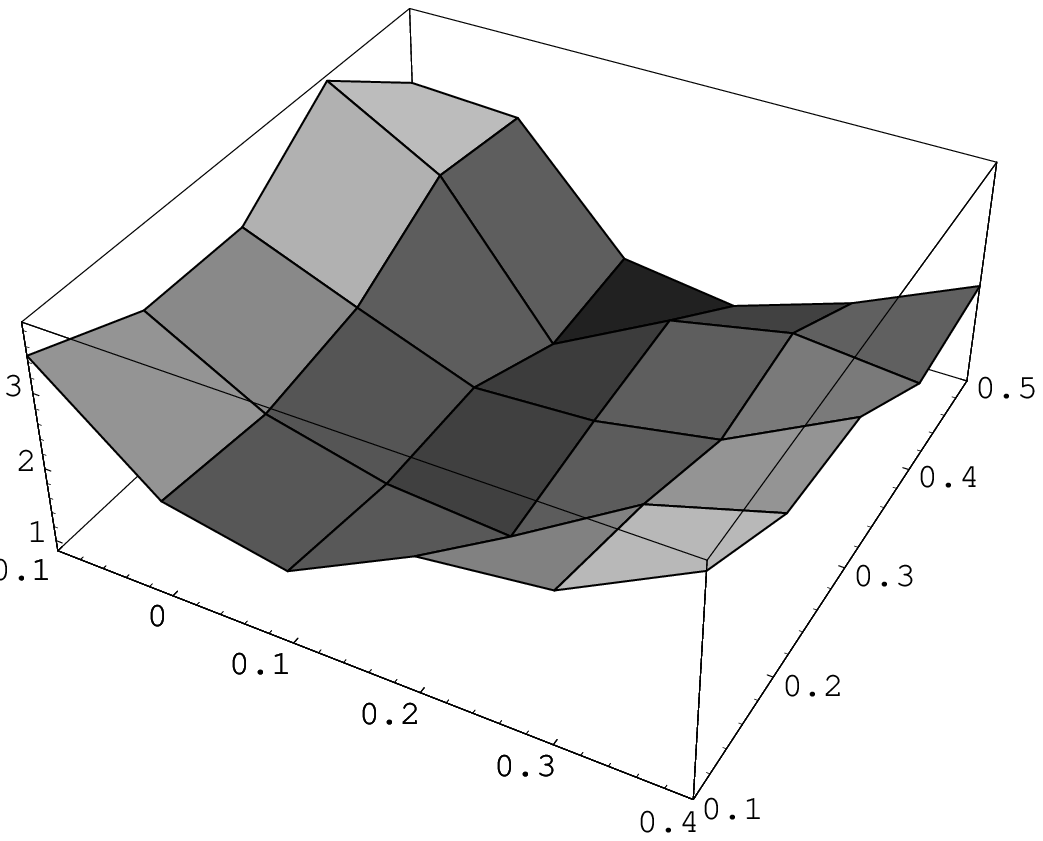 scaled 500} \hspace{0.1in} $m_b$\\[-20pt]
\hspace{-0.6in} $l_2$ \hspace{2.1in} $l_4$
\\
\caption{Charts of coupling space with a $\chi^{2}$-test
of symmetry restoration. 
\label{chart}}
\end{figure}

The bottom of the 
valley in each of these diagrams represents (a projection of) 
${\cal T}_s$, the best approximation
to a renormalised trajectory. There are a number of reasons
to believe that this feature is not accidental:  it appears
in every figure, no matter how you slice the space;
it is robust under changes of the $\chi^2$ test;
each valley has a flat direction (parameterized by $a\overline{G}$);
low-lying invariant masses ${\cal M}$ scale approximately along the valley
and  agree with large-$N$ extrapolations
of glueballs masses computed by quenched finite-$N$ Euclidean
lattice methods, that do not make a colour-dielectric expansion
\cite{teper4}.\footnote{Similar results 
have been produced for pure gauge theory
in $2+1$-dimensions \cite{dv2,dv5,teper3}, 
where greater precision is possible.} 
The glueball spectrum has also been estimated
\cite{other} for $SU(3)$ using the
light-cone methods of refs.\cite{flow}, and in the large-$N$ limit
using a conjectured relation to supergravity \cite{gravity}.

Most importantly, the light-cone wavefunctions are found to be concentrated
on particular numbers of links, meaning that the
expansion of the Hamiltonian in link-fields is self-consistent.
In the region of ${\cal T}_s$ that is accessible in the
colour-dielectric
regime, the lattice spacing $a$ is of order $0.65$ fm and varies only by about 
$10 \%$. If the 
discretization  is made too coarse, tuning the couplings in 
$U_{\bf x}(M)$ Eq.~(\ref{pot1})
is insufficient to produce a clear valley. If $a$ is made too small, the
link mass $m_b$ becomes tachyonic, requiring quantization about a new
vacuum. It is not yet known whether the above analysis again
picks out an unambiguous trajectory at higher orders of the
colour-dielectric expansion.

\subsection{\it Results for glueball boundstates}

In this section we extract some possible lessons for 
glueball phenomenology from large-$N$  calculations on the
transverse lattice. Quantitative comparison with experimental results
in this sector will require the computation of corrections to
the large-$N$ limit, so as to take account of mixing and decay
into mesons.

Because there are no quark loops in the
large-$N$ limit, the transverse lattice result for glueballs should equal the
large-$N$ limit of full QCD, not only its quenched approximation.
Fig.~\ref{alln2} illustrates, moreover, how remarkably
close the large-$N$ limit is to even $N=2$ pure glue, so far as
low-lying
masses are concerned. A priori there
is no reason to expect this. For those glueballs where the OZI rule 
is  valid, QCD itself 
should yield answers close to that of the large-$N$ limit.
If the large-$N$ limit is a good approximation to QCD, this implies 
that glueballs can be  accurately
pictured as a single connected loop of flux. This is the basis of
string \cite{string1,string2} and flux-tube models \cite{flux1,flux2} 
of glueball
dynamics. These models also predict an exponentially rising density
of states, which also occurs in the transverse lattice spectrum
\cite{dv2}, and is ultimately responsible for the
finite temperature phase transition of pure gauge theory \cite{hag}.

\begin{figure}
\centering
$\displaystyle\frac{M}{\sqrt{\sigma}}$\hspace{5pt}
\BoxedEPSF{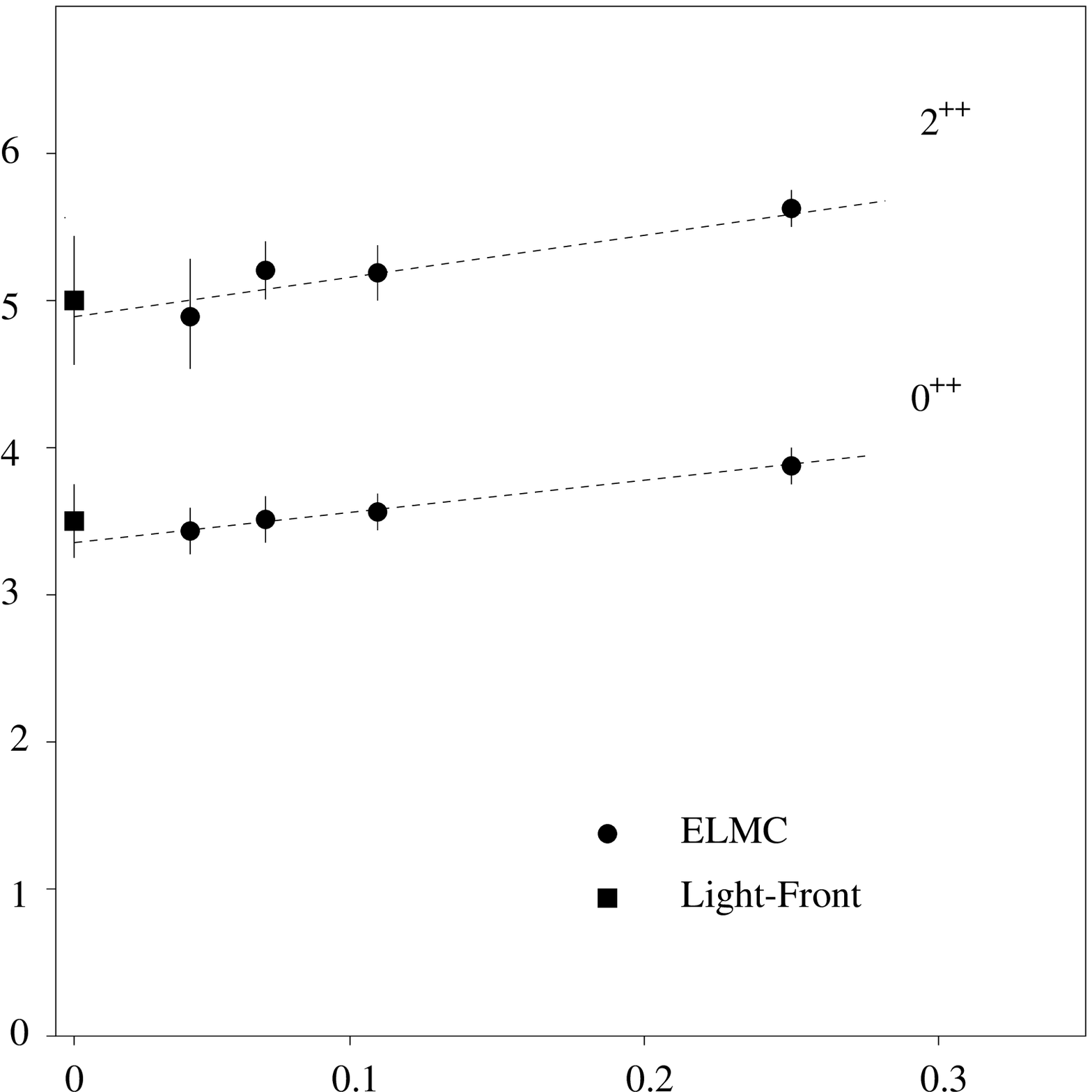 scaled 600}\\
\hspace{0.5in}$\displaystyle 1/N^{2}$
\caption{The variation of glueball masses with $N$ (pure glue). The
 transverse lattice results \cite{dv4} for scalar and tensor glueballs
are denoted by boxes. Euclidean
Lattice Monte Carlo (ELMC)
predictions are given by circles and compiled for $N=2,3,4,5$ 
\cite{star,teper4}. 
The chain lines are a fit to the form $A + B/N^2$ based on finite-$N$
data \cite{teper4}.
\label{alln2}}
\end{figure}

\begin{figure}
\centering
\begin{tabular}{c@{}c}
 $G_d$ & \BoxedEPSF{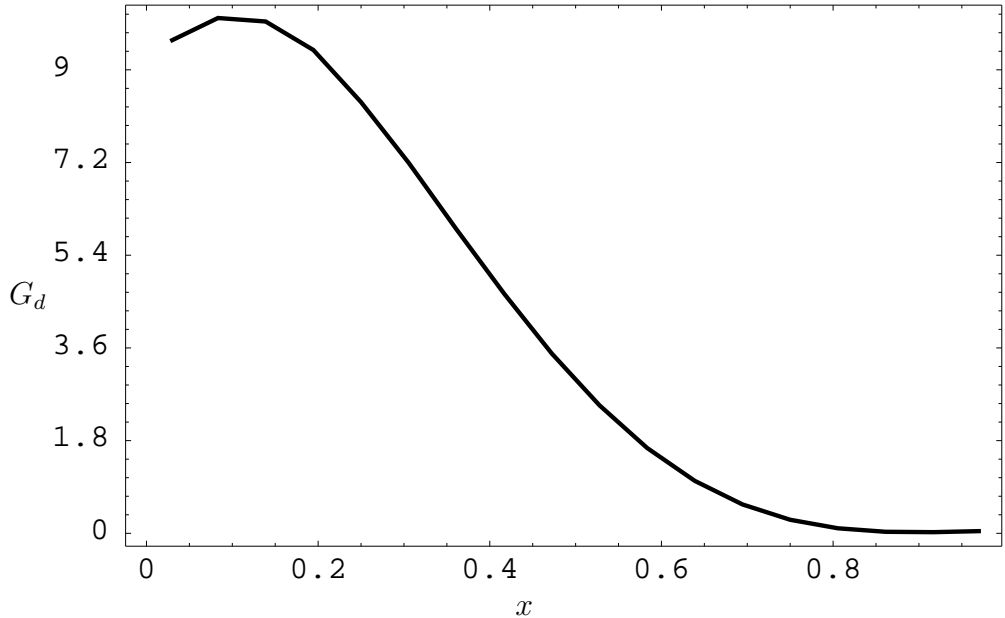 scaled 1000}\\
      & $x$ 
\end{tabular}
\caption{The distribution of $P^+$ momentum fraction $x$ in the
$0^{++}$ glueball at large $N$. 
\label{dist}}
\end{figure}

What about the structure on this single loop? In the transverse plane,
about $90 \%$ of the wavefunction in low-lying glueballs 
is concentrated in an area of order
one lattice spacing across. Even with such a coarse cut-off, any
amount of structure can be accommodated because, unlike unitary matrix
link variables, complex links can back-track any number of times.
Fig.~\ref{dist} shows the distribution of $P^+$ longitudinal
momentum fraction $x$ among the link partons in the lightest large-$N$
glueball:
\begin{eqnarray}
G_{d}(x,1/a)  & = & 
{1 \over 2\pi x P^+} \int dx^- {\rm e}^{-{\rm i} x P^+ x^-}
\langle\psi(P^+)|
        \Tr\left\{ \partial_{-} M_r \partial_{-} M^{\dagger}_r \right\}
|\psi(P^+)\rangle \nonumber \\
& = & \langle\psi(P^+)|
       a_{\lambda,ij}^{\da}(xP^{+},{\bf x}) 
a_{\lambda,ij}(xP^{+},{\bf x}) 
|\psi(P^+)\rangle
\ .
\end{eqnarray}
$G_d$ is the expectation of the number operator for 
links of momentum fraction $x$, exactly satisfies 
the momentum sum rule
\be
\int_{0}^{1}dx \  x G_d(x) = 1 \ ,
\eq
and becomes the gluon
distribution in the limit $a \to 0$. At finite $a$, $M$ is a
collective gluon excitation. The momentum sum rule would therefore 
na\"{\i}vely predict the gluon distribution 
to be even softer than $G_{d}$, since every link is a superposition of
gluons carrying only a fraction of its momentum. 
In this case, the $0^{++}$ glueball
wavefunction seems to have many gluons and 
does not resemble gluonium (two-gluon boundstate),
which would have a distribution symmetric about $x=0.5$.  
The relationship between this result and constituent gluon models
\cite{model},
where the lightest glueball {\em is} gluonium, is presumably same
as that between other hadronic structure functions and the
constituent quark model. The latter would have us believe the nucleon
is three quarks and the meson two, when in fact only half the
momentum is typically carried by quarks in a fully relativistic
description. The missing momentum is carried by gluons `in flight'
that are the true source of binding modeled by first-quantized
potentials. In glueballs, it so happens that the forces and
constituents are made of the same stuff. This result implies one
should be cautious about using a two-gluon model for glueballs
in a relativistic context; for example, 
when trying to predict branching ratios \cite{branch}.

\section{Mesons on the transverse lattice}
\label{sec:ferm}
In this section we will introduce fermions on the
transverse lattice and discuss issues that are
specific for fermions in this framework.
Some of them, such as species doubling, are
familiar from other frameworks, while others,
such as the issue of vacuum condensates, seem to
be specific to the light-cone formulation.
We then go on to construct meson boundstates in gauge theory.

\subsection{\it Formulation of the light-cone Hamiltonian with fermions}
The simplest fermion Lagrangian for the transverse
lattice, which satisfies gauge invariance and reduces
in the naive continuum limit $a \to 0$ to the continuum QCD 
action, reads 
\bea
L = a^2\sum_{\bf x}\int dx^- \left[\bar{\Psi}({\bf x})
\left(i\gamma^\alpha D_\alpha -\mu_f \right)
\Psi ({\bf x})
+i\sum_s\bar{\Psi}({\bf x})\gamma^s
\frac{M_s({\bf x})\Psi({\bf x}+a\hat{\bf s})
-M_s^\dagger({\bf x}-a\hat{\bf s})
\Psi({\bf x}-a\hat{\bf s})}{2a}\right]
\label{eq:fnaive} \ , 
\eea
where the fermion fields are defined on the sites of
the transverse lattice. Fermion fields map under
transverse lattice gauge transformations
$U({\bf x},x^+,x^-)\in SU(N)$ as
\be
\Psi({\bf x}) \quad \rightarrow \quad U({\bf x})
\Psi({\bf x}) \ .
\ee
The naive transverse derivative term in 
Eq.~(\ref{eq:fnaive}) exhibits the familiar problem of 
`species doubling', i.e.  it
yields too many low energy degrees of freedom in the
continuum limit. To see this, consider the above
transverse derivative for free fields ($M_s\equiv 1$). In this case
the Lagrangian becomes
\be
L_\perp\equiv
\sum_{\bf x} 
\int dx^- \bar{\Psi}({\bf x}) i\gamma^s
\frac{\Psi({\bf x}+a\hat{\bf s})
-\Psi({\bf x}-a\hat{\bf s})}{2a} \ .
\label{eq:ffree}
\ee
Note that we have chosen a lattice representation for
the transverse derivative that involves a difference
between fields on sites that are separated by two 
lattice spacing $\Psi({\bf x}+a\hat{\bf s})
-\Psi({\bf x}-a\hat{\bf s})$ in order to 
obtain a manifestly Hermitian Lagrangian. 
Replacing the derivative terms in the action by
expressions like  Eq.~(\ref{eq:ffree}) leads to
`species doubling' in the sense that multiplying
any solution to the free Dirac equation by 
a rapidly oscillating transverse phase factor
yields another
solution with the same energy. As a result, low
momentum solutions are degenerate with solutions
near the edge of the Brillouin zone and therefore
too many low energy degrees of freedom remain in the
continuum limit. Other, more complicated, discrete 
representations of the transverse derivatives
exhibit similar problems. This is not an accident
because the Nielsen-Ninomiya theorem \cite{nn} ---
this  states
that any local, Hermitian and chirally invariant
fermion kinetic term necessarily exhibits species
doubling ---  also applies here. 
Although the use of the colour-dielectric expansion means that
ultimately we will be interested in the transverse lattice theory 
far from the continuum, the problem that leads to species doubling 
is still relevant because it entails bad dispersion. The dispersion
induced by Eq.~(\ref{eq:ffree}) at finite lattice spacing is
considerably
worse than that of a scalar hopping term.

Similar to the situation in Euclidean and equal-$x^0$ 
Hamiltonian lattice gauge theories, there exist
several options to deal with this situation.
Here we will focus on generalizations of Wilson 
fermions to the transverse lattice.
\footnote{Staggered fermions have been investigated on the 
transverse lattice in Refs. \cite{bard1,paul}.}
The role of Wilson
fermions  on a coarse lattice will be to improve the bad dispersion.

The basic reason for the appearance of doublers in
lattice actions for fermions is the fact that the 
continuum action is first order in the derivative.   
Therefore, one natural way to remove doublers is
to add to the Lagrangian a term 
\bea
\label{eq:wilson1}
L_r &=& ra^3\int dx^- \sum_{\bf x}  \sum_s
\bar{\Psi}({\bf x})\frac{
M_s({\bf x})\Psi({\bf x}+a\hat{\bf s}) - 2 
\Psi({\bf x})
+ M_s^\dagger({\bf x}-a\hat{\bf s})
\Psi({\bf x}-{\bf \hat{s}})}{a^2}
\nonumber\\
&\stackrel{\small continuum}{\longrightarrow} &
ar \int dx^- d^2{\bf x}
\bar{\Psi} D_s^2 \Psi \ ,
\eea
where $r={\cal O}(1)$ is a dimensionless number.
In order to remove doublers
both for particles and antiparticles, the
added term has the Dirac structure of a mass term, 
which breaks chiral symmetry explicitly.

After adding such a Wilson term to the Lagrangian 
fermions on the transverse lattice, one can proceed 
with  canonical light-cone quantization.
Firstly, one eliminates the constrained fermion
component 
$\Psi_{(-)}\equiv \frac{\gamma^-\gamma^+}{2} \Psi$
using the classical equations of motion.
In  the next step, one rescales the dynamical component
$\Psi_{(+)}\equiv \frac{\gamma^+\gamma^-}{2} \Psi$ 
such that there are no dimensionful couplings
multiplying the $x^+$ derivative term, i.e. we
introduce
\be
\Phi({\bf x}) \equiv a \Psi_{(+)}({\bf x}) \ ,
\ee
in terms of which the term containing a
light-cone time derivative has the form of a 
$1+1$ dimensional field theory at each lattice site
\be
L_{kin} \equiv a^2\int dx^-\sum_{\bf x} 
i \bar{\Psi}({\bf x}) \gamma^+ \partial_+ \Psi({\bf x})
= 
\int dx^-\sum_{\bf x} 
i\bar{\Phi}({\bf x}) \gamma^+ \partial_+ \Phi({\bf x}) \ .
\ee
It is convenient to use a chiral representation,
where 
$\bar{\Psi}^\dagger = (u_{+}^*,v_{+}^*,
v_{-}^*,u_{-}^*)/
a 2^{1/4}$ decomposes into left (right) movers $v$ 
($u$) with sign of helicity $h=\pm$. The symbol $*$ means complex conjugate,
and
from now on we assume the transposition in hermitian conjugate
$\dagger$ acts on colour indices. In terms of these
one finds
\be
L_{kin} = \int dx^-\sum_{\bf x} \sum_{h=\pm}
u_{h}^\dagger({\bf x}) \gamma^+ \partial_+ u_h({\bf x}) \ , 
\ee
and therefore the canonical anti-commutation 
relations at equal light-cone times $x^+$ are
\bea
\left\{ u_h({\bf x}), u_{h^\prime}^*
({\bf y})\right\} &=&
\delta_{hh^\prime} \delta_{{\bf x}{\bf y}}
\delta(x^--y^-)\nonumber\\
\left\{ u_h({\bf x}), u_{h^\prime}
({\bf y})\right\} &=&
\left\{ u_{h}^*({\bf x}), u_{h^\prime}^*
({\bf y})\right\} = 0 \ .
\eea
For the Fock expansion we employ again a convenient
mixed representation, using longitudinal 
(continuous) momentum and transverse (discrete)
position variables
\be
u_h(x^+=0,x^-,{\bf x}) = 
\frac{1}{\sqrt{2\pi}}\int_0^\infty dk^+
\left[b_h(k^+,{\bf x}) e^{-ik^+x^-} 
+ d_{-h}^*(k^+,{\bf x}) e^{ik^+x^-}\right]
\ee
with
\bea
\left\{b_h(k^+,{\bf x}),
b^{*}_{h^\prime}(\tilde{k}^+,{\bf y})\right\}&=&
\delta_{hh^\prime}\delta_{xy}\delta(k^+-\tilde{k}^+)
\nonumber\\
\left\{b_h(k^+,{\bf x}),
b_{h^\prime}(\tilde{k}^+,{\bf y})\right\}&=&
\left\{b^{*}_h(k^+,{\bf x}),
b^{*}_{h^\prime}(\tilde{k}^+,{\bf y})\right\}= 0 \ ,
\eea
and likewise for $d^{*}_h(k^+,{\bf x})$.
$b^{*}_h(k^+,{\bf x})$ and 
$d^{*}_h(k^+,{\bf x})$ have 
the usual interpretation as creation operators for
a quark/antiquark with momentum $k^+$ and sign of helicity
$h\in \{{\pm }\}$ on the transverse site ${\bf x}$.

If we order the terms in the Hamiltonian 
$P^- = a^2 \sum_x \int dx^- {\cal H}$
according to the powers of the interactions,
to second order in the fields we find a `kinetic
energy' term
\be
{\cal H}^{(2)}({\bf x}) = 
\left(\mu_{f} + \frac{2r}{a}\right)^2 \bar{\Phi}
({\bf x})
\frac{1}{i\sqrt{2}\partial_-}\Phi({\bf x}) \ .
\ee
To third order, we obtain both a local `$1+1$ Coulomb'
term as well as a transverse hopping term
\be
{\cal H}^{(3)}({\bf x}) = 
{\cal H}^{(3)}_{Coul.}({\bf x}) +
{\cal H}^{(3)}_{hopp.}({\bf x}) \ .
\ee
The Coulomb coupling for fermions is local in the
transverse direction because both the fermions as well
as the longitudinal gauge field $A_+$ are defined
on the sites of the transverse lattice
\be
{\cal H}^{(3)}_{Coul.}({\bf x}) = \frac{G}{\sqrt{N}}
\Phi^\dagger({\bf x}) \gamma^+ A_+({\bf x})
\Phi({\bf x}) \ .
\ee
One obtains both transverse hopping terms that flip
the helicity of the fermions as well as terms that
do not flip the helicity
\be
{\cal H}^{(3)}_{hopp}({\bf x}) =
\kappa_a {\cal H}^{(3)}_{flip}({\bf x})+
\kappa_s {\cal H}^{(3)}_{noflip}({\bf x}) \ ,
\ee
where
\be
{\cal H}^{(3)}_{flip}({\bf x}) = i
\sum_r\Phi^\dagger({\bf x}) \frac{\gamma^r}
{i\partial_-} \left[M_r({\bf x}) \Phi({\bf x}
+a\hat{\bf r}) - M_r^\dagger({\bf x}-a\hat{\bf r})
\Phi({\bf x}-a\hat{\bf r})\right] \ ,
\label{eq:hopp1}
\ee
and
\be
{\cal H}^{(3)}_{noflip}({\bf x}) = 
\sum_r\Phi^\dagger({\bf x}) \frac{1}
{i\partial_-} \left[M_r({\bf x}) \Phi({\bf x}
+a\hat{\bf r}) + M_r^\dagger({\bf x}-a\hat{\bf r})
\Phi({\bf x}-a\hat{\bf r})\right] \ .
\label{eq:hopp2}
\ee
The relation between the bare couplings and the
coefficients of the hopping terms is given by
\bea
\kappa_a &=& - \frac{\mu_f +\frac{2r}{a}}{2\sqrt{2}a}\nonumber\\
\kappa_s &=&-r\frac{\mu_f +\frac{2r}{a}}
{\sqrt{2}a} \ .
\eea
We should emphasize that these hopping terms
also represent the most general terms 
bilinear in fermions and linear in the link fields
which are invariant under those symmetries that
are not broken by the transverse lattice.
Therefore, even though we have used canonical 
reasoning to derive these terms, they also
appear naturally in the colour dielectric expansion
at third order in the fields.

An interesting feature of the light-cone formulation is that
the fermion kinetic energy is quadratic in the
transverse derivative after eliminating the constrained 
component of the fermion field. For free fields in the continuum
\be 
{\cal H}_{kin} = \Phi^\dagger({\bf x}) 
\frac{\mu_{f}^2 - \partial_r^2}{i\sqrt{2}\partial_-}\Phi ({\bf x}) \ .
\ee
Therefore, if one
discretizes the free field transverse lattice Hamiltonian 
{\it after} eliminating $\Psi_{(-)}$, it appears
possible to write down a discretized Hamiltonian
for fermions which does not exhibit species doubling
\cite{hala}
\be
{\cal H}^{latt}_{kin} \sim \Phi^\dagger ({\bf x})
\frac{\mu_{f}^{2}}{i\sqrt{2}\partial_-}\Phi({\bf x})
- \sum_r \Phi({\bf x})  
\frac{1}{i\sqrt{2}\partial_-}\frac{\Phi({\bf x}+a\hat{\bf r})
-2 \Phi({\bf x}) + \Phi({\bf x}-a\hat{\bf r})}{a^2}\ .
\ee
In fact, in the light-cone formulation of the transverse 
lattice
there appears to be a whole class of light-cone Hamiltonians
that do not exhibit species doubling and yet they
are chirally invariant when $\mu_f=0$. To see this, let
us introduce a modified $r$-term \cite{hala}
\bea
\tilde{L}_r &=& a\tilde{r}\int d^2{\bf x} \int dx^- 
\sum_s\bar{\Psi}\frac{\gamma^+}
{2i\partial_-}D_s^2\Psi \nonumber\\
&\stackrel{lattice}{\longrightarrow}&
a \tilde{r} \sum_{\bf x}\int dx^-
\sum_s\bar{\Psi}({\bf x}) \frac{\gamma^+}
{2i\partial_-} \left[M_s({\bf x}) \Psi({\bf x}
+a\hat{\bf s}) - 2 \Psi({\bf x}) +
M_s^\dagger({\bf x}-a\hat{\bf s})
\Psi({\bf x}-a\hat{\bf s})\right] \ .
\eea
Note that $\tilde{r}$ has dimensions of mass because of the
extra $1/\partial_{-}$.
This modified $r$-term is chirally invariant 
because of the $\gamma^+$ matrix in 
between the bilinear $\bar{\Psi}\Psi$. It is 
Hermitian as well as local in the transverse direction.
However, it is non-local in the $x^-$ 
direction, but the non-locality is of the same type
$\sim \frac{1}{i\partial_-}$ as many other terms
that one has to deal with in the light-cone formulation, and
therefore one should consider this as a possible 
alternative to the standard Wilson approach. 

If one repeats the light-cone
Hamiltonian quantization with such a modified $r$ term, one
obtains the same hopping terms 
as in the usual Wilson approach Eqs.~(\ref{eq:hopp1},
\ref{eq:hopp2}). It should not be surprising that
these operators appear also here since they 
represent the most general hopping terms that are
permitted by symmetry to this order.
However, what does change is the dependence
of the coefficients of these terms on $\mu_f$ and 
$\tilde{r}$. One finds \cite{hala}
\bea
\kappa_a &=& - \frac{\mu_f}{2\sqrt{2}a}\nonumber\\
\kappa_s &=&-\frac{\tilde{r}}
{2\sqrt{2}a} \ .
\label{eq:ctilde}
\eea
 Eq.~(\ref{eq:ctilde}) also shows explicitly that,
even in the presence of the (modified) $r$ term,
fermion helicity is conserved for $\mu_f\rightarrow 0$.
This reflects the fact that fermion helicity is
conserved in the chiral limit of $QCD$. These issues
are discussed in further detail in the appendix.

\subsection{\it Numerical studies of light mesons}
We now apply the ideas to the practical calculation of light meson
boundstates.
The Hamiltonians we have discussed so far 
for fermions on the transverse lattice
contain the following parameters (in addition to
pure glue parameters)
\begin{enumerate}
\item A (kinetic) mass term for the fermions.
In numerical calculations that employ
a truncation of the Fock expansion, it is in general necessary
to make this mass term Fock sector dependent. For
example, if one truncates the Fock expansion above
three particles ($q$, $\bar{q}$, plus at most one 
link quantum) then the mass of fermions in the
two particle Fock component gets renormalised due
to the `dressing' of fermions with one link quantum. 
However, the truncation of Fock space would not 
permit the same dressing for a fermion that is
already in the three particle Fock component ---
a process that is also generated by the 
Hamiltonian if no Fock space truncation is used.
In order to compensate for this effect, it is 
necessary to allow for Fock sector dependent masses
in the Hamiltonian.
\footnote{We should also emphasize that these kinds
of issues also arise if the only cutoff used
is a DLCQ cutoff --- even if no Fock space 
truncation or any other truncation is applied.
In this case one would need to allow for a
momentum  dependent mass in the Hamiltonian 
\cite{dlcqisbad}.}
\item The helicity flip and noflip couplings
$\kappa_a$ and $\kappa_s$
\item The coupling $G$ of fermions to the longitudinal
component of the gauge field $A_+$. This coupling
constant is the same for fermions and link
fields because of gauge invariance.
\end{enumerate}
We should also comment here on the issue of
dynamical (spontaneous) chiral symmetry breaking.
In the light-cone formulation, physical states are 
constructed by applying creation operators to the
Fock vacuum. The Fock vacuum is an eigenstate of
the light-cone Hamiltonian, which at first appears to 
exclude any possibility for a complex vacuum 
structure and spontaneous symmetry breaking 
phenomena. However, first of all the trivial
Fock vacuum does not contradict nonzero vacuum
condensates. In fact, recently it has been 
demonstrated that nontrivial vacuum condensates are
obtained within the light-cone framework, if the operator
products appearing in those condensates are 
defined through a point-splitting procedure 
\cite{flenz}. Many model studies over the last years
have demonstrated that nontrivial vacuum structure
does not disappear on the light-cone, but is rather shifted
from the states to the operators. Therefore, on the
transverse lattice one expects that the coefficients of
operators in the Hamiltonian are renormalised in a non-trivial way
due to such ``vacuum effects''. 

The first study of meson spectra and structure on
the transverse lattice was done in Ref. \cite{hala},
where one can also find a more detailed discussion
of species doubling within this framework.
In that work, the form that we described above 
was chosen for the Hamiltonian.
The numerical work in Ref. \cite{hala} employed 
a `femtoworm approximation',
where the meson Fock space is truncated above three
particles (i.e. at most one link quantum in addition
to the $q\bar{q}$ pair). This is the  smallest
Fock space that allows transverse propagation of mesons
and hence a dependence of the energy $P^-$ on 
${\bf P}$. Transverse propagation proceeds by
hopping of a quark (or antiquark) by one link, which
implies creation of a link quantum on the link
connecting the $q\bar{q}$ pair. The antiquark (quark)
then follows, absorbing the link quantum in the
process (Fig.~\ref{fig:femto}). 
\begin{figure}
\unitlength1.cm
\begin{picture}(15,5)(0,22)
\includegraphics{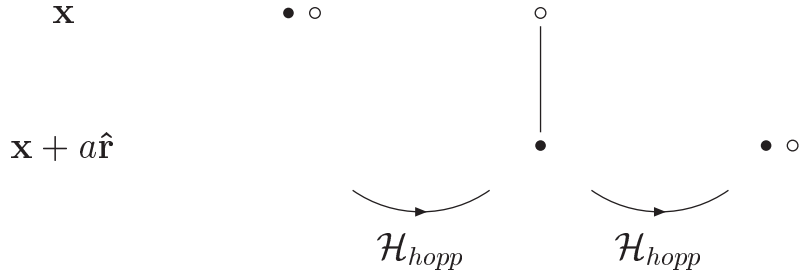}
\end{picture}
\caption{From left to right: sequence of `hopping'
interactions that leads to
transverse progagation (`hopping') of meson
by one lattice unit from site ${\bf x}$ to site
${\bf x}+a\hat{\bf r}$. The black and open dot 
represent the quark and antiquark respectively.}
\label{fig:femto}
\end{figure}
Transverse propagation over several links
proceeds by repeating the above sequence, which
resembles the stretching/contracting motion of an
inchworm (on a 1 fermi scale) --- hence the name.

The Hamiltonian that was introduced in Ref. 
\cite{hala} was later revisited in Refs. 
\cite{sd:mes,sudip1}. As a major improvement over 
Ref. \cite{hala}, Lorentz invariant $\pi$ and
$\rho$ dispersion relations were required as 
renormalisation conditions --- analogously to the 
pure glue studies described in Section 
\ref{sec:glue}. Furthermore, physical
transverse lattice spacings, as determined from the 
meson dispersion relations, were required to agree 
with the
lattice spacings obtained in the pure glue sector
($a \sim \frac{2}{3} \ {\rm fm}$). Finally, the
mass splitting within the $\rho$ multiplet, which
is another measure for the degree of rotational
symmetry breaking, was also considered.

In general, the Fock space for mesons in the large
$N$ limit on the transverse lattice consists of
open strings of link fields with quarks and
antiquarks at the ends.
If one truncates the Fock space at two particles
then mesons are described as bound states of a quark
and an anti-quark on the same site. Since transverse
propagation is not possible within this 
approximation, different transverse sites completely
decouple from each other and on each transverse site
one obtains the bound state equation and spectrum
that are identical to the ones found in large $N$
QCD in 1+1 dimensions \cite{hoof2} where one finds
a discrete spectrum of meson states.

The numerical calculations that have been performed
so far go one step beyond this lowest order 
approximation. They are performed in a Fock
space that was truncated above three particles (two quarks and one link).
For states with vanishing transverse momentum 
${\bf P}$, the wavefunction can be written
\begin{eqnarray}
|\psi(P^+)\rangle &= &  
 {2 a \sqrt{\pi} \over \sqrt{N} } \sum_{\bf x}  \sum_{h,h'}
\int_{0}^{P^+} dk^+ \left\{ \psi_{hh'}(x,1-x) 
\ b_{h}^{\dagger}(k^+, {\bf x})d_{h'}^\dagger(P^+ -k^+,
{\bf x}) |0\rangle \right\}\nonumber \\
&& + 
 {2 a \sqrt{\pi} \over N } \sum_{\bf x}  \sum_{h,h',r}
\int_{0}^{P^+} {dk_{1}^{+} dk_{2}^{+} \over P^+}  \label{link} \\ 
&& \times \left\{ \psi_{h(r)h'}(x_1,x_2,1-x_1-x_2) 
 \  b_{h}^{\dagger}(k_{1}^{+}, {\bf x})a^{\dagger}_{r}(k_{2}^{+},{\bf x})
d_{h'}^*(P^+ -k_{1}^{+}-k_{2}^{+}, {\bf x}+ a\hat{\bf r}) 
|0\rangle \right. \nonumber \\
&& \left. +  \psi_{h(-r)h'}(x_1,x_2,1-x_1-x_2)  
\  b_{h}^{\dagger}(k_{1}^{+}, {\bf x}+ a\hat{\bf r})
a^{\dagger}_{-r}(k_{2}^{+},{\bf x})
d_{h'}^*(P^{+}-k_{1}^{+}-k_{2}^{+}, {\bf x}) 
|0\rangle \right\} \nonumber \ . 
\end{eqnarray}
In this expression $\dagger$ acts on gauge indices
and  $x = k^{+}/P^+$ etc.. 
This truncated expansion for the states simplifies
the calculations considerably,
but as a result one should 
not expect restoration of Lorentz
symmetry to be as good as in the pure glue calculations of 
Section~\ref{sec:sol}, where no truncation was made.

For the wavefunction at nonzero transverse momentum 
$|\psi(P^{+},{\bf P})\rangle$, an
expansion very similar to the one above is
made. The only difference is that each term in the
summation over transverse lattice sites ${\bf x}$
is multiplied by an appropriate phase factor.
In Ref. \cite{sudip1} the phase factor chosen for
each term was $e^{i{\bf x}.{\bf P}}$ 
 in the two particle sector, while the
terms in the three particle Fock sector were
multiplied by a phase
$e^{i({\bf x}+\frac{a}{2}\hat{\bf r}).
{\bf P}}$, i.e. each state was assigned an 
effective transverse mean location equal to the
midpoint between the quark and the antiquark.
In Ref. \cite{sd:mes}, a slightly different phase
conventioned was made in the 3 particle Fock 
component; namely, for a state with the quark located 
on site ${\bf x}$ and the antiquark on site
${\bf x}+a\hat{\bf r}$ a phase factor $e^{i\bar{\bf x}.{\bf P}}$
was chosen, where
$\bar{\bf x}= x_1{\bf x}+ x_{3}  
({\bf x}+a\hat{\bf r})+ 
(1-x_1-x_{3})({\bf x}+
\frac{a}{2}\hat{\bf r})$. The latter choice 
is physically motivated by use of
the  transverse boost-rotation operators $M_{-r}$
(c.f. Eqn.~(\ref{boost})), 
but we should emphasize that both choices are equivalent. The
only difference is in the phase convention
for the Fock space amplitude of extended states.
As long as one uses either choice consistently
for the evaluation of physical observables, the
phase difference drops out.
The wavefunction is normalised
covariantly 
\be
\langle \psi(P^{+}_{1},{\bf P}_{1})|\psi(P^{+}_{2},{\bf P}_2)\rangle =
2P^{+}_{1} 
(2\pi)^3 \delta(P^{+}_{1} - P^{+}_{2}) \delta({\bf P}_{1} - {\bf P}_{2})
\eq
if
\be
1  =  \int_{0}^{1} dx \sum_{h,h'} | \psi_{hh'}(x,1-x)|^2  
 + \int_{0}^{1} dx_1 dx_2 \sum_{h,\lambda,h'}  
|\psi_{h(\lambda)h'}(x_1,x_2,1-x_1-x_2)|^2 \ . \label{normed}
\eq

In QCD, dimensionful parameters are the quark masses
and the string tension. The string tension is 
implicitly used if one inputs  from the pure
glue calculation the numerical value of the 
longitudinal gauge coupling $\bar{G}$ as well as the
transverse lattice spacing in physical units.
The $\pi$ meson mass can be  used to fix  the
quark masses. 
In field theory, the mass of the $\pi$ is related to
spontaneous chiral symmetry breaking. A massless pion,
in a Lorentz covariant theory, usually means chiral 
symmetry (realised in the Goldstone mode).
The mass {\em difference} between 
the $\pi$ and $\rho$ mesons is also directly related to chiral
symmetry breaking in light-cone formalism. One would hope that 
this scale was generated dynamically once Lorentz covariance
is imposed, holding the pion mass at (or near) zero. 
That is, helicity-violating interactions in the
coarse transverse lattice Hamiltonian should be forced to have
non-zero couplings, at least in a truncated Fock space,
in order to obtain covariance and a massless
pion.\footnote{With no truncation of Fock space the scale could be
generated by wee parton effects \cite{casher}.}
Unfortunately, Lorentz symmetry is sufficiently
badly violated in the one-link approximation that this
is difficult to verify practice. Other problems of this approximation
in implementing
this reasoning are that only a single helicity violating
interaction ${\cal H}_{flip}^{(3)}$ is available 
and stability of the groundstate is
difficult to assess because of the truncation of many-particle
states.
Calculations so far have therefore modelled the chiral
symmetry breaking aspects by fixing the $\pi$-$\rho$
mass difference, and hence the helicity violating coupling to
${\cal H}_{flip}^{(3)}$ 
from experiment. A more fundamental
understanding of the origin of this term must await relaxation
of the one-link approximation.

Using the dimensionful scale $\bar{G}=G\sqrt{(N^2-1)/N}$ 
we introduce the following renormalised dimensionless parameters
\be
{\mu_b \over \bar{G}} \to m_b \ \ ; \ \ {\mu_{f} \over \bar{G}} \to
m_f \ \ ;   {\kappa_a \sqrt{2N} \over \mu_f \bar{G}} \to k_a \ \ ; \ \
{\kappa_s \sqrt{2N} \over \mu_f \bar{G}} \to k_s \ . \label{coup} 
\eq
Using the truncated Fock expansion,
we can project the eigenvalue equation
$2P^+ P^- |\psi (P^+)\rangle = {\cal M}^2  |\psi
(P^+)\rangle$ for ${\bf P} =0$ 
onto Fock basis states, to derive the following 
set of coupled integral
equations for individual Fock amplitudes:
\begin{eqnarray}
{{\cal M}^2 \over \overline{G}^2} \psi_{hh'}(x_1,x_2) & = & 
\left( {m_{f}^{2} \over x_1}+ {m_{f}^{2} \over x_2} \right)
\psi_{hh'}(x_1,x_2) + K(\psi_{hh'}(x_1,x_2))
\nonumber \\
&& + {(k_{a}^{2} + k_{s}^{2}) \over \pi} \left( {1 \over x_1}
\int_{0}^{x_1} {dy \over y} +  {1 \over x_2} \int_{0}^{x_2} {dy \over y}
\right) \psi_{hh'}(x_1,x_2)              \nonumber \\
& & - \sum_{\lambda} \left\{ {m_f k_s \over 2 \sqrt{ \pi}} \int_{0}^{x_1}
        \frac{dy}{\sqrt{y}} \left(\frac{1}{x_1-y} + \frac{1}{x_1}
\right) \psi_{h(\lambda)h'}(x_1-y,y,x_2) \right. 
 \label{int1}    \\
& & \left. + {m_f k_s \over 2 \sqrt{  \pi}} \int_{0}^{x_2}
        \frac{dy}{\sqrt{y}} \left(\frac{1}{x_2-y} + \frac{1}{x_2}
\right) \psi_{h(\lambda)h'}(x_1,y,x_2-y) \right.
             \nonumber \\
& & + \left.  {{\rm Sgn}(\lambda) m_f k_a(h {\rm i} 
\delta_{|\lambda |1} + \delta_{|\lambda | 2}) \over 2 \sqrt{ \pi}} 
\int_{0}^{x_1}
        \frac{dy}{\sqrt{y}} \left(\frac{1}{x_1-y} - \frac{1}{x_1}
\right) \psi_{-h(\lambda)h'}(x_1-y,y,x_2) \right. 
  \nonumber    \\
& & \left. - {{\rm Sgn}(\lambda) m_f k_a( h'{\rm i} 
\delta_{|\lambda |1} +   \delta_{|\lambda |2})
  \over 2 \sqrt{ \pi}} \int_{0}^{x_2}
        \frac{dy}{\sqrt{y}} \left(\frac{1}{x_2-y} - \frac{1}{x_2}
\right) \psi_{h(\lambda)-h'}(x_1,y,x_2-y) \right\}     \nonumber \ ,
\end{eqnarray}
\begin{eqnarray}
{{\cal M}^2 \over \overline{G}^2} \psi_{h(\lambda)h'}(x_1,x_2,x_3) & = & \left(
{m_{b}^{2} \over x_2}  
+  {m_{f}^{2} \over x_1}+ {m_{f}^{2} \over x_3} \right)
\psi_{h(\lambda)h'}(x_1,x_2,x_3) + K(\psi_{h(\lambda)h'}(x_1,x_2,x_3))
\nonumber \\
& & - {m_f k_s  \over 2 \sqrt{  \pi x_2}} \left(
        \frac{1}{x_1} + \frac{1}{x_1 + x_2} \right) \psi_{hh'}(x_1+x_2,x_3) 
\nonumber \\
& & - {m_f k_s  \over 2 \sqrt{  \pi x_2}} \left(
        \frac{1}{x_3} + \frac{1}{x_2+x_3} \right) \psi_{hh'}(x_1,x_2+ x_3) 
\label{int2} \\
& &  - {\rm Sgn}(\lambda) {(h{\rm i}\delta_{|\lambda |1}+ 
\delta_{|\lambda |2}) m_f k_a  
\over 2\sqrt{  \pi x_2}} \left(
        \frac{1}{x_1} - \frac{1}{x_1 + x_2} \right) \psi_{-hh'}(x_1+x_2,x_3) 
\nonumber \\
& & + {\rm Sgn}(\lambda){(h'{\rm i}\delta_{|\lambda |1} + 
\delta_{|\lambda |2}) 
m_f k_a  \over 2 \sqrt{  \pi x_2}} \left(
        \frac{1}{x_3} - \frac{1}{x_2+x_3} \right) \psi_{h-h'}(x_1,x_2+
x_3)
\ . \nonumber 
\end{eqnarray}
As stated earlier, for full generality one should also allow
the kinetic masses $m_{f}^{2}$ to be Fock sector dependent.
The conventions
of ref.\cite{fran} have been adopted for the instantaneous gluon kernels
\be
K(\psi_{hh'}(x_1,x_2)) =  {1 \over 2 \pi} \int_{0}^{1} dy
\left\{ {\psi_{hh'}(x_1,x_2) - \psi_{hh'}(y,1-y) \over (y-x_1)^2}
	\right\} \ ,
\eq
\begin{eqnarray}
K(\psi_{h(\lambda)h'}(x_1,x_2,x_3)) & = & {1 \over 2 \pi} 
\int_{0}^{x_2+ x_3} dy  {(x_3 +2x_2 -y) 
\over 2(x_3 -y)^2\sqrt{x_2(x_2 + x_3 -y)}}\left\{
\psi_{h(\lambda)h'}(x_1,x_2,x_3) \right. \nonumber \\
&& \left.  - \psi_{h(\lambda)h'}(x_1,x_2 +x_3-y,y)  \right\}   
\nonumber \\
&&
+ {1\over 2 \pi x_3} \left(  \sqrt{1 + {x_3 \over x_2}} -1 \right)
\psi_{h(\lambda)h'}(x_1,x_2,x_3)
  \nonumber \\
&& + {1 \over 2 \pi} \int_{0}^{x_1+x_2} dy {(x_1 +2 x_2 -y)
\over
2(x_1-y)^2 \sqrt{x_2 (x_1 + x_2 -y)}} 
\left\{ \psi_{h(\lambda)h'}(x_1,x_2,x_3)
\right.  \nonumber \\ &&
\left.  -  \psi_{h(\lambda)h'}(y,x_1 + x_2-y,x_3) \right\}
\nonumber \\
&& + {1\over 2 \pi x_1 }\left( \sqrt{1 + {x_1\over x_2} } -1 \right) 
\psi_{h(\lambda)h'}(x_1,x_2,x_3)  \ . 
\end{eqnarray}
These coupled integral equations have to be solved
numerically. Note that iterating the link-field
emission terms, i.e. those terms 
in Eqs. (\ref{int1}) and (\ref{int2})
that mix different Fock components, one obtains
a (logarithmically) divergent self-energy correction.
In order to cancel this divergence, a divergent
counter-term has been added to the second line of 
Eq. (\ref{int1}).
In numerical calculations this divergence has to be
regularized. In Ref. \cite{sd:mes} this divergence
is automatically regularized since in DLCQ integrals
are replaced by summations, which has a similar 
effect on small $x$ divergences as introducing a
lower bound on the $x$ integration.
In Refs. \cite{hala,sudip1} an alternative regulator
has bees used, where link-field emmission and
absorption vertices are multiplied by a factor
$x^\varepsilon$, where $x$ is the momentum of the 
link field and $0< \epsilon $ is a small 
regularization parameter. 

In Ref. \cite{sd:mes}, a DLCQ basis was used to
cast these coupled integral equations into matrix
form. However, for theories with fermions, no
improvement techniques \cite{dv1,dv2} have been developed yet to
accelerate the convergence in the DLCQ parameter $K$, which is very
slow as a consequence.
If basis functions are used instead, a  
convenient choice is
\be
\psi(x_{f_1},x_{f_2},...;x_{b_1},x_{b_2},...)
= x_{f_1}^\alpha x_{f_2}^\alpha\cdot \cdot \cdot 
x_{b_1}^{(\alpha+\frac{1}{2})} 
x_{b_2}^{(\alpha+\frac{1}{2})}\cdot \cdot \cdot 
P(x_{f_1},x_{f_2},...;x_{b_1},x_{b_2},...) \ ,
\label{eq:ansatz}
\ee
where $x_{f_i}$ and $x_{b_i}$ denote longitudinal  momentum fractions of 
fermions and bosons respectively and
$P$ belongs to a complete set of polynomials. 
In this ansatz the
exponent for bosons (link fields), 
$\alpha+\frac{1}{2}$, is chosen in order
to cancel the factors of $\frac{1}{\sqrt{x_b}}$
that typically arise at vertices involving bosons.
That way, all necessary integrals can be 
performed analytically with such an ansatz,
using again  Eq.~(\ref{eq:integral}).

In actual calculations using Eqn.~(\ref{eq:ansatz}), $\alpha$ was not
taken from  an equation analogous to Eq.~(\ref{wform}) available in the
two-particle truncation, 
because the naive end-point behaviour
gets renormalized anyway, due to mass 
renormalization for example. In all but the lowest
Fock component, the wavefunction is in fact finite
as the fermion momentum approaches zero \cite{fran}.
However, as long as one uses a complete 
set of functions $P$ in the expansion, the actual 
ansatz for the end-point behaviour does not  
matter in principle. 
It only matters from a practical point of 
view, since convergence in the 
Hilbert space expansion will be degraded by using
a basis of functions with the wrong small momentum behaviour.
The sensitivity to the $x^{\epsilon}$ regulator  
can be studied in final results, which should be finite in 
the limit that the regulator is removed (although convergence in the
number of basis functions will be degraded as the regulator is made
smaller). 
The  basis function method was used
to produce the results \cite{sudip1,sudip2} that 
we show later. After errors due to  cut-off 
extrapolation have been quantified, these results are consistent with
those obtained by DLCQ \cite{sd:mes}, indicating that both methods are
giving reasonable results.

In the numerical calculations, $\alpha \in (0,1)$
was picked and for a fixed value of $\alpha$ the
convergence of physical observables as a function
of the number of polynomials $P$ was studied. 
For typical coupling choices that led to the correct meson masses
and relativistic dispersion, this convergence was found to be
sufficiently rapid for values of 
$\alpha \approx 0.5$. 
Typical results, which demonstrate the size of 
residual Lorentz symmetry breaking in the light mesons are displayed
in Fig.~\ref{fig:disp}.
\begin{figure}
\unitlength1.cm
\begin{picture}(15,0)(-19.2,7.5)
\includegraphics{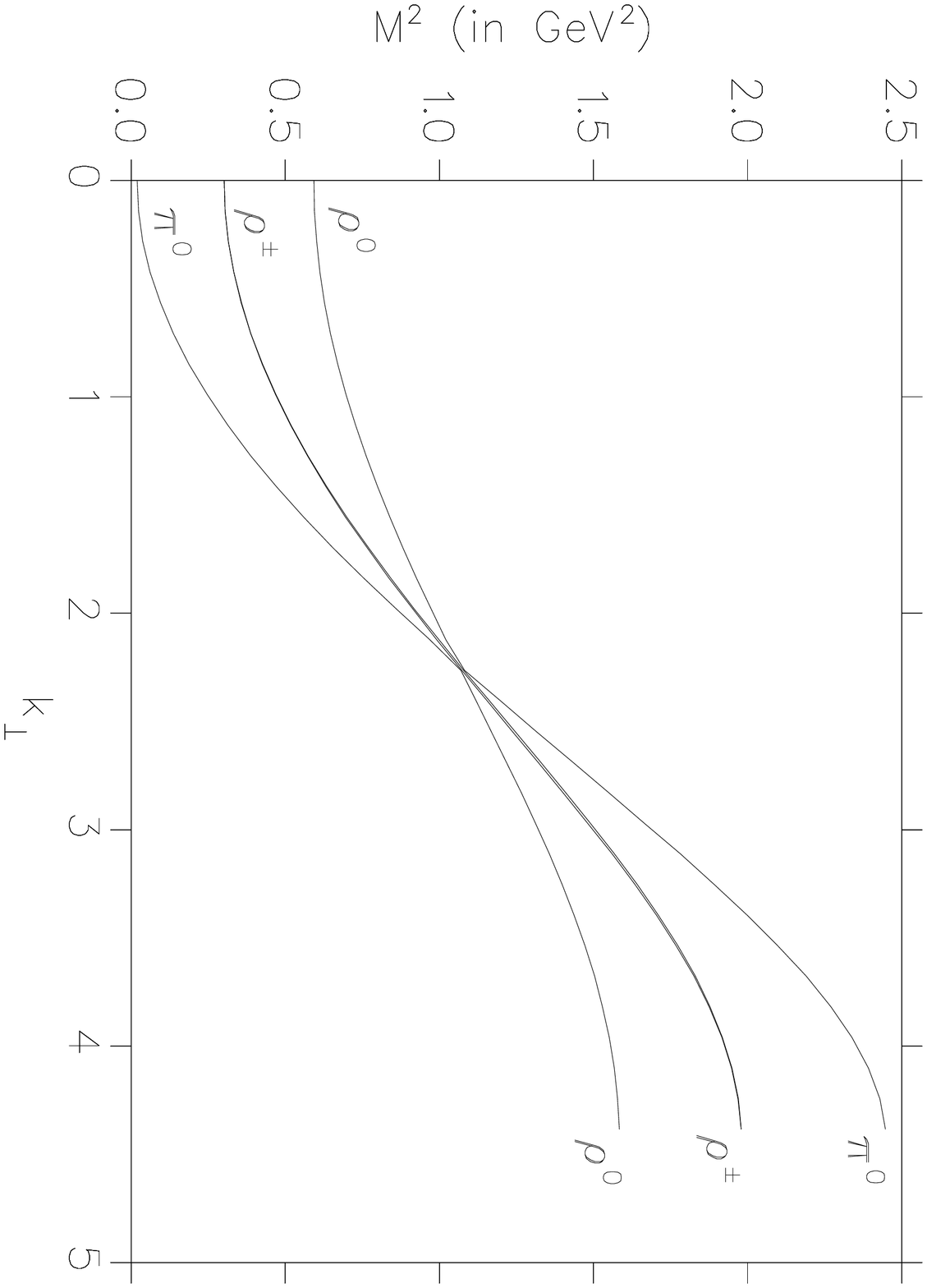}
\end{picture}
\begin{picture}(15,6.7)(-9.9,.8)
\includegraphics{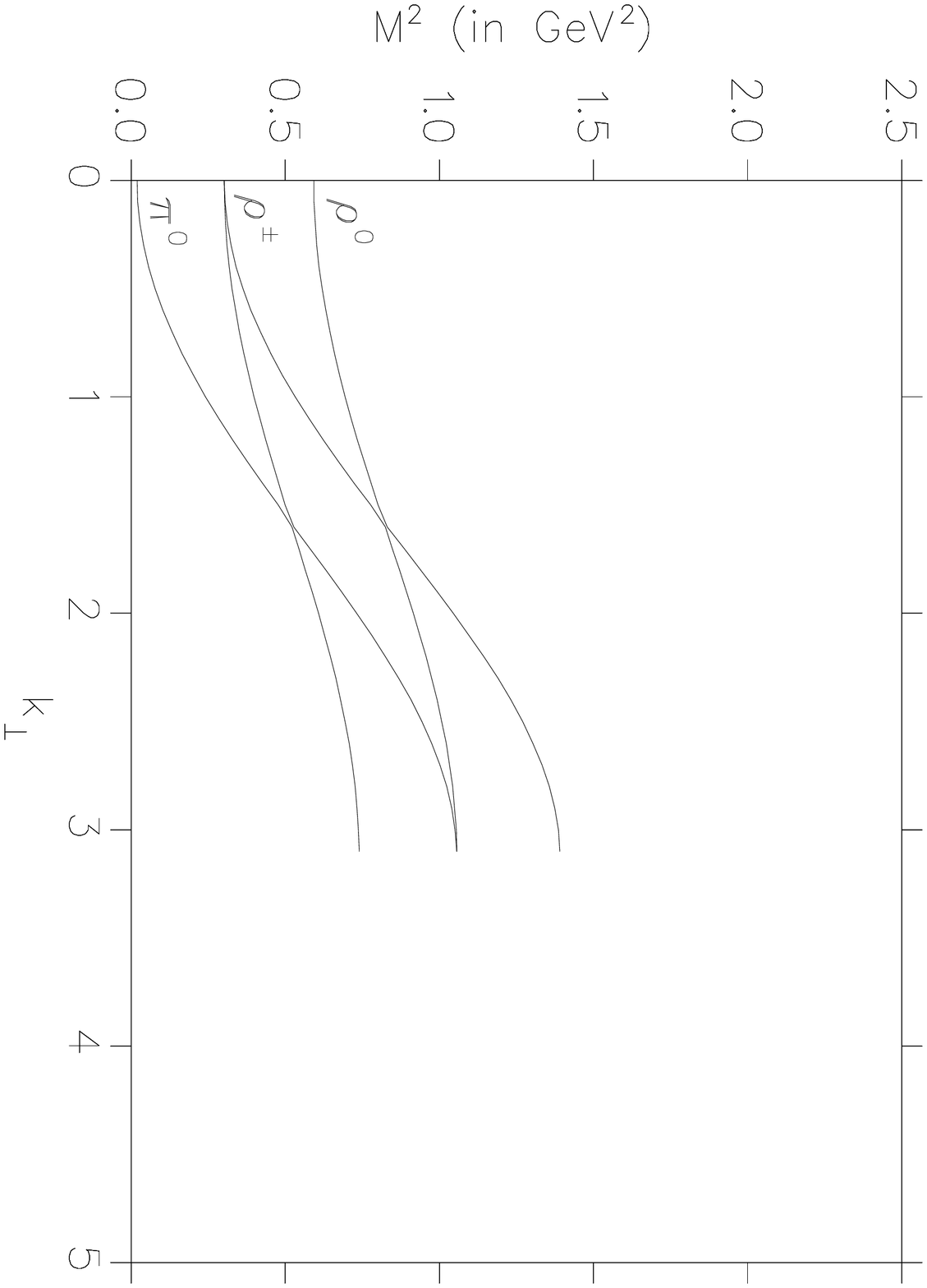}
\end{picture}
\caption{Dependence of  
$M^2 \equiv 2P^+P^-$ of $\pi$ and $\rho$ mesons on
their transverse momentum $k_{\perp}$, for momenta along a lattice
axis and along a diagonal respectively. The
superscripts $\pm$, $0$ refers to the helicity at 
$k_{\perp}=0$. An exactly relativistic result should follow
the form $M^2 = {\cal M}^2 + k_{\perp}^{2}$.}
\label{fig:disp}
\end{figure}
Future calculations including 
additional terms in $P^-$ as well as higher
Fock components can improve this situation.

\begin{figure}
\unitlength1.cm
\begin{picture}(15,8.5)(-15.5,1.2)
\includegraphics{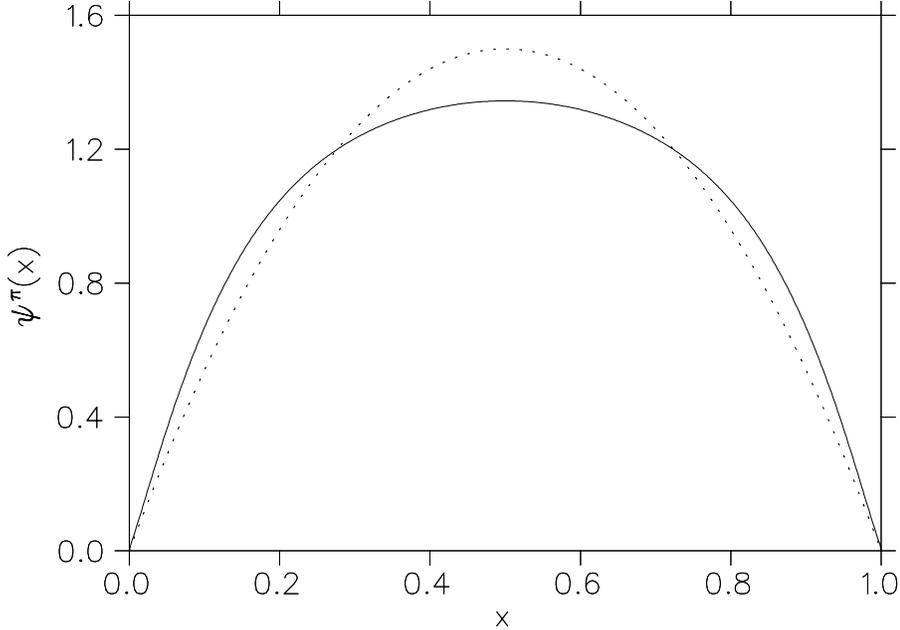}
\end{picture}
\caption{Pion distribution amplitude calculated on
the transverse lattice. For comparison, the 
asymptotic shape is shown as a dashed line. Both curves are 
normalised to area one.}
\label{fig2}
\end{figure}

Twist-2 distribution amplitudes for the helicity-zero $\pi$ and 
$\rho$ mesons are defined by (covariant normalisation)
\bea
f_\pi \phi_\pi(x) &=& \int \frac{dx^-}{2\pi}
\left\langle 0 \left| \bar{\Psi}(0)\gamma^+\gamma_5
\Psi(x^-)\right|\psi_\pi(P)\right\rangle e^{ixP^+x^-}
\nonumber\\
f_{\rho_0}\phi_{\rho_0}(x) &=& \int \frac{dx^-}{2\pi}
\left\langle 0 \left| \bar{\Psi}(0)\gamma^+
\Psi(x^-)\right|\psi_\rho(P)\right\rangle e^{ixP^+x^-}
\ .\label{eq:distampl}
\eea
Because of the 
appearance of the `good' $\gamma^+$ components
of the currents  in Eq.~(\ref{eq:distampl}), one can express these
distributions entirely in terms of the
two particle Fock component of these mesons
\bea
f_\pi
\phi_\pi(x) &=& \frac{2}{a} \sqrt{\frac{
N}{\pi}}\psi^\pi_{+-}(x)\nonumber\\
f_{\rho_0}
\phi_{\rho_0}(x) &=& \frac{2}{a} \sqrt{\frac{
N}{\pi}} \psi_{+-}^{\rho_0}(x)
\label{eq:fpi}
\eea 
where the distribution amplitudes are normalised
to one; for example, $\int_0^1dx \phi_\pi(x)=1.$
\footnote{Note that the explicit expression given for
$f_\pi$ in Ref. \cite{sd:mes} has an incorrect
normalisation factor.}

Using $N=3$ in these formulae, the decay constants one  calculates are 
$f_\pi \approx 300\, {\rm MeV}$ and 
$f_{\rho_0} \approx 270\, {\rm MeV}$ in the one-link
approximation, which are larger
than the experimental results of $f_\pi=132\, {\rm MeV}$ and
$f_{\rho_0}=216\, {\rm MeV}$ \cite{pdg}.
This discrepancy is most likely caused by the Fock 
space truncation, since including  higher Fock 
components tends to decrease the probability to find a
hadron in its lowest Fock component and therefore
also the normalization of the distribution amplitude.
For the $\rho$ meson the discrepancy is significantly
smaller, suggesting perhaps that it is less sensitive
to the chiral symmetry breaking aspects, that have only been
crudely treated.

Results for the $\pi$ distribution amplitude 
are shown in Fig.~\ref{fig2}. Although $\phi_\pi(x)$ has some
resemblance to the asymptotic distribution
$\phi^\pi_{as}(x)=6x(1-x)$ \cite{lepage}, the transverse lattice 
result is somewhat broader, but clearly
does not exhibit any `double hump' feature, found in early QCD sum
rule analysis \cite{cz}. This distribution has recently been studied
experimentally for the first time directly \cite{ashery}. Conventional 
Euclidean lattice QCD simulations have yielded estimates of the second
moment of $\phi_\pi(x)$ \cite{lattice}. These results, together with the more
recent `non-local' QCD sum rule calculations \cite{nlcsr}, 
are all consistent with the single
humped form.
The $\rho$ meson distribution amplitude from the transverse lattice looks
similar, although it is slightly more peaked, which
probably reflects the weaker binding of the quarks in the 
$\rho$.

In the light-cone framework, parton distributions
are also very easy to evaluate, since they are
momentum densities summed over all Fock components.
For example, for the unpolarized distribution 
function for quarks one finds
\be
q(x) = \sum_{h,h^\prime} 
\left|\psi_{hh^\prime}(x,1-x)\right|^2
+ \sum_{h,\lambda,h^\prime}
\int_{0}^{1-x} dy  \left|\psi_{h(\lambda)h^\prime}(x,y,1-x-y)
\right|^2 + \cdots \ , \label{eq:pdf}
\ee
where $\cdots$ indicates contributions from
higher Fock components that were not included in
the present calculations. As an example, the
distribution function from the transverse lattice for
the $\pi$ meson in the one-link approximation is shown in Fig.~\ref{fig:pdf}.
The peak around $x\approx 0.5$
is mostly due to the $q\bar{q}$ Fock component, while
contributions from higher Fock components dominate
at smaller values of $x$. 
This distribution can inferred from a fit to 
pion-nucleon Drell-Yan scattering \cite{grv}, has been recently
investigated
in Dyson-Schwinger models \cite{craig}, and has had the lowest moments
measured in Euclidean lattice QCD \cite{mom}. More precise experimental
data will soon be provided by the Jefferson laboratory, which will
allow to compare more meaningfully the various theoretical predictions.
Polarized parton distributions can
be investigated in a similar manner, for mesons with spin \cite{sd:mes}.

\begin{figure}
\unitlength1.cm
\begin{picture}(15,8.5)(-15.5,0.8)
\includegraphics{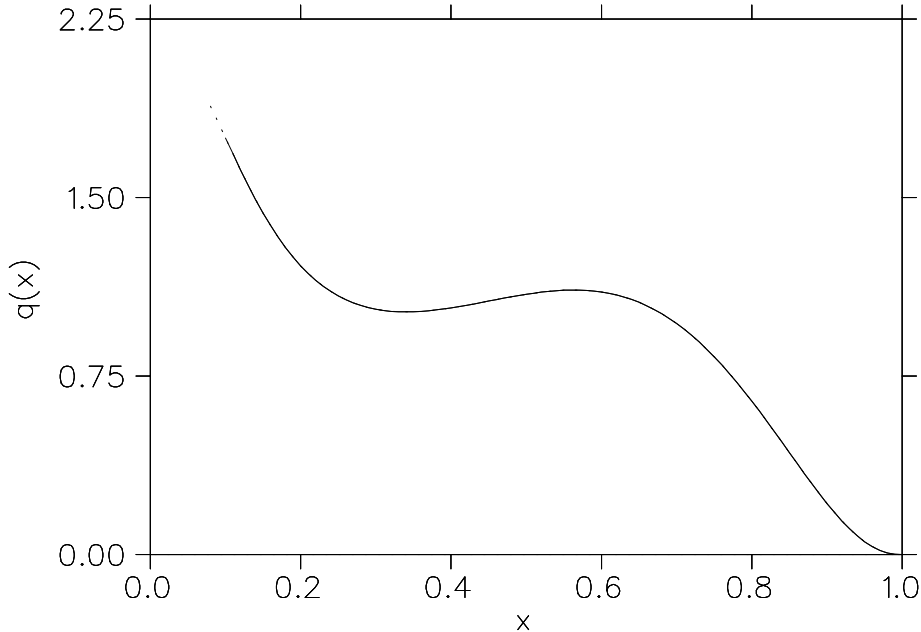}
\end{picture}
\caption{Quark distribution function Eq.~(\ref{eq:pdf})
for the pion.}
\label{fig:pdf}
\end{figure}
Parton distribution functions are scale dependent.
For the parton distribution functions calculated on
the transverse lattice the relevant scale should
be $Q_0\sim \frac{1}{a}$, where $a \sim 0.5-0.7$ fm. Note that the
precise relationship between this scale
and those used in perturbative renormalisation schemes is
unknown so long as one cannot match onto perturbation theory. 
Because of this uncertainty in scale
setting and also because the use of perturbation 
theory at these low momentum scales is questionable,
we do not show `evolved' parton distribution 
functions here.
 
Another very interesting class of observables that 
are accessible on the transverse lattice are hadron
form factors. For example, the pion elastic electromagnetic
form factor defined by
\be
\langle\psi(P')|  j^{\mu}_{\rm em}(q) |\psi(P) \rangle \equiv 
F(-q^2)(P'+P)^{\mu}\ .
\eq 
In a light-cone framework, it is 
usually preferable to evaluate form factors in the 
Drell-Yan frame \cite{dy}
($q^+=0$, where $q$ is momentum transfer). Only terms that are 
diagonal in Fock space (no pair creation terms) 
contribute to matrix elements of the `+' component of current 
operators in this frame. From a practical point of 
view, this is a great advantage since one can then 
express the form factor entirely in terms of 
convolution integrals of Fock-space amplitudes for 
the initial and final state hadron.
For space-like elastic form factors $F(-q^2)$, $q^2=2q^+q^--
{\bf q}^2<0$, it is always possible to choose 
a frame with $q^+=0$ and therefore ${\bf q}^2
= -q^2$. In terms of the 2 and 3 particle Fock space
amplitudes on the transverse lattice one  finds
\begin{eqnarray}
F(-q^2) & = & \int_{0}^{1} dx_1 dx_2
\sum_{h,h'} \delta (x_1 + x_2 -1) |\psi_{hh'}(x_1,x_2)|^2  \\
&& + 2\sum_{h,h',r} \int_{0}^{1}dx_1 dx_2 dx_3 \delta (x_1 + x_2 +x_3 -1)
\cos \left( (x_1 + {x_2 \over 2})a {\bf
q}.\hat{\bf r}
\right) 
|\psi_{h(r)h'}(x_1,x_2,x_3)|^2 \ , \nonumber 
\end{eqnarray}
where we used the quadratic expression for the electromagnetic
current $e \overline{\Psi} \gamma^{\mu} \Psi$ and used
G-parity and transverse reflection symmetries 
to simplify the final expression. Note that the  normalisation condition 
$F(0) = 1$ is ensured by the wavefunction normalisation Eq.(\ref{normed})

The calculations in the one-link approximation
yield a value for the root mean square radius of the pion, determined from
$r_\pi^2 \equiv 6 \frac{d}{dq^2} \left( F(-q^2)
\right)_{q^2=0}$, of  $r_{\pi} =(0.3\pm .05 
\,{\rm fm})$, where the main source of uncertainty is
the transverse lattice spacing. This result is much 
smaller than the experimental value
$r_\pi= 0.663\, {\rm fm}$ \cite{rpi}. 
Because we have truncated the Fock space
above the three particle Fock component, the quark
and antiquark cannot separate from each other by more
than one lattice spacing, i.e. the quark or antiquark
can seperate from the center of mass\footnote{
Actually, it is the separation from the center of
momentum that matters here, but for the qualitative
discussion below this distinction is not very 
significant numerically.} only by about
half the lattice spacing. This limit also  
represents a rough upper bound for the possible 
values of the rms radius and the numerical results come
close to that limit.
Another  difficulty with 
evaluating form factors on the transverse
lattice, using a large transverse momentum transfer, is that
this will resolve the coarse lattice
in the transverse direction.

We should emphasize that the restriction to the Drell
Yan frame ($q^+=0$) is not absolutely necessary.
One can evaluate meson form factors using purely 
longitudinal momentum transfers. 
Of course, since this 
necessarily implies $q^+\neq 0$, one has to include
matrix elements of the current operator that are
off-diagonal in Fock space (change the number of partons). 
As long as no Fock space
truncation is made, it is straightforward to
include those off-diagonal terms. 
The off-diagonal contributions
may be significant even if the probability to find
additional $\bar{q}q$ pairs in the state is small as
the following example illustrates: for $N$ large,
the amplitude to find two $\bar{q}q$ pairs in a
meson meson scales like $\frac{1}{\sqrt{N}}$, i.e.
the probability is $\frac{1}{N}$ suppressed.
Nevertheless, since the vacuum to meson matrix 
element of the vector current operator scales like
$\sqrt{N}$ (while diagonal matrix elements are
${\cal O}(1)$), the off-diagonal terms still 
contribute to the same order in $N$ as the 
diagonal terms and hence survive the large $N$
limit. Therefore, even though higher $\bar{q}q$ 
Fock components are unimportant for the energy of the
meson, they are still important in order to properly
describe off-diagonal contributions to the form
factor when $q^+\neq 0$. This observation may also 
provide a clue as to how one could estimate these 
off-diagonal contributions to the form factor 
without explicitly enlarging the Fock space in the 
Hamiltonian, namely
by calculating the mixing of the current with 
virtual meson states and then estimating the
coupling of those virtual meson states to the
target hadron perturbatively. In fact, such a 
procedure has been successfully applied to derive
and study exact expressions for off-diagonal 
constributions to form factors and generalised 
parton distributions in 
$QCD_{1+1}(N\rightarrow \infty)$ \cite{einhorn}.

In summary, we have found that a wide range of observables have simple
formulae in terms of the light-cone wavefunctions computed on the transverse
lattice. Of course, the accuracy of the final results are then limited by
the accuracy of the computed wavefunctions; the one-link approximation
employed so far is not quantitatively  realistic.
Nevertheless, it is in principle straightforward to systematically relax this
approximation.
Alternatively, one could use larger quark masses, since mesons containing
heavier quarks tend to be smaller and therefore
may be affected less by a Fock space truncation.
In addition, one would intuitively expect that
a constituent picture works better for heavier 
quarks.
Of course, for mesons containing both a heavy quark
and a heavy antiquark one may run into the opposite
problem, namely that large intrinsic transverse
momenta start to resolve the very coarse transverse
lattice. For quarks with masses on the order
of strange quark masses this may not yet be a 
serious problem. 

\subsection{\it B mesons on the transverse lattice}
In the limit where the $b$ quark is infinitely heavy,
it acts as a static colour source to which the light
quark is bound. The structure of the light degrees
of freedom becomes independent of the mass of the
heavy source in this limit. Therefore, even though
the dynamics of the light degrees of freedom in such
a `heavy-light' system is still as complex as in 
light mesons, a new symmetry (the heavy quark 
symmetry) emerges, which allow one to relate matrix 
elements from different heavy-light systems to one 
another.
symmetry to decays of $B$ and $D$ mesons play an
important role in the determination of Standard Model
parameters. Of course, even though the heavy quark
symmetry allows to relate many hadronic matrix 
elements to one another, it still leaves many
matrix elements as unknown parameters, which must
be determined either by fit to experimental data or
by nonperturbative QCD calculations.

A general method to introduce static sources into
the light-cone formalism has been introduced in Ref.
\cite{mb:zako} and is very similar to the methods 
that were used in calculations of the static quark 
antiquark potential that were described in Section 3.
However, the extension of the light meson
calculations to such a heavy-light system can also 
be done by working with heavy quarks that have a
finite mass and simply making that mass very large.
Since the static source that the heavy quark 
represents does not propagate, no new parameters,
such as hopping parameters, 
appear in the Hamiltonian for such a system.
The latter was the approach chosen in Ref. 
\cite{sudip2}, for a calculation in the one-link 
approximation that we now briefly review.

An important parameter in $B$ meson phenomenology
is the `binding energy' $\bar{\Lambda}=m_B-m_b
+{\cal O}\left(\frac{1}{m_b}\right)$. 
Different possibilities for its calculation exist.
For example, one can directly use the numerically
calculated energy eigenvalues, or one can extract
$\bar{\Lambda}$ from the average light-cone momentum 
carried by the light degrees of freedom. In either
case, the resulting value of $\bar{\Lambda}\approx
1.0\, {\rm GeV}$, found in ref.\cite{sudip2} in the 
one-link
approximation, is larger than results obtained
using other methods. For example, using 
Schwinger-Dyson techinques, one finds \cite{roberts}
$\bar{\Lambda}\approx 0.7\, {\rm GeV}$ and
Euclidean lattice gauge theory calculations yield
\cite{elmc:Lambda} $M_B-M_b \approx 0.9\, {\rm GeV}$.
It is not yet clear what
causes $\bar{\Lambda}$ to be  large in these transverse
lattice calculations, but it is conceivable that higher
Fock components  will
lead to a partial screening of the longitudinal
string tension and hence a lowering of 
$\bar{\Lambda}$. 

For the decay constant, which also plays an 
important role in B-meson phenomenology, 
$f_B\approx 240\,{\rm MeV}\pm 20\,{\rm MeV}$ is found in the 
same approximation,  also larger than those obtained using other 
methods. For example, 
Euclidean lattice gauge theory \cite{elmc:fB} and 
and Dyson-Schwinger models \cite{roberts}
yield $f_B=175\,{\rm MeV}$ and $f_B=182\,{\rm MeV}$
respectively (it has yet to be accurately measured experimentally).
However,  the discrepancy is much smaller
than for $f_\pi$. This result is consistent with the
observation that the Fock expansion also seems to 
converge much more rapidly for $B$ mesons.
One should not be surprised to find more realistic
results for $B$-mesons than for the $\pi$ in the one-link
approximation. On the basis of vector meson dominance,
both hadrons are expected to have an rms radius 
of about $\frac{2}{3}\, {\rm fm}$. In the case of the
$B$ meson however, the rms radius is the average
distance between the static source and the light
quark, while the rms in the $\pi$ is given by
the average distance from the center of mass
(or momentum), which is about half the separation
between $q$ and $\bar{q}$. Since the Fock space
truncation did not allow for separations
between $q$ and $\bar{q}$ of more than $0.7\, {\rm fm}$
in both calculations, the quark in the $\pi$
could not separate from the center of mass by more
than $0.35\, {\rm fm}$ while the $q$ in the B-meson
can separate from the center of mass by $0.7\, { \rm fm}$.
The more rapid convergence of the Fock expansion
is also not unexpected. The static quark in the
$B$ meson does not `hop' and therefore it is 
not dressed with virtual link quanta. Therefore,
already if one neglects interference effects
due to exchanged link quanta, one would expect
about twice as many link quanta in light-light
mesons as compared to heavy-light mesons, from
dressing of quarks.

The B-meson (twist two) distribution amplitude, which
plays an important role in exclusive $B$ decays is
shown in Fig.~\ref{fig:Bdist}. In the limit 
$m_b\rightarrow \infty$, the $b$ quark carries 100\%
of the momentum of the $B$-meson. For finite, but
large $m_b$ the wave function is localized near 
the region where the $b$ quark carries momentum
fraction $1-\frac{\bar{\Lambda}}{m_b} < x_b <1$.
We therefore rescale the $B$ meson distribution
amplitude
in such a way that it has a finite $m_b\rightarrow
\infty$ limit
\be
\phi_\infty(z) \equiv \lim_{m_b\rightarrow \infty}
\frac{1}{\sqrt{m_b}}
\psi_{m_b}\left(1-\frac{z}{m_b}\right) \ ,
\label{eq:resc}
\ee
where $x_b=1-\frac{z}{m_b}$ is the momentum fraction
carried by the $b$ quark. The peak of $\phi_\infty(z)$
is localized at comparatively large 
momenta, which is consistent with the large value
$\bar{\Lambda}\approx 1.0\,GeV$
that was found numerically from the
`binding energy' and from the momentum carried by
the light degrees of freedom.

\begin{figure}
\unitlength1.cm
\begin{picture}(15,8.5)(-15.5,0.8)
\includegraphics{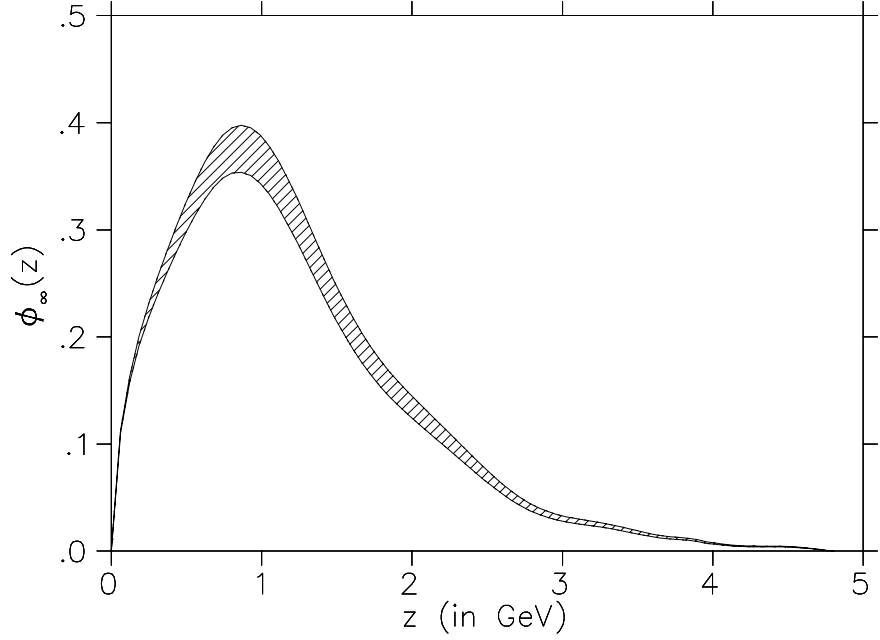}
\end{picture}
\caption{Heavy meson distribution amplitude
Eq.~(\ref{eq:resc}) in the heavy quark limit. The shaded
area reflects numerical uncertainties in the
extrapolation $m_b \to \infty$.}
\label{fig:Bdist}
\end{figure}
For phenomenological applications \cite{moments}, it 
is most important to know the moments, where one finds
\bea
\frac{\int_0^\infty dz \phi_\infty(z) z}
{\int_0^\infty dz \phi_\infty(z)}&\approx& 1.22\, {\rm GeV} 
\approx 1.2\,
\bar{\Lambda} \nonumber\\
\frac{\int_0^\infty \frac{dz}{z} 
\phi_\infty(z)}
{\int_0^\infty dz \phi_\infty(z)}&\approx& 1.51\, {\rm GeV}^{-1} 
\approx 1.5\, \bar{\Lambda}^{-1} \ .
\eea

\begin{figure}
\unitlength1.cm
\begin{picture}(15,9)(-15.5,0.8)
\includegraphics{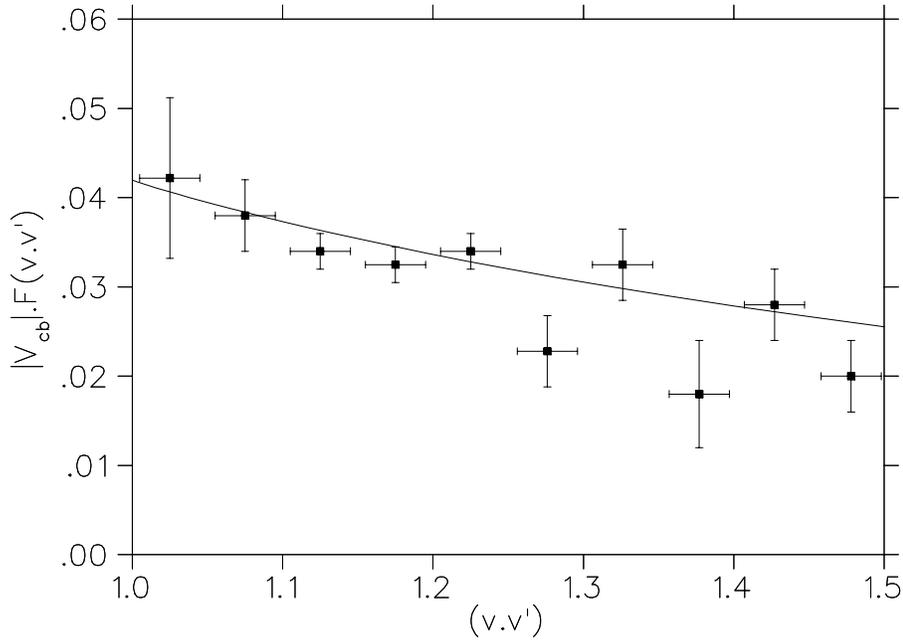}
\end{picture}
\caption{Isgur-Wise form factor Eq.~(\ref{eq:iwff}) in the heavy quark
limit. The experimental data is from Ref. \cite{exp:iwff}.}
\label{fig:iwff}
\end{figure}
Another important observable in $B$-physics is the
Isgur-Wise (IW) form factor \cite{isgur}, because of its use for 
extracting the CKM matrix element $V_{bc}$ from 
decays like $B\rightarrow \bar{D}^*l \nu$.
We work in the limit $m_c$, $m_b \rightarrow \infty$,
where matrix elements like
$\langle B^\prime |\bar{b}\gamma^\mu b|B\rangle$,
$\langle D^* |\bar{c}\gamma^\mu b|B\rangle$ are
all described by the same universal form factor
\be
\langle B^\prime |\bar{b}\gamma^\mu b|B\rangle
= m_B \left({v^\prime}^\mu+v^\mu\right)
F(v\cdot v^\prime) \ ,
\ee
with $v$ the velocity of the heavy quark.
For $m_b, m_c \gg \Lambda_{QCD}$, heavy quark pair
creation is suppressed, i.e. the relevant matrix 
element is diagonal in Fock space and an overlap 
representation exists for $F(v\cdot v^\prime)$ \cite{sudip2} 
\be
F(v\cdot v^\prime) = F^{(2)} (v\cdot v^\prime) + F^{(3)} (v\cdot v^\prime)
\ ,
\label{eq:iwff}
\ee
where 
\bea
F^{(2)} (x)&=& \frac{2}{2-x}\sum_s \int^1_x dz
\psi_s(z)\psi_s^*\left(\frac{z-x}{1-x}\right)
\\
F^{(3)} (x)&=&\frac{2}{2-x}\frac{1}{\sqrt{1-x}}\sum_s
\int_x^1dz
\int_0^{1-z}dw\psi_s(z,w)
\psi_s^*\left(\frac{z-x}{1-x},\frac{w}{1-x}\right) \ .
\nonumber
\eea
If the heavy quarks were not infinitely heavy then
pair creation terms, i.e. contributions where the
current acts off-diagonally in Fock space, 
would contribute to form factors in addition to 
these overlap terms. In fact, the natural suppression
of such terms in the limit where $b$ quarks are
static was one of the motivations for treating them
as static sources in Ref. \cite{sudip2}. 

Results for the shape of the IW form 
factor, obtained from the numerically determined 
eigenstates on the transverse lattice, are shown in 
Fig.~\ref{fig:iwff}. The overall normalisation, 
which involves the CKM
matrix element $V_{bc}$, was adjusted to agree with
the data near zero recoil $v\cdot v^\prime=1$.
As one can see in Fig.~\ref{fig:iwff}, the slope and
shape of the nonperturbative transverse lattice 
results are consistent with experimental results.
The functional form of the form factor obtained on
the transverse lattice (which is also shown in Fig.~(\ref{fig:iwff}) 
reads 
\bea
F(v\cdot v^\prime) = 1- 1.40 (v\cdot v^\prime-1)
+ 1.72 (v\cdot v^\prime-1)^2 
- 0.20 (v\cdot v^\prime-1)^3
- 1.82 (v\cdot v^\prime-1)^4 \ .
\eea

Although theoretical constraints on the shape and
slope of the form factor are very useful for
extrapolating the experimental data to zero recoil,
it would also be very useful to determine the
$\frac{1}{m_{b}^2}$ corrections at zero recoil 
nonperturbatively, since those corrections represent
a major source of uncertainty in extracting $V_{cb}$
from the data. In order to estimate those corrections
on the transverse lattice, it is first of all 
necessary to allow the heavy quarks to propagate.
The most straightforward procedure would be to 
repeat the same
method that was used to determine the hopping
parameters for the light quarks.
A less trivial complication will be the inclusion of
those terms in the $b\rightarrow c$ current operator
that are off-diagonal in the Fock space, i.e. which
correspond to a virtual $\bar{b}c$ meson in the
intermediate state.

\section{Summary and Outlook}
Light-cone variables allow  a very physical 
approach towards a nonperturbative first principles
description of many high-energy scattering 
observables, such as parton distributions measured in
deep-inelastic scattering experiments.
More generally, quantization of QCD on a null plane offers many advantages
when tackling relativistic bound state problems.
The transverse lattice is an extremely powerful
framework to formulate and numerically solve QCD 
quantized in this way. Keeping the longitudinal 
$x^\pm$ directions continuous preserves manifest 
boost invariance in one direction, while employing a
spatial lattice regulator together with flux 
link-fields in the transverse direction
maintains some gauge invariance. The transverse 
lattice thus combines advantages of lattice gauge 
theories with those of light-cone quantization.
In the colour-dielectric formulation of transverse 
lattice QCD, linearized link field variables 
facilitate the formulation of an approximation scheme
for the light-cone Hamiltonian which realises a
constituent approach to hadrons.
In this work, we have reviewed  results from studies
of glueballs and mesons within such a framework.

For the effective link-field interactions in 
pure 
gauge theory,
a colour-dielectric  expansion was made that included all terms up to
fourth order that are consistent with the
symmetries that are not broken by the transverse 
lattice. In the large $N$ limit this theory dimensionally
reduced to $1+1$ dimensions, greatly simplifying the calculation.
The  coefficients (`coupling 
constants') in this expansion were
determined by seeking regions of enhanced Lorentz
symmetry in coupling constant space. That way,
the light-cone Hamiltonian for pure glue can be accurately determined
from first principles, using only the string tension
as input to fix the overall scale.
Empirically, a one-dimensional subspace in coupling
constant space --- the renormalised trajectory ---
is found along which Lorentz symmetry is greatly
enhanced and physical observables are nearly 
constant. Along this trajectory, observables are
evaluated and numerical results for 
$N \rightarrow \infty$ 
glueball spectra obtained on the transverse lattice 
agree with extrapolations of finite $N$ Euclidean
lattice gauge theory calculations. These results not
only provide strong evidence for the validity of the 
$\frac{1}{N}$ expansion of glueball masses, but
at the same time provide an important consistency 
test for the colour-dielectric formulation of gauge 
theories on a transverse lattice. The corresponding 
light-cone wavefunctions suggest much more complex 
glueball structure than naive extrapolation of 
constituent gluon ideas would indicate, even at 
resolution scales of order 1 GeV.

Calculations with fermions have so far only been
performed in a truncated Fock space basis,
where no more than one link quantum was allowed.
Since this implies that the transverse separation
between $q$ and $\bar{q}$ in a meson can never
exceed one lattice spacing, i.e. $0.5-0.7$ fm  in
the present calculations, this is clearly too
little separation between $q$ and $\bar{q}$
to accomodate event the smallest
mesons, with an rms radius of about $0.7$ fm.
Therefore, it is not very surprising that the
numerical results for mesons exhibited a larger
violation of Lorentz symmetry than 
in the studies of glueballs.

In the glueball calculations on the transverse 
lattice the only input parameter was the confinement 
scale. Ideally, an extension of these calculations 
to mesons should have used only quark masses as 
additional parameters. However, because of the 
severe approximation of the Fock space involved and 
because the leading approximation of the Hamiltonian 
does not obviously incorporate spontaneous chiral 
symmetry breaking effects properly, one 
phenomenological input parameter (a chiral symmetry 
breaking scale) had to be used in order to obtain 
reasonable solutions.

Despite these drawbacks, qualitatively interesting results 
for the structure of mesons were obtained in the one-link
approximation.
For example, the pion distribution amplitude was
found to be single humped and its shape was not far 
from the asymptotic shape, although the normalisation, set by
$f_{\pi}$, is much larger than the experimental value. About $25\%$ of the
momentum of $\pi$ and $\rho$ mesons is carried by
gluonic degrees of freedom (the link fields) at the
transverse resolution scale of order 1 GeV. This is somewhat smaller
than experiment. The quantitative discrepancies are all consistent
with the fact that, because of probability conservation, the 
one-link approximation overestimates the pure $\bar{q}q$ contribution
to Fock space. About half
the $\rho$ meson helicity is found to be carried 
by quark and antiquark helicity \cite{sd:mes}. The rest resides in
orbital angular momentum and gluon spin.

Including a static heavy quark into the above
formalism is straightforward and requires no 
additional parameters in the Hamiltonian.
The parameters in the effective Hamiltonian
can be taken from the light meson and glueball
calculations. 
Both $\bar{\Lambda}$, the binding 
energy of B-mesons, and the decay constant $f_B$
come out relatively large compared to other 
calculations and phenomenological models,
although the discrepancy is less than what it
is for $f_\pi$.  Excellent agreement between the 
calculated slope of the Isgur-Wise form factor 
and experimental data from $B\rightarrow D^*$
decays is obtained.

For many phenomenological applications, it will be necessary  to
extend the analysis to baryons; at the time of writing, no
comprehensive work has yet been performed on the transverse lattice,
although in principle it is straightfoward. Since this will be a very
important direction for the future, let us briefly indicate the
new features that will arise. 
In the studies of mesons presented in 
Section \ref{sec:ferm}, we formally made use of a 
large $N$ approximation when classifying operators,
in order to be consistent with the pure-glue studies.
However, in a Fock space truncated to $q\bar{q}$
and $q M\bar{q}$, numerical results are independent
of $N$. For example, non-planar diagrams which would 
involve crossing of link fields would require at 
least two intermediate link fields. For baryons, of 
course, the value of $N$ is determined by the number 
of quarks in the valence component.
For finite $N$, there is an additional class of
operators, compatible with the (gauge) symmetries on 
the transverse lattice, that one should consider;
namely,  operators involving the determinant of the 
link fields. For $N=3$, such a term 
$\propto {\rm Re \ det}[M]$ is cubic in the fields 
and, since we have otherwise included all possible 
terms up to $4^{th}$ order, we should include such 
a term in the renormalised Hamiltonian
\be
P^-_{det}\equiv \sum_{\bf x}
\int dx^-{\cal H}_{det}(M) = 
\sum_{\bf x}\int dx^- \sum_{r}{\rm det}
[M_r] + {\rm det}[M_r^\dagger]\ .
\label{eq:det}
\ee
All other additional operators that one can include 
are of higher order.
The richer Fock space structure, as well as the 
additional mixing of Fock sectors induced by the 
determinant interaction Eq.~(\ref{eq:det}), requires 
previous ($N=\infty$) calculations to be repeated in 
order to redetermine the coefficients of all terms 
in the renormalised Hamiltonian.
One of the most important roles played by 
${\cal H}_{det}$ 
is in the lattice propagation of baryons, which is 
also crucial for the spin splitting between the 
nucleon  and the $\Delta$.

Compared to many other techniques, the transverse lattice is still
largely uninvestigated. The next applications in hadronic physics
will require  the extension of Fock space 
truncation beyond one link to allow lighter mesons 
to expand nearer their physical size, 
the study of 
baryons, 
finite quark mass corrections
in $B$ meson studies, 
and eventually the inclusion of $\bar{q}q$
pair production effects. 
At the formal level, it  will be interesting to see 
if symmetry (including parity \cite{mb:parity}) 
is enough to fix all parameters in 
the quark sector and beyond the leading order
of the colour-dielectric expansion. In the quark 
sector, this ultimately entails a first principles
understanding of how spontaneous chiral symmetry 
breaking 
manifests itself in the renormalised Hamiltonian. 
Even if symmetry is not enough, perhaps only a discrete 
phenomenological input needs to be introduced
to determine all couplings accurately. The predictive power is 
still enormous, since entire functions of momentum 
are determined. 
Although a typical coarse transverse lattice  Hamiltonian contains many
parameters, (almost) all of them are determined by
symmetry alone. 

{\bf Acknowledgments} M.B. was supported by the
Department of Energy (DE-FG03-96ER40965)
and by Deutsche Forschungsgemeinschaft and would 
like to thank W.Weise and the nuclear
theory group at TU M\"unchen for their hospitality
while finishing this manuscript.
S.D. was supported by
the Particle Physics and Astronomy Research Council grant
no. GR/LO3965.

\appendix

\section{Miscellaneous Remarks about Fermions}
There are a number of subtle issues related to
fermions within the light-cone framework that need to
be addressed also by transverse lattice calculations.
These are issues that arise because of small $k^+$
divergences caused by the particular momentum 
dependence of the coupling of fermions to 
link-fields. Of course, small $k^+$ divergences 
also arise in the pure glue formulation of the
transverse lattice. However, in that case it seems that
adding the appropriate one loop mass counterterms
(often called self-induced inertias within the
light-cone framework) takes care of these problems.
Furthermore, if one maintains gauge invariance 
many divergences at small $k^+$ cancel among each
other.
When fermions are included, new small
$k^+$ divergences arise from Yukawa-type couplings to
the transverse gauge field degrees of freedom. Below
we discuss the two most important complications
and ways to deal with them. 

\subsection{\it Momentum dependent mass counterterms}
The coupling of quarks to link fields on the transverse
lattice has a longitudinal momentum dependence that 
is similar to the one for Yukawa theories, e.g.
\be
{\cal H}^{(3)} 
\propto i \kappa \left(\frac{1}{k^+_{in}}-
\frac{1}{k^+_{out}}\right)\frac{1}{\sqrt{k_{b}^+}},
\label{three}
\ee
where $k^+_{in}$ and $k^+_{out}$ are initial and final 
fermion momenta, and $k_{b}^+$ the boson momentum.
In second order perturbation theory, this type
of interaction leads to logarithmically 
divergent self-energies for the fermion. For a fermion of momentum
$p^+$ the shift in light-cone energy is
\be
\Delta^{(2)}P^- \sim \kappa^2
\int_0^{p^+} \frac{\left(\frac{1}{k^+}
-\frac{1}{p^+}\right)^2}
{P^{-} - \frac{m^2}{2k^+} - \frac{m_{g}^2}
{2(p^+-k^+)}}\ dk^+ \ .
\ee
Because of this divergence, a regulator for small
longitudinal momenta needs to be introduced and
counterterms need to be added to compensate the
divergence.

The issues that arise in this context are the
Fock sector and momentum dependence of mass 
counterterms. The Fock sector dependence is easy
to understand. For example, it one truncates the
Fock space above states with one boson (link
quantum) `in flight', then the above mass
renormalization only effects quark masses in the
sector without the boson, since only in that sector
would the allowed Fock space permit adding another
boson. Therefore, within the colour dielectric
expansion one needs to treat the masses of the
quarks in different Fock sectors as independent
parameters. Of course, ultimately they are not
independent, but the relation between the masses
in different Fock sectors is dynamical and 
nonperturbative. 

A more subtle issue is the momentum dependence
of mass counterterms. If one employs a regularization
scheme that breaks the manifest boost invariance
in the longitudinal direction, then one should 
not expect momentum independent mass counterterms
 either. A popular example for 
such a regularization scheme is DLCQ \cite{dlcq}, where  
(anti-) periodicity
conditions in the $x^-$ direction lead to momenta
that are integer (or half integer) multiple of
some momentum unit. In such a scheme the self mass
$P^+\Delta^{(2)}P^-$ depends on $P^+$ because
the phase space that is allowed for the intermediate
states does depend on the value of integer momentum
of the quark.
In order to properly compensate for this fact, it is
in general necessary to introduce a momentum 
dependent mass counterterm in DLCQ.
\footnote{Although contemporary DLCQ calculations
usually include the momentum dependent 
${\cal O}(\kappa^2)$ mass counterterm, higher order
momentum dependent counterterms have usually not been 
 treated in this framework.}
Transverse lattice calculations with fermions that 
have been performed so far (using DLCQ as well as
continuum basis functions) included a maximum of one
 link
quantum in flight. As a result the momentum
dependence of the self-energy is easily calculable
and the divergent part is cancelled by adding the
so-called self-induced inertia terms \cite{dlcq}.
However, future calculations which
include more Fock components will have to deal with
this issue. A couple of options might be practical:
\begin{itemize}
\item Using regulators that maintain manifest boost
invariance in the longitudinal direction. For 
example, if one expands states using complete sets of
continuous basis functions, one ends up with matrix 
elements in the Hamiltonian that involve overlap 
integrals of basis functions and operators.
Those integrals can be regulated in a boost invariant
manner, for example by introducing cutoffs on ratios of 
momenta. The integrand is mutliplied by a 
factor $\left( k^+_{after}/
k^+_{before} \right)^\varepsilon$, where 
$k^+_{before/after}$ is the quark momentum
before/after emmiting a link quantum and 
$\varepsilon$ is some small number. Calculations
are performed at finite $\varepsilon$ and observables
are extrapolated to $\varepsilon \rightarrow 0$ at
the end of the calculation.
\item One of the reasons for the popularity of DLCQ
is that it is relatively easy to develop computer
code for non-perturbative calculations. This 
advantage also applies to transverse lattice 
calculations and it is therefore worthwhile to
investigate if one can modify DLCQ in such a way that
one can cope with the momentum dependent 
counterterms. 
In more 
ambitious studies that include  higher Fock 
components one may imagine allowing the bare mass
(the term in the Hamiltonian) to depend on the
integer momentum and to fix this momentum dependence 
of this bare mass nonperturbatively by some means. For example,
one could demand 
that hadron masses are approximately independent
of the total (integer) momentum of the hadron.
\end{itemize}

\subsection{\it Vertex mass renormalisation}
In the common light-cone quantization procedure for field
theories involving fermions one normally uses the
constraint equation to eliminate
the non-dynamical spinor components $\Psi_{(-)}$.
Even if the original Lagrangian involves only
couplings of fermions to bosons that are   linear
in the boson field, the effective Lagrangian for
the dynamical component $\Psi_{(+)}$ usually also
involves terms where the fermions couple to the
second power of the boson field 
(Fig.~\ref{fig:inst}a).
\begin{figure}
\unitlength1.cm
\begin{picture}(15,7.5)(1.5,19)
\includegraphics{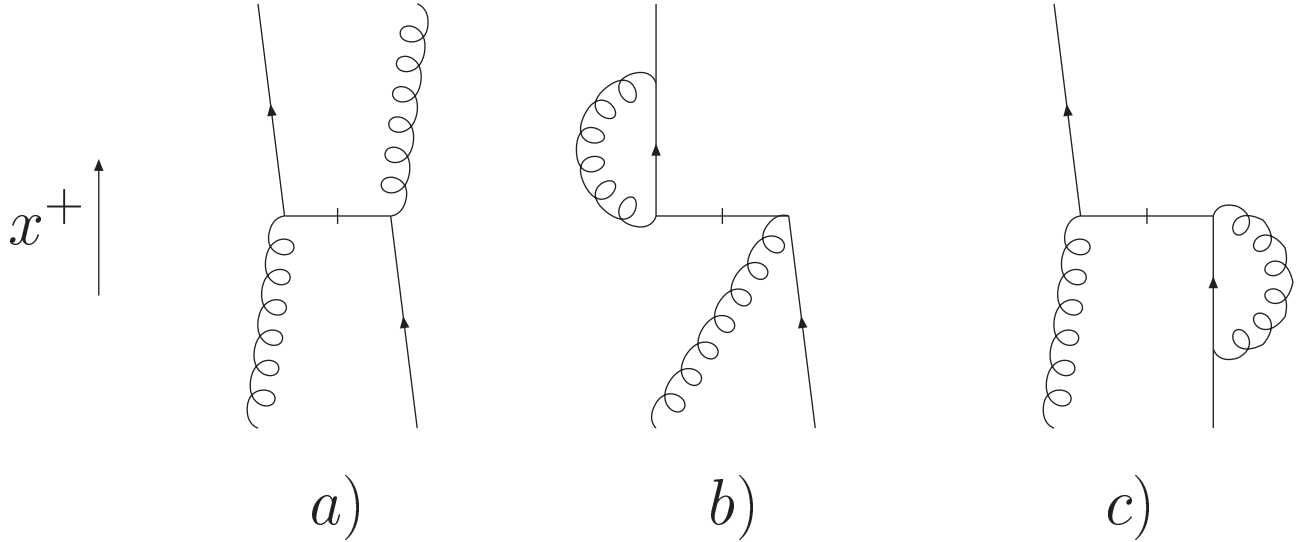}
\end{picture}
\caption{a) interaction for the dynamical
fermion component that is induced by eliminating
the constrained component of the fermions 
$\Psi_{(-)}$. The slashed horizontal ($x^+$ instantaneous)
fermion line represents the eliminated degrees of 
freedom. b) higher order correction to the
three point function (boson absorption), which 
involves this four point coupling. c) same as b),
but with a different time ordering of the 
interactions.}
\label{fig:inst}
\end{figure}
In Ref. \cite{mb:dvmg}, it was shown that this
four-point interaction can lead to large 
non-perturbative corrections of the three point 
coupling of the fermions to bosons. Loop
corrections, such as the ones depicted in Fig.~\ref{fig:inst} 
b and c, may effectively enhance
the three-point couplings. This may have important
effects. For example, the higher order corrections to the tree level  
three point interaction Eqn.~(\ref{three})
lead to an effective
enhancement $\kappa\rightarrow \kappa_{ren}$, where
$\kappa_{ren}$ can differ significantly from 
$\kappa$.
These fundamental observations have several
important practical consequences.
\subsubsection{\it Three point vertex in the chiral
limit}
In the chiral limit of QCD, helicity is conserved
in perturbation theory. This reflects itself in the
fact that the three point quark helicity flip
vertex in the canonical light-cone Hamiltonian for QCD is 
linear in the quark mass. If one would simply set
this term to zero in the chiral limit, the dynamics
would become completely independent on the quark 
helicity --- yielding  $\pi$'s that are
degenerate with $\rho$'s and nucleons that are
degenerate with $\Delta's$. The resolution to this
apparent paradox lies in the abovementioned
observation that non-perturbative effects (from very
high Fock components) can give rise to an 
enhancement of $\kappa$ such that the helicity-flip
vertex amplitude for dressed quarks may not vanish
--- even if it does so at any finite order in
perturbation theory.

In a truncated Fock space, such nonperturbative 
enhancements would be suppressed and it is thus
necessary to allow the three-point coupling to
remain finite in the chiral limit in order to
mimic this enhancement which would be present 
without Fock space truncation. 
This observation explains why it is fallacious
in calculations with Fock space truncation to
argue that the chiral limit corresponds to the 
subspace in parameter space where the quark 
helicity flip coupling for the quarks vanishes.
\subsubsection{\it Fock sector dependent
vertex renormalisation}
Imposing a cutoff on the Fock space also leads to an
asymmetric treatment of loop corrections to the
vertex in the initial and final state.
For example, consider the process of boson absorption
by a quark (Fig.~\ref{fig:inst} b,c).
If the Fock space truncation is such that only
one boson is allowed `in flight', then vertex
corrections in the initial state (Fig.~\ref{fig:inst}
c) are suppressed. The practical consequence of
such an asymmetric renormalisation is that
one generates an effective three-point coupling that
involves fermion momenta in an asymmetric way \cite{mb:dvmg,hala}
\be
{\cal H}^{(3)} 
\propto i \kappa \left(\frac{1}{k^+_{in}}-
\frac{1}{k^+_{out}}\right)\frac{1}{\sqrt{k_{b}^+}}
\longrightarrow
 i \left(\frac{\kappa}{k^+_{in}}-
\frac{\kappa_{ren}}{k^+_{out}}\right)
\frac{1}{\sqrt{k_{b}^+}}\ .
\ee
Since, as we emphasized above, the enhancement
$\kappa_{ren}$ vs. $\kappa$ can be significant, one 
would in
principle have to introduce Fock sector dependent
three point couplings,  for example
\be
{\cal H}^{(3)}
\to i\left(\frac{\kappa_{in}}{k^+_{in}}-
\frac{\kappa_{out}}{k^+_{out}}\right)
\frac{1}{\sqrt{k_{b}^+}} \ ,
\ee
where `in' and `out' are defined in reference to the
absorbed boson, and treat them as independent 
parameters.

In addition to this proliferation of three-point couplings, that is
necessary once four-point interactions of Fig.~\ref{fig:inst}
are included, there are many other types of 
four-point interaction between quarks and link fields 
allowed.
Therefore, before enlarging the space of couplings once should
have some physical criterion for selecting these four-point couplings.
In all transverse lattice calculations to date, only the four-point interaction
mediated by the instantaneous gluons $A_{+}$ has been included, since
this is responsible for confinement.

Given the complications that are introduced
by cutting off the Fock space, it becomes clear
that one should ultimately avoid such truncations.
Of course, strictly speaking it is impossible
to work without any Fock space restrictions. 
However, with appropriate cutoffs on energy
(or invariant mass) differences at each vertex,
one may succeed to achieve that  truncating
very high energy Fock components has only a 
negligible impact on the low energy dynamics.
If there is a cutoff $\Delta E$ on 
`allowed' energy (or invariant mass) differences
at a vertex, then very high energy states that have an energy
that is several $\Delta E$ higher than the ground 
state are connected to the ground state only
by several orders of the interaction.
Therefore, cutting off such high energy states
should have only little effect on low energy
dynamics and one should thus be able to 
ignore problems that may in principle arise due to 
Fock space truncation.

\end{document}